\documentclass[trackchanges, twocolumn]{aastex7}

\usepackage{gensymb}
\usepackage{xcolor}
\usepackage{booktabs, multirow}
\usepackage{tablefootnote}
\usepackage[print-unity-mantissa=false, separate-uncertainty=true, multi-part-units=single, range-units=single]{siunitx}
\DeclareSIUnit\erg{erg}
\DeclareSIUnit{\angstrom}{\textup{\AA}}

\begin{document}

\title{The First Dedicated Survey of Atmospheric Escape from Planets Orbiting F Stars}

\correspondingauthor{Morgan Saidel}
\email{msaidel@caltech.edu}

\author[0000-0001-9518-9691]{Morgan Saidel}
\affiliation{Division of Geological and Planetary Sciences, California Institute of Technology, Pasadena, CA 91125, USA}
\email{msaidel@caltech.edu}

\author[0000-0003-2527-1475]{Shreyas Vissapragada}
\affiliation{Carnegie Science Observatories, 813 Santa Barbara Street, Pasadena, CA 91101, USA}
\email{TBD}

\author[0000-0002-5375-4725]{Heather A. Knutson}
\affiliation{Division of Geological and Planetary Sciences, California Institute of Technology, Pasadena, CA 91125, USA}
\email{TBD}

\author[0000-0003-2872-760X]{Ethan Schreyer}
\affiliation{Department of Astronomy and Astrophysics, University of California, Santa Cruz, CA 95064, USA}
\email{TBD}

\author[0000-0002-0371-1647]{Michael Greklek-McKeon}
\affiliation{Earth and Planets Laboratory, Carnegie Institution for Science, Washington, DC 20015, USA}
\email{TBD}

\author[0000-0002-0672-9658]{Jonathan Gomez Barrientos}
\affiliation{Division of Geological and Planetary Sciences, California Institute of Technology, Pasadena, CA 91125, USA}
\email{TBD}

\author[0000-0002-1422-4430]{W. Garrett Levine}
\affiliation{Department of Earth, Planetary, and Space Sciences, University of California, Los Angeles, CA 90095, USA}
\affiliation{Department of Astronomy, Yale University, New Haven, CT, 06511, USA}
\email{wglevine@epss.ucla.edu}

\author[0000-0001-5097-9251]{Carlos Gascón}
\affiliation{Center for Astrophysics, Harvard \& Smithsonian, 60 Garden Street, MS-16, Cambridge, MA 02138, USA}
\affiliation{Institut d'Estudis Espacials de Catalunya (IEEC), 08860 Castelldefels, Barcelona, Spain}
\email{TBD}

\author[0000-0002-3641-6636]{George W. King}
\affiliation{Department of Astronomy, University of Michigan, Ann Arbor, MI 48109, USA}
\email{TBD}

\author[0000-0002-1417-8024]{Morgan MacLeod}
\affiliation{Center for Astrophysics, Harvard \& Smithsonian, 60 Garden Street, MS-16, Cambridge, MA 02138, USA}
\email{TBD}

\author[0009-0006-0871-1618]{Haedam Im}
\affiliation{Department of Physics, Massachusetts Institute of Technology, Cambridge, MA 02139, USA}
\email{TBD}

\author[0000-0001-9686-5890]{Nick Tusay}
\affiliation{Department of Astronomy, University of Washington, 3910 15th Avenue NE, Seattle, WA 98195, USA}
\email{ntusay@uw.edu}

\begin{abstract}

Hydrodynamic escape can strip the envelopes of close-in exoplanets, but most observations of atmospheric mass loss to date have been confined to planets orbiting K and M dwarfs.
A growing body of detections of atmospheric escape from planets orbiting early-type stars indicates that they may have significantly stronger and more extended outflows than planets orbiting cooler stars. However, it is unclear whether this limited sample of planets is representative of all gas giants orbiting early-type stars. 
Motivated by this question, we initiated the first dedicated survey of atmospheric escape from gas giants orbiting F stars in order to understand how their distinct radiation environments shape planetary outflows. We observed ten transits of six planets in an ultra-narrowband filter centered on the metastable helium line using Palomar/WIRC. We report strong ($>3\sigma$) detections of atmospheric escape for WASP-12~b and WASP-180~A~b, tentative ($>2\sigma$) detections for WASP-93~b and HAT-P-8~b, and non-detections for WASP-103~b and KELT-7~b. We fit these measurements with a 1D Parker wind model to derive corresponding mass-loss rates, and combine our results with literature measurements to obtain an updated picture of mass loss from planets orbiting early-type stars. Our results indicate that the observed variation in mass-loss rates can be explained by a combination of Roche filling factor and XUV luminosity, and disfavors NUV-driven escape models.

\end{abstract}

\section{Introduction}\label{sec:intro}

Exoplanets on close-in orbits are exposed to intense irradiation from their host stars, powering mass loss that can catastrophically alter their evolution. Mass loss is thought to play a central role in sculpting the population of close-in exoplanets, and is likely responsible for the creation of both the `evaporation valley' \citep[a period-dependent separation between rocky super-Earths and sub-Neptunes; e.g.,][]{FultonPetigura2017,Rogers2024} and the lower boundary of the `Neptune desert' \citep[a deficit of Neptune-sized planets on close-in orbits; e.g.,][]{OwenLai2018}. To quantify the effect of atmospheric mass loss on the population of close-in planets, we must be able to accurately predict mass-loss rates for a wide range of stellar environments. By measuring the amount of absorption in strong atomic lines including Ly$\alpha$, H$\alpha$, and the \SI{1083}{\nano\meter} line of metastable helium (He$^*$) when the planet passes in front of its host star, we can detect atmospheric outflows from close-in transiting planets and characterize their corresponding mass-loss rates, outflow geometries, and kinematics \citep[see reviews by][]{Owen2019, DosSantos2023}. This in turn allows us to test predictions from current mass-loss models, and guide the development of a new generation of improved models. 

To date, there have been more than 25 published detections ($>3\sigma$) of atmospheric escape using Ly$\alpha$, H$\alpha$, and He$^*$
\citep[for a recent list, see the review by][]{Kempton2024}. Unlike Ly$\alpha$, which can only be observed from space and suffers from interstellar medium absorption in the line core, He$^*$ and H$\alpha$
are easily accessible from the ground. H$\alpha$ requires a significant population of excited hydrogen, and as a result the small number of H$\alpha$ detections that have been published to date are limited to the subset of planets with the highest XUV fluxes, many of which orbit early-type stars \citep[e.g.,][]{Jensen2012,Yan2018,CasasayasBarris2019,Wyttenbach2020,CzeslaLampon2022,Yan2021,Yan2022}.

\begin{figure}[h!]
\centering
\includegraphics[width=0.45\textwidth]{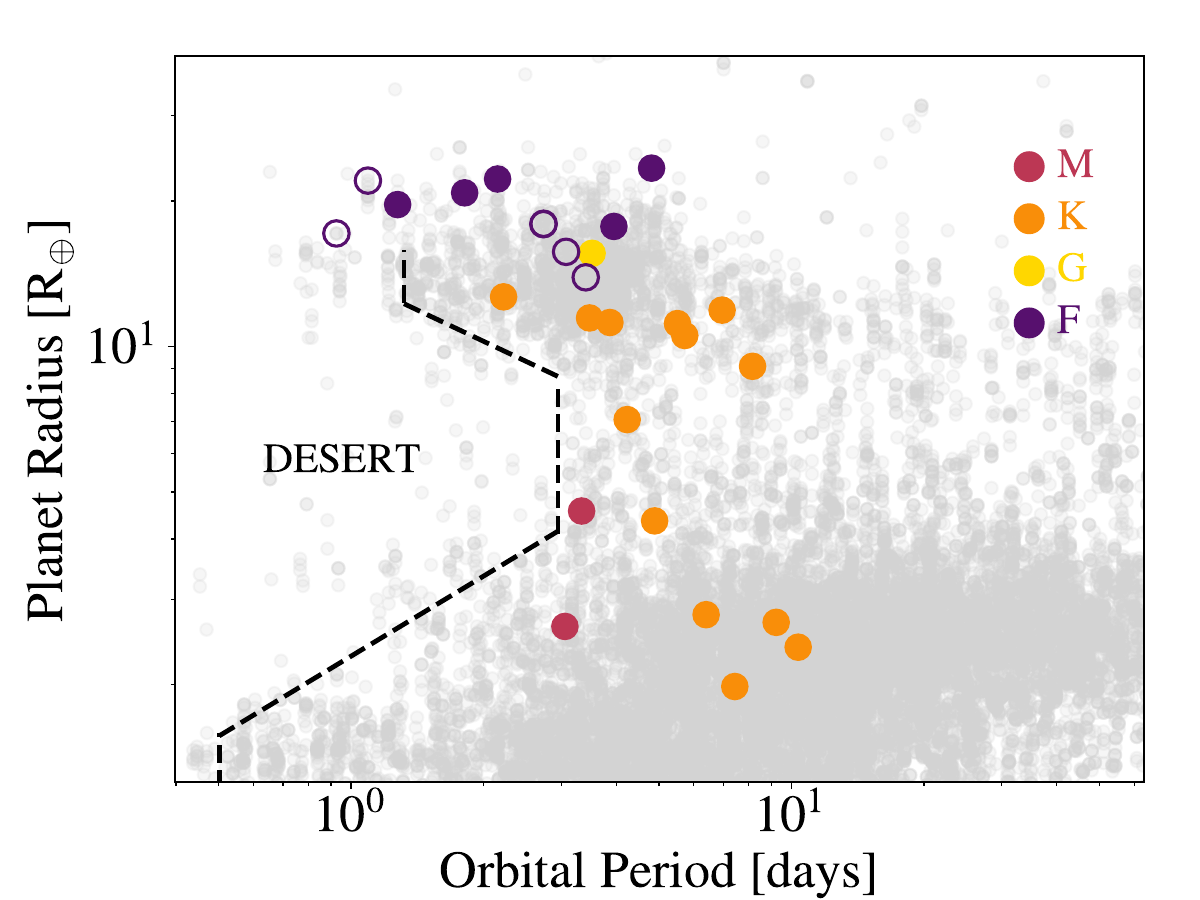}
\caption{Confirmed transiting planet radii and periods drawn from the NASA Exoplanet Archive on August 26th, 2025 \citep{AkesonChen2013, ps}. 
Closed colored circles represent He$^*$ detections of atmospheric escape as a function of spectral type. Open circles denote survey targets. The dashed lines indicate the Neptune desert boundary from \citet{Castro-GonzalezBourrier2024}. }
\label{fig:nepdes}
\end{figure}

In contrast, He$^*$ acts as a strong tracer of atmospheric outflows over a relatively broad range of temperatures and mass loss rates \citep{OklopcicHirata2018}. This has made He$^*$ the most successful method for detecting atmospheric outflows, with over 20 published detections to date (see Figure~\ref{fig:nepdes}). Most of these detections have been for planets orbiting K dwarfs, likely because He$^*$ is populated by extreme ultra-violet (EUV) radiation and
depopulated by mid-ultra-violet (mid-UV) radiation, and K dwarfs provide the optimal radiation balance \citep{OklopcicHirata2018, Oklopcic2019}. However, an ever-growing list of He$^*$ detections for planets orbiting F stars \citep[HAT-P-32~b, HAT-P-67~b, WASP-76~b, WASP-94~A~b, WASP-121~b;][]{CzeslaLampon2022, ZhangMorley2023,BelloArufe2023AJ,Gully-SantiagoMorley2024,Czesla2024,Mukherjee2025,AllartCoulombe2025} indicates that hotter stars can also be suitable targets for mass loss studies using this feature. For planets whose outflows are also detectable in H$\alpha$, simultaneous measurements of the excess absorption in both H$\alpha$ and He$^*$ can be used to directly constrain the H/He ratio of the outflow \citep[e.g.,][]{Yan2022,CzeslaLampon2022,Yan2024}. 

There are currently six planets orbiting F or A stars with published mass loss rates derived from He$^*$, H$\alpha$, or a combination of the two. Four of these planets (HAT-P-32~b, HAT-P-67~b, KELT-9~b, and WASP-121~b) exhibit unusually high mass-loss rates ($>$\SI{1e12}{\gram\per\second}), and three (HAT-P-32~b, HAT-P-67~b, WASP-121~b) have very spatially extended outflows \citep[$>50\times$ the planet radius;][]{ZhangMorley2023,Gully-SantiagoMorley2024,Czesla2024}. The reported mass loss rate for WASP-33~b is also just barely below this threshold \citep{Yan2021}.  In contrast, WASP-94~A~b appears to have a more typical mass loss rate of \SI{7e10}{\gram\per\second} \citep{Mukherjee2025}, and \cite{Bennett2023AJ} reported a $3\sigma$ upper bound on the mass-loss rate of WASP-48~b of \SI{2.5 e11}{\gram\per\second}. WASP-48 is an old ($\sim$7.9 Gyr) slightly evolved F star and will have a lower XUV flux than younger, less evolved stars of the same spectral type, likely contributing to the non-detection and low upper bound on the mass loss rate reported in \citet{Bennett2023AJ}. WASP-76~b does not currently have a reported mass-loss rate for its He$^*$ absorption signal \citep{OrellMiquel2025}. 

Given this limited sample size, it is unclear whether the strong outflows from HAT-P-32~b, HAT-P-67~b, KELT-9~b, WASP-33~b, and WASP-121~b are broadly representative of most gas giants orbiting early-type stars, or if they instead reflect characteristics particular to these systems. For example, most of these planets fill $40-50\%$ of their Roche lobes
\citep[`Roche filling factor''; calculated based on their white light radius and using the Hill radius as an approximation for the size of the Roche lobe, as is done in][for ease of comparison]{MacLeodOklopcic2025}, making them particularly susceptible to enhanced atmospheric loss via Roche lobe overflow \citep[RLO; e.g.,][]{Jackson2017,Koskinen2022}.

Alternatively, the planets with the highest measured mass loss rates might have enhanced X-ray and EUV luminosities that could drive increased mass loss. This is supported by \textit{XMM-Newton} observations of HAT-P-32 and WASP-121, which revealed X-ray (\SIrange[range-phrase={--}]{5}{100}{\angstrom}) luminosities of \SI{2.3e29}{\erg\per\second} \citep{CzeslaLampon2022} and \SI{1.3e29}{\erg\per\second} \citep{Czesla2024}, respectively. However, KELT-9 has an upper bound of \SI{3e29}{\erg\per\second} \citep{SanzForcada2025}, and there are currently no published X-ray or EUV constraints for HAT-P-67, WASP-33, or WASP-94~A.

Lastly, it is also possible that the high NUV luminosities of early-type stars might contribute additional power that could drive massive outflows. \citet{GarciaMunozSchneider2019} found that the high NUV fluxes of these stars can lead to Balmer-driven escape, resulting in atmospheric mass-loss rates as much as two orders of magnitude larger than the corresponding rates from EUV and X-ray emission alone. If correct, this would suggest that gas giants orbiting early-type stars with high NUV fluxes should experience systematically higher atmospheric mass-loss rates than their counterparts around cooler stars.

\begin{figure}[h!]
    \centering
    \hspace*{-0.6cm}\includegraphics[width=0.5\textwidth]{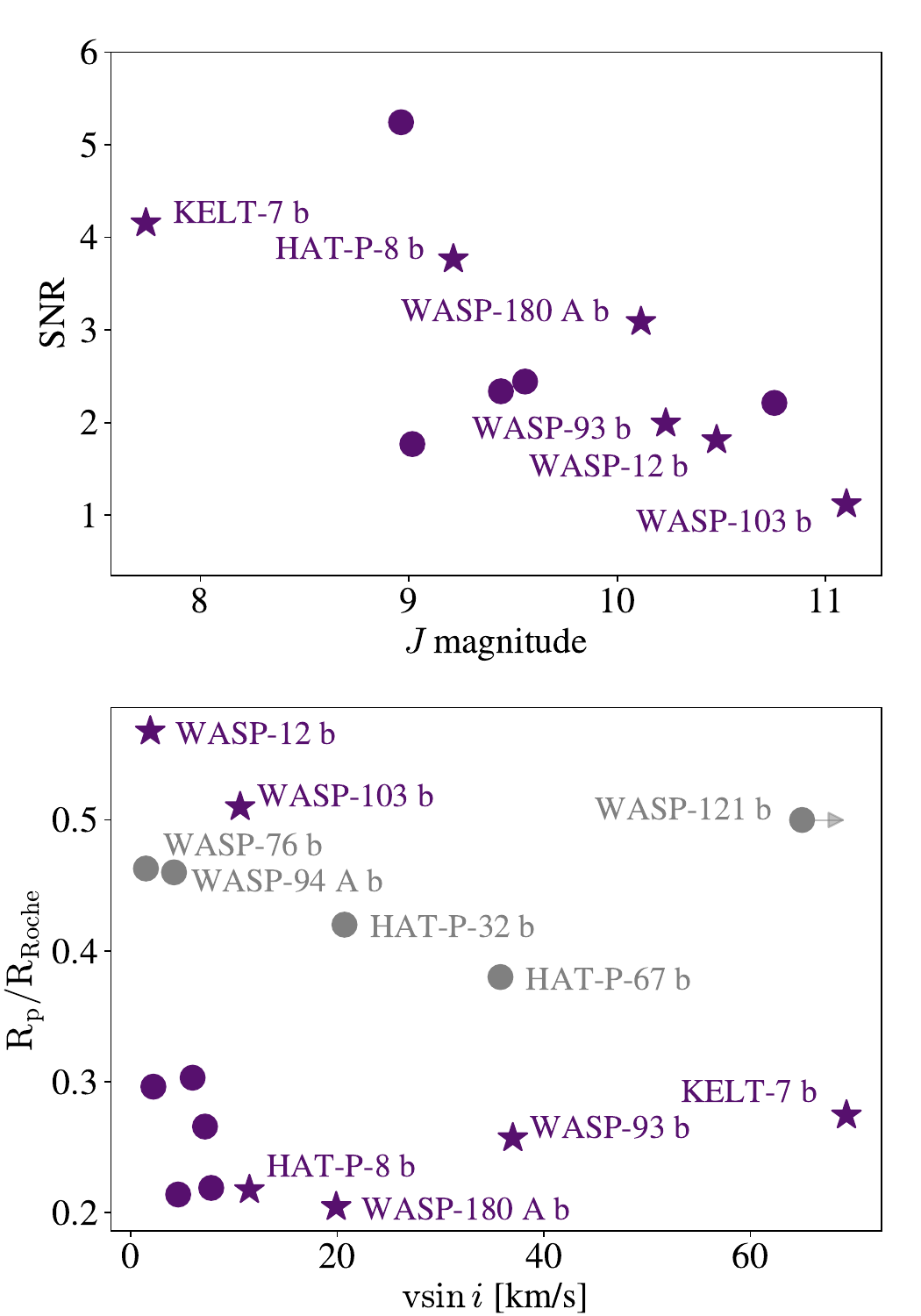}
    \caption{Top: Predicted signal-to-noise ratios (SNR) for two WIRC transit observations of each F-star planetary system observable from Palomar Observatory after excluding those with published mass loss detections. Labeled stars indicate our survey targets. The highest SNR planet is KELT-3 b, which was initially a survey target but was lost to poor weather.  Bottom: Roche filling factor versus stellar rotational velocity for all F-star planetary systems observable from Palomar Observatory (purple) and F-star planetary systems with detected outflows (gray). WASP-121 is viewed pole-on and we therefore plot the minimum equatorial stellar rotational velocity constraint from \cite{Bourrier2020} as a horizontal arrow.}
    \label{fig:targetselection}
\end{figure}

In order to quantify the importance of these various factors in determining mass-loss rates for planets orbiting early-type stars, we require additional detections of atmospheric mass loss. With this goal in mind, we designed a dedicated ground-based survey to search for atmospheric escape from six gas giants orbiting F stars: 
HAT-P-8, WASP-93, WASP-180 A, WASP-103, KELT-7, and WASP-12 (see \S\ref{sec:targetselect} and Figure~\ref{fig:targetselection} for more details on target selection). These systems have varying Roche filling factors,  projected stellar rotational velocities ($v{\sin}i_*$; see Table~\ref{table:stars} and bottom panel of Fig.~\ref{fig:targetselection}) which we used as a proxy for XUV activity, as well as varying NUV fluxes, allowing us to explore the importance of these parameters for determining the outflow properties. In Section~\ref{sec:obsdatared} we detail our target selection, observational setup and data reduction process. We detail the results of our observational survey in Section~\ref{sec:Results} and retrieve mass-loss rates based on these results in Section~\ref{sec:massloss}. We
discuss the implication of our results on the broader landscape of atmospheric escape observations in Section~\ref{sec:disc}. Finally, we summarize our conclusions in Section~\ref{sec:conc}.   

\section{Observations \& Data Reduction}\label{sec:obsdatared}

\subsection{Target Selection}\label{sec:targetselect}

We selected targets for our survey using the NASA Exoplanet Archive \citep{AkesonChen2013,ChristiansenMcElroy2025}. We identified an initial sample of transiting gas giant planets ($R>4$ $R_{\oplus}$) orbiting bright ($J < 12$) F stars ($T_\mathrm{eff}=6000-7935$~K) that are accessible from Palomar Observatory (declinations $>-20^{\circ}$). For each planet, we then calculated a predicted mass-loss rate using an energy-limited mass-loss equation \citep{CaldiroliHaardt2022}. We fixed the heating efficiency $\eta$ to 0.40, as \citet{CaldiroliHaardt2022} notes that most heating efficiencies for thermal escape are between 0.3 to 1. While high gravity planets can have lower efficiencies than 0.3, many of our targets have high Roche filling factors, meaning the atmospheres of these planets should therefore not be in a high gravity regime. We conservatively assumed a solar EUV spectrum \citep{DosSantosVidotto2022} scaled by $(R_\star/R_\odot)^2$ for every star in the sample, as there are relatively few measurements of EUV spectra for stars in this temperature range. For all planets, we estimated a thermosphere temperature ($T_0$) of the outflow based on the planet's orbital period following \cite{SalzCzesla2016}: $P >$~4 days: $T_0=6000$ K; 1~$< P \leq$~4 days: $T_0=8000$ K; $P \leq$~1 days: $T_0=10,000$ K). We combined these thermosphere temperatures and mass-loss rates to calculate the predicted He$^*$ absorption signals during transit using the open-source \texttt{p-winds} package \citep{DosSantosVidotto2022}, which models escaping planetary atmospheres as one-dimensional hydrodynamic Parker winds. We then integrated the predicted He$^*$ excess absorption spectrum for each planet over the transmission of the Palomar/WIRC He$^*$ filter using Eqn.~3 in \citet{VissapragadaKnutson2020} to obtain a predicted mid-transit excess absorption value (i.e., the difference between the expected transit depths for a planet with and without an atmospheric outflow). Lastly, we calculated the expected photometric noise and corresponding SNR for a detection of excess absorption during transit by scaling the photometric noise from previous observations \citep[e.g.,][]{VissapragadaKnutson2022, SaidelVissapragada2025} to the $J$ magnitudes of each stellar host. We rank-ordered our targets by SNR and selected the six highest SNR targets, additionally checking to ensure that our chosen sample spanned a range of $v{\sin}i_*$ (\SIrange[range-phrase={--}]{2}{70}{\kilo\meter\per\second}) and Roche filling factors ($0.20-0.55$, see Fig.~\ref{fig:targetselection}). Our six survey targets are: HAT-P-8, WASP-93, WASP-180~A, WASP-103, KELT-7, and WASP-12. We note that our survey originally included another target with a high predicted SNR, KELT-3, but the weather on our scheduled nights for this target was poor and we were unable to obtain any usable observations.

\begin{deluxetable*}{cccccccccc}[ht]
\tabletypesize{\scriptsize}
\tablecaption{Summary of target stellar and planetary properties. \label{table:stars}}
\tablehead{\colhead{Planet} &  \colhead{$R_p$ ($R_J$)} & \colhead{$M_p$ ($M_J$)} &\colhead{$R_\star$ ($R_\odot$)}  & \colhead{$M_\star$ ($M_\odot$)} & \colhead{$T_\mathrm{eff}$ (K)} & \colhead{$v{\sin}i_*$ (\unit{\kilo\meter\per\second})} & \colhead{$R_p/R_\mathrm{Roche}$} & \colhead{$\lambda$ ($^{\circ}$)}& \colhead{References}}
\startdata
HAT-P-8~b & $1.500\pm0.070$ & $1.354\pm0.035$ & $1.580^{+0.080}_{-0.060}$ & $1.280\pm0.040$ & $6200\pm80$ & $11.50\pm0.50$ & $0.23$ & $-17^{+9.2}_{-11.5}$ & L09, M11, B17 \\
WASP-93~b & $1.597\pm0.077$ & $1.47\pm0.29$ & $1.524\pm0.040$ & $1.334\pm0.033$ & $6700\pm100$ & $37\pm3$ &  $0.27$ & - & H16 \\
WASP-180~A~b & $1.24\pm0.04$ & $0.9\pm0.1$ & $1.19\pm0.06$ & $1.17\pm0.01$ & $6500\pm150$ & $19.9\pm0.6$ &  $0.20$ & $-157\pm2$ & T19\\
WASP-103~b & $1.528^{+0.073}_{-0.047}$ & $1.49\pm0.22$ & $1.436^{+0.052}_{-0.031}$ & $1.220^{+0.039}_{-0.036}$ & $6110\pm160$ & $10.6\pm0.9$ &  $0.51$ & $3\pm33$ & G14, A16 \\
KELT-7~b & $1.60\pm0.06$ & $1.28\pm0.18$ & $1.732^{+0.043}_{-0.045}$ & $1.535^{+0.066}_{-0.054}$ & $ 6789\pm50$ & $69.3\pm0.2$ &  $0.26$ & $-2.7\pm0.6$ & B15, Z16, S17 \\
WASP-12~b & $1.965^{+0.088}_{-0.087}$ & $1.465\pm0.079$ &  $1.690^{+0.019}_{-0.018}$ & $1.325^{+0.026}_{-0.018}$ & $6265\pm50$ & $1.66^{+0.8}_{-0.4}$ & $0.55$ & $59^{+15}_{-20}$ & A12, CS19, L24 \\
\hline
Literature Measurements & & & & & & & & \\
\hline
HAT-P-32~b & $1.980\pm0.045$ & $0.68^{+0.11}_{-0.10}$ &  $1.367^{+0.031}_{-0.030}$ & $1.132^{+0.051}_{-0.050}$ & $6207\pm88$ & $20.7\pm0.5$ &  $0.42$ & $85.0\pm1.5$ & A12, B17, W19\\
HAT-P-67~b & $2.085^{+0.096}_{-0.071}$ & $0.34^{+0.25}_{-0.19}$ &  $2.546^{+0.099}_{-0.084}$ & $1.642^{+0.155}_{-0.072}$ & $6406^{+65}_{-61}$ & $35.8\pm1.1$ &  $0.38$ & $2.9^{+6.4}_{-4.9}$ & Z17\\
KELT-9~b & $1.783\pm0.009$ & $2.44\pm0.70$ &  $2.178\pm0.011$ & $1.978\pm0.023$ & $9550^{+194}_{-132}$ & $111.4\pm1.3$ &  0.34 & $-84.8\pm1.4$ & G17, H19, S19\\
WASP-33~b & $1.679^{+0.019}_{-0.030}$ & $3.28\pm0.73$ &  $1.444\pm0.034$ & $1.495\pm0.031$ & $7430\pm100$ & $90\pm10$ &  0.43  & $251.2\pm1.0$ & CC10, T16\\
WASP-94~A~b & $1.72^{+0.06}_{-0.05}$ & $0.452^{+0.035}_{-0.032}$ &  $1.62^{+0.05}_{-0.04}$ & $1.45\pm0.09$ & $6153^{+75}_{-76}$ & $4.2\pm0.5$ &  0.46  & $151^{+16}_{-23}$ & NV14\\
WASP-121~b & $1.753\pm0.036$ & $1.157\pm0.070$ &  $1.458\pm0.030$ & $1.358^{+0.075}_{-0.084}$ & $6459\pm140$ & $65-120$\tablenotemark{a} &  0.50 & $87.20^{+0.41}_{-0.45}$ & D16, B20\\
\enddata
\tablenotetext{a}{This star is seen nearly pole-on, so we instead quote the estimated equatorial velocity from B20.}
\tablecomments{
The lower part of the table lists system parameters for F or A star planets with published measurements of atmospheric escape rates from the literature as a point of comparison. We do not include WASP-76~b as there is no reported mass loss constraint for this He$^*$ detection.
Roche filling factor is a derived property and refers to the fraction of the Roche lobe filled by the planet radius. $\lambda$ is the sky-projected obliquity, ``-'' signifies an unknown quantity. 
In the references column, L09 refers to \citet{LathamBakos2009}, M11 is \citet{MoutouDiaz2011}, B17 is \citet{BonomoDesidera2017}, H16 is \citet{HayCollier-Cameron2016}, T19 is \citet{TempleHellier2019}, G14 is \citet{GillonAnderson2014}, A16 is \citet{AddisonTinney2016}, B15 is \citet{BierylaCollins2015}, Z16 is \citet{ZhouLatham2016}, S17 is \citet{StassunCollins2017}, A12 is \citet{AlbrechtWinn2012}, CS19 is \citet{ChakrabartySengupta2019}, L24 is \citet{LeonardiNascimbeni2024}
, W19 is \citet{WangWang2019}, Z17 is \citet{ZhouBakos2017}, G17 is \citet{Gaudi2017}, S17 is \citet{StassunCollins2017}, H17 is \citet{Hoeijmakers2019}, CC10 is \citet{CollierCameron2010}, T16 is \citet{Turner2016}, NV14 is \citet{NeveuVanMalle2014}, D16 is \citet{Delrez2016}, and B20 is \citet{Bourrier2020}.}
\end{deluxetable*}

\subsection{Palomar/WIRC Data Collection \& Reduction}\label{sec:datareduc}

We obtained a total of ten photometric transit observations for our six survey targets between 2023-2025, with twelve additional transit opportunities lost to poor weather. These observations used an ultra-narrowband helium filter (FWHM: \SI{0.635}{\nano\meter}) centered on the
\SI{1083.3}{\nano\meter} He$^*$ feature \citep{VissapragadaKnutson2020}. We followed a similar procedure to that outlined in \cite{VissapragadaKnutson2020, ParagasVissapragada2021, VissapragadaKnutson2022,LevineVissapragada2024,Perez-GonzalezGreklek-McKeon2024, SaidelVissapragada2025}, which we summarize briefly here. All observations began with a helium arc lamp calibration image, which allowed us to position each target on the detector location where the position-dependent transmission function was centered on the He$^*$ line. Images were obtained using a custom-beam shaping diffuser which spreads incoming light into a top-hat shaped point-spread function (psf) with a FWHM of 3$\arcsec$, improving the duty cycle of our observations and minimizing time-correlated systematics \citep{StefanssonMahadevan2017, VissapragadaJontof-Hutter2020}. For all observations, we obtained at least five dithered frames for background correction as noted in Table~\ref{table:datasumm}.

We first dark-subtracted and flat-fielded all images, and then corrected for bad pixels and subtracted the sky background following \cite{VissapragadaKnutson2020}. Given that the bandpass of the helium filter changes across the detector field of view, different comparison stars will undergo different time-varying water absorption and OH emission \citep{VissapragadaKnutson2020, VissapragadaKnutson2022}. We therefore correct for telluric emission lines and the time-varying water absorption by first sigma-clipping each science frame to remove sources and then median-scaling the dithered background frame to match the science frame in 10-pixel radial steps from the zero point of the filter (where light passes through the filter at normal incidence). While this process eliminates most of the telluric background, we can use the scaling factors from this procedure to also correct for the time-varying telluric water absorption following the procedure outlined in \citep{ParagasVissapragada2021}. In particular, we take the ratio of the scaling factors for telluric OH lines contaminated by water absorption to the telluric OH lines uncontaminated by water absorption. The ratio of these scaling factors serves as a proxy for the time-varying water absorption proxy that we can use for decorrelation in our transit lightcurve fits.

We used \texttt{photutils} \citep{BradleySipocz2023} to perform aperture photometry on both the target and nearby comparison stars that fell within WIRC's $8\times8\arcmin$ field of view. As part of this step, we also subtracted any residual local sky background using annular apertures centered on the target and comparison stars with inner radii of 25 pixels (6\farcs25) and outer radii of
50 pixels (12\farcs50). For each observation, we determined the optimal aperture size by first normalizing the target star's light curve using an average of the comparison star light curves and removing $4\sigma$ outliers using a moving-median filter. We then tested apertures ranging from 5 to 20 pixels in radius in 1 pixel steps (pixel scale: 0\farcs25/px), and selected the aperture that minimized the variance in the normalized and outlier-corrected target photometry. The optimal apertures for each target in our survey are reported in Table~\ref{table:datasumm}.

\begin{deluxetable*}{ccccccccccc}[t!]
\tabletypesize{\scriptsize}
\tablecaption{Summary of Palomar/WIRC observations analyzed in this work.\label{table:datasumm}}
\tablehead{\colhead{Planet}  & \colhead{Date (UT)}  & Transit Fraction & \colhead{$t_\mathrm{exp}$ (s) } & \colhead{Start/Min/End Airmass} & \colhead{$n_\mathrm{comp}$} & \colhead{$n_\mathrm{dither}$} & \colhead{$r$ (px)} & \colhead{$\sigma$ [\%]} & \colhead{$\sigma/\sigma_\mathrm{phot}$} & \colhead{Covariates}}
\startdata
HAT-P-8~b & 2023 Sep 21 & Partial & 90 & 1.16/1.16/2.91 & 3 & 9 & 12 & 0.75 & 2.3 & w, a\\
 & 2024 Aug 09 & Partial & 90 & 1.72/1.00/1.00 & 2 & 9 & 7 & 0.92 & 3.4 & w  \\
WASP-93~b & 2024 Jan 08 & Full & 90 & 1.13/1.13/2.41 & 4 & 5 & 14 & 0.49 & 1.4 & p, a \\
 & 2024 July 29 & Full & 90 & 1.54/1.05/1.06 & 3 & 5 & 12 & 0.59 & 1.1 & -\\
WASP-180~A~b & 2024 Dec 06 & Full & 90 & 1.98/1.22/1.70 & 3 & 5 & 11 & 0.62 & 2.0 & a\\
WASP-103~b & 2024 May 16 &  Full & 120 & 2.07/1.12/1.71 & 3 & 5 & 10 & 0.85 & 1.0 & a\\
 & 2024 May 17 & Full & 120 & 2.75/1.11/1.71 & 3 & 5 & 10 & 0.89 & 1.2 & a\\
 & 2025 Apr 30 & Full & 120 & 2.79/1.16/1.16 & 3 & 5 & 10 & 0.85 & 1.2 & -\\
KELT-7~b & 2023 Dec 09 & Full & 90 & 1.09/1.00/1.13 & 3 & 9 & 13 & 0.17 & 1.9 & w, a\\
& 2023 Dec 28 & Partial & 90 & 1.02/1.02/1.70 & 3 & 9 & 9 & 0.27 & 1.8 & -\\
WASP-12~b & 2025 Feb 20 & Full & 90 & 1.08/1.00/2.92 & 3 & 9 & 7 & 0.56 & 1.2 & a\\
\enddata
\tablecomments{In the column headers, ``Transit fraction'' refers to the observation of either a full or partial transit, $t_\mathrm{exp}$ is the exposure time, $n_\mathrm{comp}$ is the number of comparison stars used in the reduction, $n_\mathrm{dither}$ is the number of dithered frames used to construct the sky background frame, $r$ is the aperture radius used in the photometric extraction, $\sigma$ is the rms scatter of the residuals to our final light-curve fit,  and $\sigma/\sigma_\mathrm{phot}$ is the ratio of the rms scatter to the photon noise limit. The ``Covariates'' column displays the optimal detrending vectors in the final versions of our fits, in which ``w'' refers to the time-varying telluric water proxy, ``a'' represents airmass, ``p'' is psf width, and ``-'' refers to no covariates.}
\end{deluxetable*}

\begin{deluxetable*}{ccccccc}[ht]
\tabletypesize{\scriptsize}
\tablecaption{Priors for the transit light-curve fits. \label{table:priors}}
\tablehead{\colhead{Planet}  & \colhead{$P$ (days)}  & \colhead{$T_0$ (BJD)} & \colhead{$b$} & \colhead{$a/R_\star$} & \colhead{References}}
\startdata
HAT-P-8~b & $\mathcal{N}(3.07634347, 0.00000058)$ & $\mathcal{N}(2456052.75596, 0.00024)$ & $\mathcal{N}(0.32, 0.19)$ & $\mathcal{N}(6.29, 0.06)$ & L09, K23 \\
WASP-93~b & $\mathcal{N}(2.7325321, 0.0000020)$ & $\mathcal{N}(2456079.56501, 0.00026)$ & $\mathcal{N}(0.9036, 0.0090)$ & $\mathcal{N}(5.938, 0.133)$ &   H16, Y24 \\
WASP-180~A~b & $\mathcal{N}(3.409264, 0.000001)$ & $\mathcal{N}(2458206.519399, 0.000049)$ & $\mathcal{N}(0.29, 0.02)$ & $\mathcal{N}(8.75, 1.13)$ &  T19, K23\\
WASP-103~b & $\mathcal{N}(0.925545386, 0.000000056)$ & $\mathcal{N}(2457308.324538, 0.000030)$ & $\mathcal{N}(0.19, 0.13)$ & $\mathcal{N}(3.013, 0.027)$ &  G14, K23 \\
KELT-7~b & $\mathcal{N}(2.734770, 0.000004)$ & $\mathcal{N}(2458827.45748, 0.00009)$ & $ \mathcal{N}(0.593, 0.0550)$ & $\mathcal{N}(5.50, 0.17)$ & S17, PE22 \\
WASP-12~b & $\mathcal{N}(1.091418901, 0.000000018)$ & $\mathcal{N}(2457607.519305, 0.000032)$ & $\mathcal{N}(0.424, 0.013)$ & $\mathcal{N}(3.04, 0.03)$ &  L24, K23 \\
\enddata
\tablecomments{ 
In the references column, L09 refers to \citet{LathamBakos2009}, K23 is \citet{KokoriTsiaras2023}, H16 is \citet{HayCollier-Cameron2016}, Y24 is \citet{YalcinkayaEsmer2024}, T19 is \citet{TempleHellier2019}, G14 is \citet{GillonAnderson2014}, S17 is \citet{StassunCollins2017}, PE22 is \citet{PatelEspinoza2022}, and L24 is \citet{LeonardiNascimbeni2024}.
}
\end{deluxetable*}

\subsection{Light-Curve Modeling}\label{sec:lightcurve}

We fit each target's light curve with a model consisting of a transit light curve multiplied by a systematics model. 
We calculated the transit light curve using \texttt{exoplanet} \citep{Foreman-MackeyLuger2021, Foreman-MackeyLuger2024}, following a procedure similar to \cite{VissapragadaKnutson2022} and \cite{SaidelVissapragada2025}. Briefly, we modeled limb-darkened versions of our transit light curves using \texttt{starry} \citep{LugerAgol2019}, where each transit light curve is parameterized by the orbital period $P$, the ratio of the planetary to stellar radius $R_{p}/R_\star$, the epoch $T_{0}$, the impact parameter $b$, the ratio of the semi-major axis to stellar radius  $a/R_\star$, the quadratic limb darkening coefficients $u_{1}$, $u_{2}$, and a jitter term log($\sigma_{extra}$) which we add in quadrature to the photon noise for each data point. We placed Gaussian priors on $P$, $T_{0}$, $b$, and $a/R_\star$ using published values from the literature as detailed in Table \ref{table:priors}, and utilized uniform priors for $R_{p}/R_\star$ (prior range: \SIrange[range-phrase={--}]{0}{0.25}{}), $u_{1}$, $u_{2}$ \citep[sampled using the approach detailed in][]{Kipping2013}, and log($\sigma_{extra}$) (prior range: \SIrange[range-phrase={--}]{1e-6}{1e-2}). 

We accounted for time-varying telluric and instrumental effects by fitting a linear trend in time along with a linear combination of comparison star light curves, where the weight of each comparison star light curve is a free parameter in the fit \citep[e.g.,][]{SaidelVissapragada2025}. We also considered four additional covariates for each observation, including the airmass, the distance from the median centroid, the psf width, and the time-varying telluric water
absorption proxy. We fit each transit light curve using all possible combinations of these four covariates and selected the version of the fit that minimized the Bayesian Information Criterion \citep[BIC;][]{Schwarz1978}. The final sets of covariates selected for each individual observation are listed in Table~\ref{table:datasumm}. 

After selecting the optimal detrending vectors, we jointly fit the individual nights of data for each planet assuming common values for $R_{p}/R_\star$, $u_{1}$, and $u_{2}$ while allowing for night-specific systematics models and noise terms.  We then repeated these joint fits with fixed limb darkening coefficients calculated using \texttt{ldtk} \citep{ParviainenAigrain2015} and checked to see if our posteriors for $R_{p}/R_\star$ were consistent with the values from the free limb darkening fits.
We also checked to see if the joint probability distribution for $u_{1}$ and $u_{2}$ was consistent with the predicted values from \texttt{ldtk}. We illustrate how our retrieved limb darkening coefficients compare to the \texttt{ldtk}-calculated values in Appendix~\ref{limbdarkappendix}. For fits that were $1\sigma$ consistent in both of these parameters (HAT-P-8~b, WASP-93~b, WASP-103~b, WASP-12~b), we adopted the fixed limb darkening coefficients in the final version of fits, as these values are in good agreement with our empirically measured limb-darkening coefficients. For cases where the two fits were inconsistent (WASP-180 A b, KELT-7~b), we adopted the fits with limb-darkening as a free parameter. Our results are generally in good agreement with those of \cite{VissapragadaKnutson2022}, who found that their retrieved limb-darkening coefficients were consistent with model predictions for most stars in their survey of He$^*$ absorption from gas giants transiting K dwarfs. 

Although He$^*$ nominally probes the chromosphere, \citet{SchmidtKnoelker1994} found that the Sun is not limb-brightened in this
feature, and that the difference in the shape of the limb-darkening between the He$^*$ feature and adjacent wavelengths is relatively modest. We therefore expect that any discrepancies between the model-predicted and true limb-darkening coefficients for our F-star targets should have a negligible effect on our retrieved planet-star radius ratios, which are largely insensitive to our choice of free versus fixed limb-darkening parameterizations.

We used the Hamiltonian MCMC No U-Turn Sampler \citep[NUTS; ][]{HoffmanGelman2011} in \texttt{pymc3} \citep{SalvatierWiecki2016} to sample the posterior probability distributions. All fits were run using four chains with 1,500 tuning steps and 2,500 draws per chain. For all fits, we confirmed that the Gelman-Rubin parameter $\hat{R} < 1.01$ for all sampled parameters, indicating the fit has converged \citep{GelmanRubin1992}. We list the priors and posteriors for the transit light curve model fits in Table~\ref{table:priors} and \ref{table:posteriors} respectively, and discuss results for each individual planet in more detail below. In Table~\ref{table:posteriors} we also report values for the He$^*$ excess absorption at mid-transit, calculated following the methodology
outlined in \citet{VissapragadaKnutson2022}.
The normalized and binned light curves for all planets are shown in Fig. \ref{fig:lightcurves}.

\section{Individual Planet Results}\label{sec:Results}

\subsection{HAT-P-8~b}\label{secc:HP8}

HAT-P-8~b \citep{LathamBakos2009} is an inflated hot Jupiter (1.35 $M_J$, 1.50 $R_J$) with a relatively low Roche filling factor (0.23) on a 3.08 day orbit (equilibrium temperature $T_{eq}=1700$~K) around a late F star \citep[$6200\pm80$~K, $v\sin$i$_*=11.50\pm0.50$~\unit{\kilo\meter\per\second};][]{BonomoDesidera2017} near the Kraft break \citep{Kraft1967}.  This star is the central component of a hierarchal triple system, with a pair of stellar companions with effective temperatures of $\sim3100$~K and $\sim3200$~K located at a projected separation of 1.0\arcsec~\citep{Bechter2014,Ngo2015}.  The two companions have $\Delta J$ of $6.59\pm0.12$ and $7.16\pm0.15$, respectively, and therefore contribute negligible flux in our unresolved WIRC He$^*$ photometry. We confirm this by using PHOENIX stellar models  \citep{Husser2013} to scale the reported $\Delta J$ magnitudes measured using Keck/NIRC2 to a corresponding flux ratio in the Palomar/WIRC He$^*$ bandpass. We find a combined flux contamination ratio of 0.0031 from these two stars, which corresponds to a change of 0.02$\sigma$ in our best-fit planet-star radius ratio reported below.

We observed He$^*$ transits of HAT-P-8~b on UT August 9 2023, UT September 21 2023, and UT August 9 2024. The night of UT August 9 2023 suffered from poor weather conditions that prevented us from detecting the transit, and we therefore excluded this night from our subsequent analysis. On UT September 21 2023 we were able to observe the second half of the transit. On UT August 9 2024 night we successfully observed the first half of the transit, but deteriorating weather conditions resulted in a telescope closure that prevented us from observing the second half. For this night we excluded the last fourteen science frames from our analysis, as these were taken during the poor weather conditions immediately before the telescope closure and had relatively low fluxes. 

\begin{figure*}
\gridline{
\fig{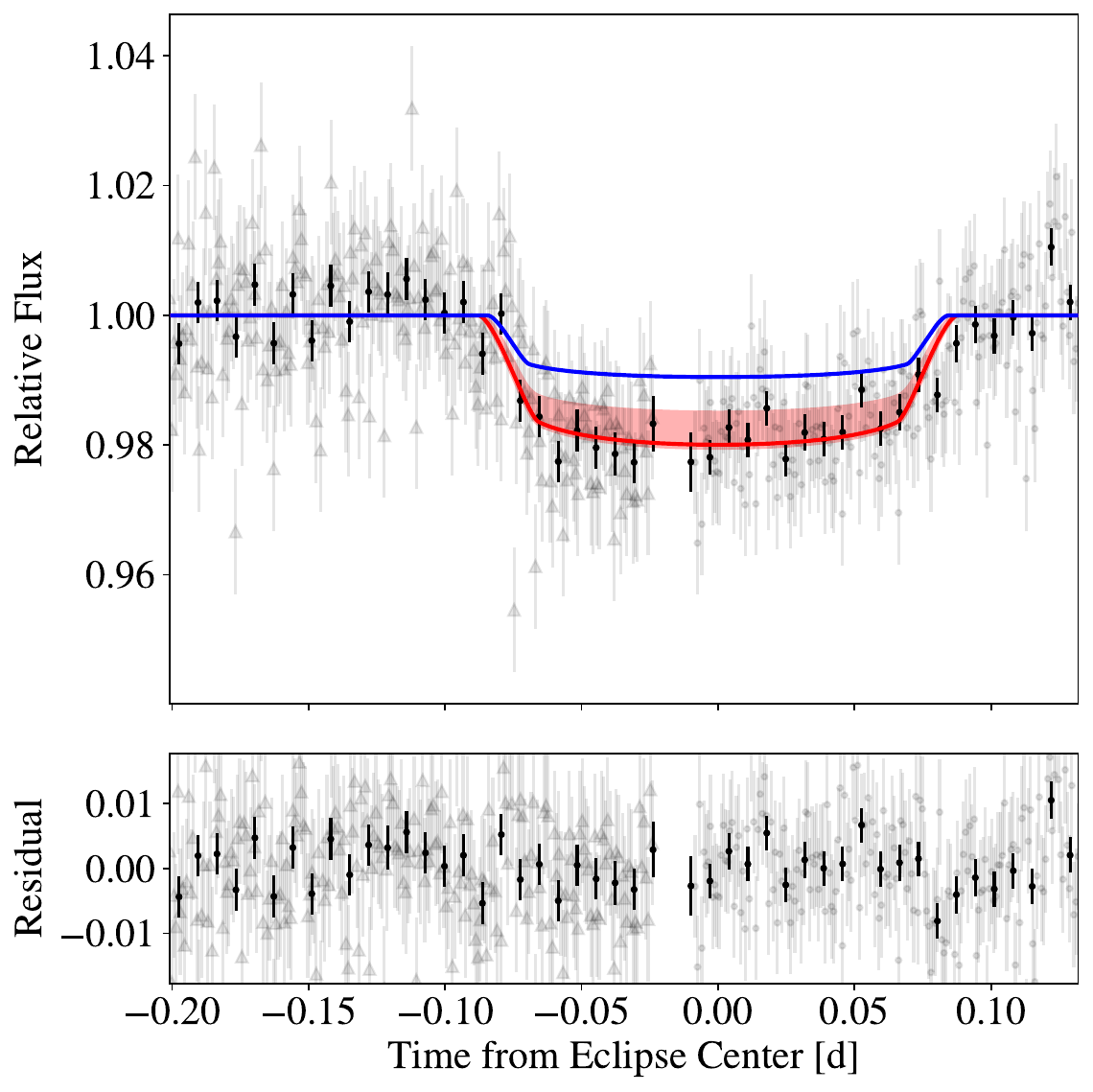}{0.45\textwidth}{(a) HAT-P-8~b}
\fig{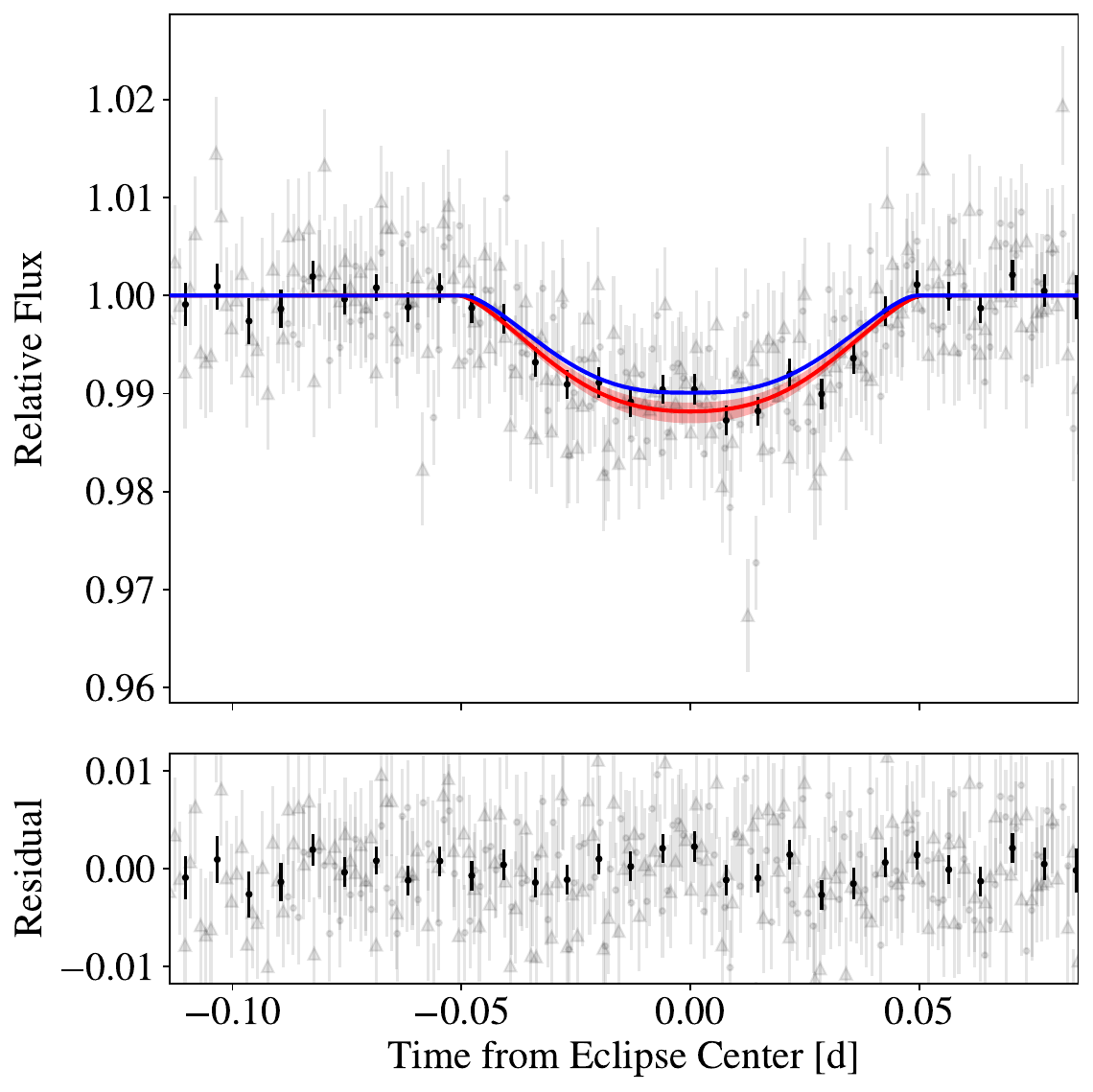}{0.45\textwidth}{(b) WASP-93~b}
}
\gridline{
\fig{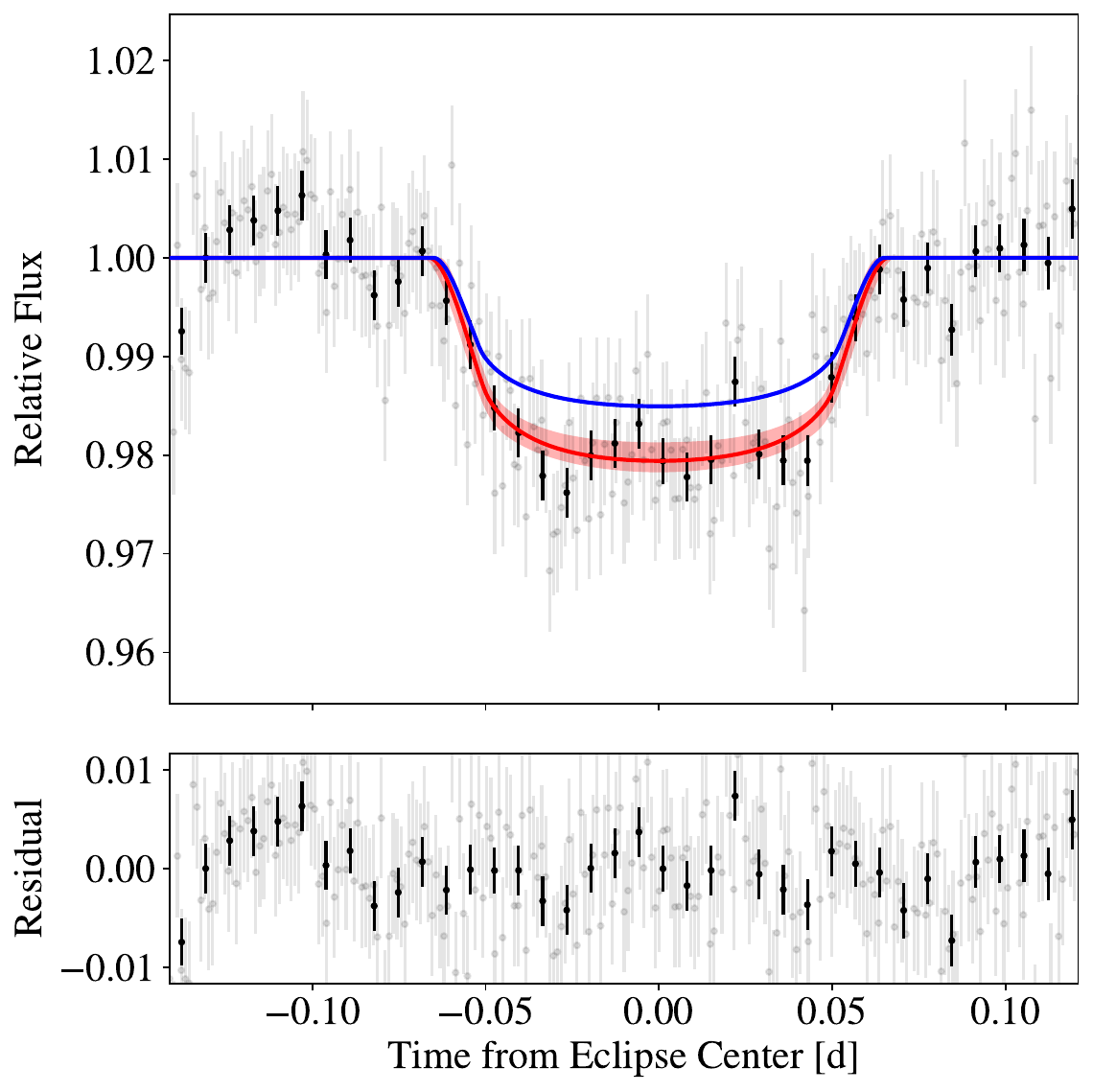}{0.45\textwidth}{(c) WASP-180~A~b}
\fig{{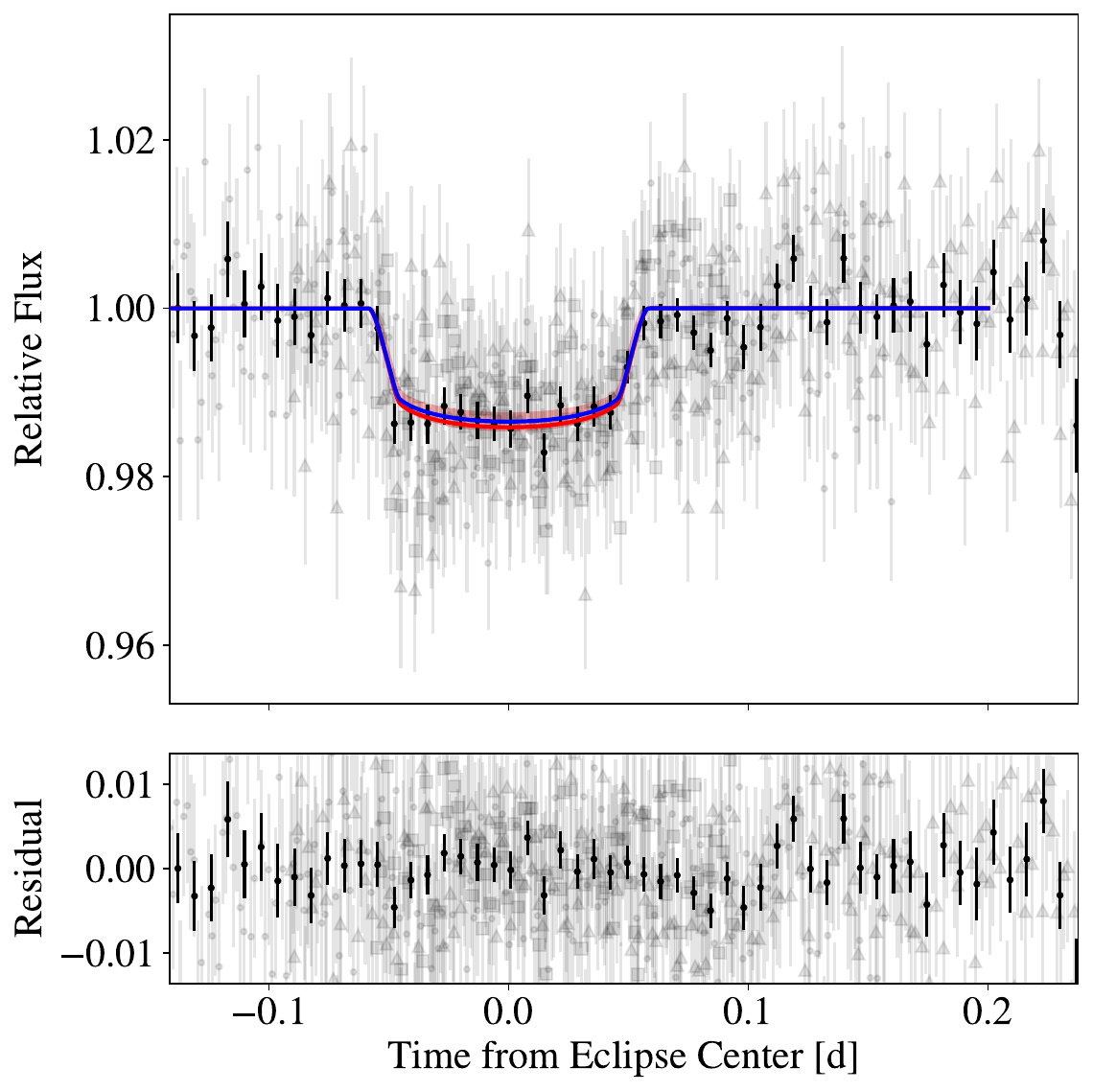}}{0.45\textwidth}{(d) WASP-103~b}
}
\end{figure*}%
\begin{figure*}
\gridline{
\fig{{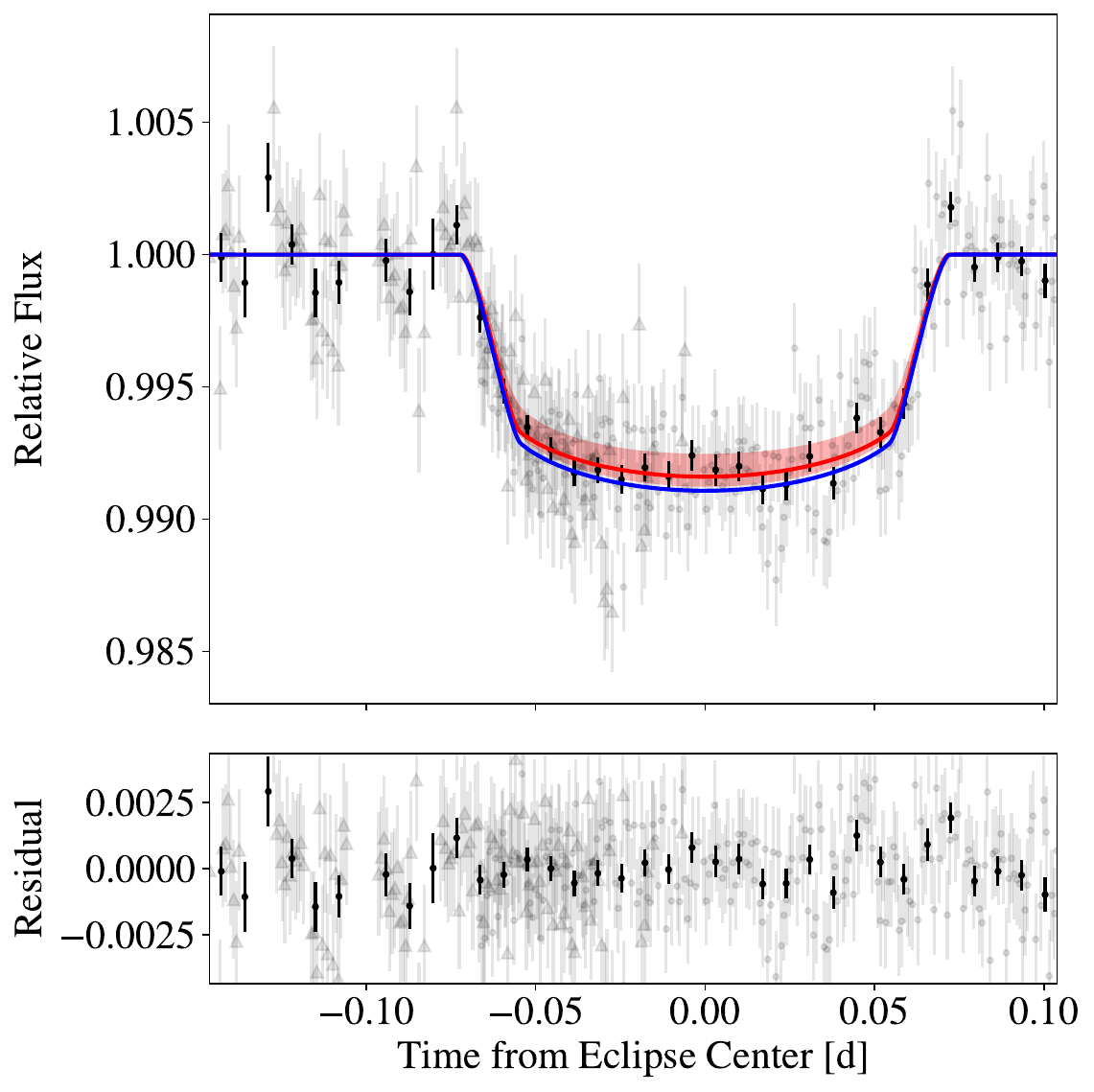}}{0.45\textwidth}{(e) KELT-7~b}
\fig{{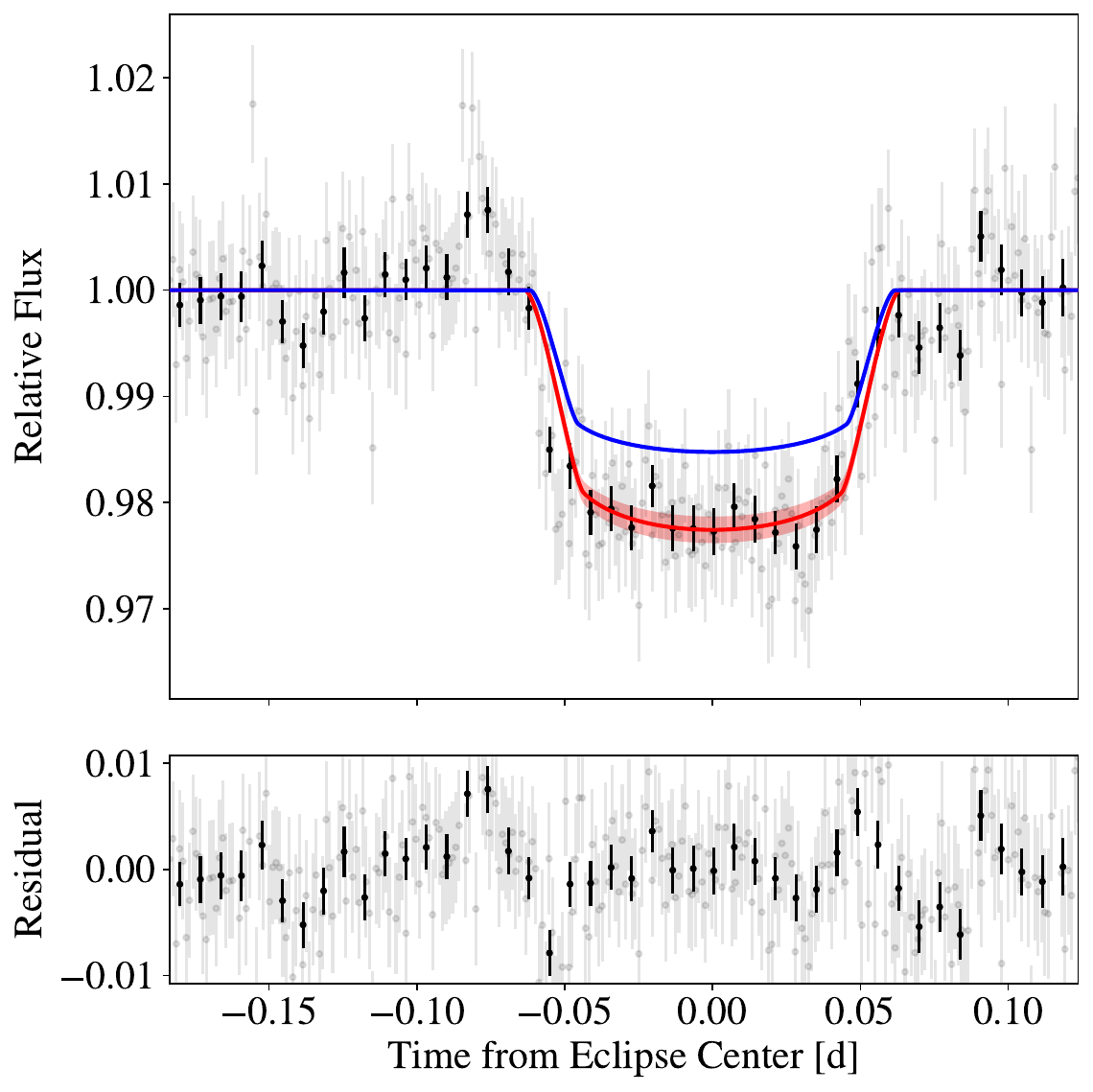}}{0.45\textwidth}{(f) WASP-12~b}
}
\caption{Transit light curves (top) and residuals (bottom) for the six planets in our survey. The detrended data (grey points, different nights have different symbol shapes) are binned to 10 minute cadence and overplotted as black circles. Our best-fit transit models are overplotted as red lines, with red shading indicating the corresponding 1$\sigma$ uncertainty. The blue curves show the predicted light curve models for the case where there is no outflow. \label{fig:lightcurves}}
\end{figure*}

We found that that the optimal covariates for the night of UT September 21 2023 are the time-varying telluric water absorption proxy and airmass, while the water absorption proxy alone is preferred on the UT August 9 2024 night. 
When fitting each night separately we found discrepant radius ratios of $0.035_{-0.023}^{+0.028}$ and $0.161_{-0.011}^{+0.011}$. The partial transit obtained on the first night had a shorter out-of-transit baseline and higher noise and as a result suffered from significant degeneracies that resulted in widely varying transit depths. We jointly fit both partial transits to achieve full transit coverage and minimize these degeneracies.
We jointly fit both nights of data and upon comparing the retrieved transit depth to the nominal depth reported in \citet{KokoriTsiaras2023}, we find a mid-transit excess depth of $0.82_{-0.30}^{+0.30}\%$. For this fit we adopted fixed limb darkening coefficients in the final version of our fits, as the retrieved limb darkening coefficients agree with the model limb darkening coefficients within $1\sigma$. The results of our joint fit are shown in Figure~\ref{fig:lightcurves}. Our mid-transit excess absorption corresponds to a marginally significant $2.7\sigma$ detection of an outflow, and we conclude that additional transit observations are needed to confirm this detection.

\subsection{WASP-93~b}\label{sec:WASP93}

WASP-93~b is a close-in ($P = 2.73$ days, $T_{eq}=1942$~K) 
1.47 $M_J$, 1.60 $R_J$ hot Jupiter orbiting a mid-F star \citep[\SI{6700(100)}{\kelvin};][]{HayCollier-Cameron2016}.
This planet has a relatively modest Roche filling factor (0.27), but its star has a high $v$sin$i_*$ \citep[\SI{37(3)}{\kilo\meter\per\second};][]{HayCollier-Cameron2016} comparable to that of HAT-P-32 and HAT-P-67.  \cite{HayCollier-Cameron2016} report a nearby mid-K dwarf companion with a projected separation of 0.69\arcsec and $\Delta J=3.70\pm0.18$.  
The companion is unresolved in our WIRC imaging, but contributes a negligible amount of flux in the He$^*$ band. To confirm this, we again used PHOENIX models to scale the measured $\Delta J$ value from 
the near-infrared camera INGRID on the William Herschel Telescope to a corresponding flux ratio in the He$^*$ bandpass.
\citet{HayCollier-Cameron2016} do not report an effective temperature for the K-dwarf companion, so we
adopted a value of \SI{4450}{\kelvin} for our calculation. We find a corresponding flux contamination ratio of 0.028 in the Palomar/WIRC He$^*$ bandpass. This translates to a change of 0.2$\sigma$ in our best-fit planet-star radius ratio.

We observed transits of WASP-93~b on UT January 8 2024 and UT July 29 2024. For the night of UT January 8 2024, we found that the BIC was minimized when we fit using the airmass and psf width as additional covariates. For the night of UT July 29 2024, we did not prefer any additional covariates. We adopted fixed limb darkening coefficients in our final version of our fits, as we found that the calculated limb darkening coefficients agreed well with the joint distribution of retrieved quadratic limb darkening coefficients, and our choice did not make a substantial difference for the retrieved transit depth. When fitting each night individually we retrieved radius ratios of $0.1078_{-0.0074}^{+0.0076}$ and $0.1192_{-0.0078}^{+0.0081}$, which corresponded to mid-transit excess depths of $0.052_{-0.125}^{+0.124}\%$ and $0.25_{-0.13}^{+0.13}\%$ respectively
when compared to the updated transit depth reported in \citep{KokoriTsiaras2023}. 
We note that while these two depths are consistent at the $1\sigma$ level, there does appear to be correlated noise at approximately mid-transit on both nights that is likely contributing to the variations in the retrieved depths. In our final joint fit, we find a final radius ratio of $0.1172_{-0.0065}^{+0.0068}$ corresponding to a mid-transit excess depth of $0.22_{-0.11}^{+0.11}\%$. This is a tentative ($2.0\sigma$) outflow detection. We emphasize that this detection is tentative and requires follow-up for confirmation given that the correlated noise present in both of our observations could bias our retrieved transit depths.

\begin{deluxetable*}{cccccccccc}[ht]
\tabletypesize{\scriptsize}
\tablecaption{Posteriors for the light-curve fits. \label{table:posteriors}}
\tablehead{\colhead{Planet} & \colhead{LDC} & \colhead{$P$ (days)}  & \colhead{$T_0$ (BJD $-$ 2450000)} & \colhead{$b$} & \colhead{$a/R_\star$} & \colhead{WIRC $R_\mathrm{p}/R_\star$} & \colhead{$u_1$} & \colhead{$u_2$} & \colhead{$\delta_\mathrm{mid}$ (\%)}}
\startdata
HAT-P-8~b & fixed & $3.07634378_{-0.00000055}^{+0.00000055}$ & $6052.75600_{-0.00024}^{+0.00024}$ & $0.15_{-0.13}^{+0.11}$ & $6.271_{-0.058}^{+0.058}$ & $0.126_{-0.011}^{+0.010}$ & $0.28$ & $0.13$ & $0.82_{-0.30}^{+0.30}$
 \\ 
WASP-93~b & fixed & $2.73253553_{-0.00000099}^{+0.00000097}$ & $6079.56504_{-0.00025}^{+0.00026}$ & $0.8975_{-0.0090}^{+0.0093}$ & $5.87_{-0.12}^{+0.12}$ & $0.1172_{-0.0065}^{+0.0068}$ & $0.27$ & $0.12$ & $0.22_{-0.11}^{+0.11}$ \\
WASP-180~A~b & free & $3.40926356_{-0.00000089}^{+0.00000087}$ & $8206.519398_{-0.000049}^{+0.000048}$ & $0.29_{-0.02}^{+0.02}$ & $9.07_{-0.36}^{+0.38}$ & $0.1306_{-0.0051}^{+0.0050}$ & $0.28_{-0.20}^{+0.32}$ & $0.37_{-0.37}^{+0.33}$ & $0.56_{-0.15}^{+0.15}$ \\
WASP-103~b & fixed &  $0.92554513_{-0.00000037}^{+0.00000036}$ & $7308.324538_{-0.000030}^{+0.000031}$ & $0.137_{-0.108}^{+0.096}$ & $2.905_{-0.062}^{+0.063}$ &  $0.1092_{-0.0047}^{+0.0046}$ & $0.29$ & $0.12$ & $-0.0024_{-0.1129}^{+0.1151}$ \\
KELT-7~b & free &  $2.7347652_{-0.0000015}^{+0.0000016}$ & $8827.457473_{-0.000087}^{+0.000091}$ & $0.587_{-0.036}^{+0.034}$ & $5.56_{-0.14}^{+0.14}$ & $0.0865_{-0.0033}^{+0.0032}$ & $0.33_{-0.23}^{+0.28}$ & $0.19_{-0.35}^{+0.35}$ & $-0.086_{-0.063}^{+0.061}$ \\
WASP-12~b & fixed &  $1.091418897_{-0.000000018}^{+0.000000018}$ & $7607.519301_{-0.000033}^{+0.000031}$ & $0.420_{-0.013}^{+0.012}$ & $3.017_{-0.028}^{+0.028}$ & $0.1434_{-0.0041}^{+0.0039}$ & $0.29$ & $0.12$ & $0.73_{-0.13}^{+0.12}$ \\
\enddata
\tablecomments{The ``LDC" column denotes the choice of either free or fixed limb darkening coefficients. The mid-transit excess depth $\delta_\mathrm{mid}$ is a derived parameter. }
\end{deluxetable*}

\subsection{WASP-180~A~b}\label{sec:WASP180}

WASP-180~A~b is an inflated hot Jupiter (0.9 $M_J$, 1.24 $R_J$) in a retrograde orbit ($P=3.41$ days, $T_{eq}=1540$~K) that transits the primary star of a visual binary with a projected separation of 5\arcsec \citep{TempleHellier2019}. WASP-180~A has a $T_\mathrm{eff}=~$\SI{6500\pm150}{\kelvin}, placing it comfortably above the Kraft break.  When combined with the presence of a binary companion, WASP-180~A~b's significantly misaligned orbit suggests that this planet may be a compelling candidate for high-eccentricity migration \citep[e.g.,][]{Petrovich2015,AndersonStorch2016, StorchLai2017}. 
This star has a moderately rapid $v\sin$$i_*$ \citep[\SI{19.9\pm0.6}{\kilo\meter\per\second};][]{TempleHellier2019} and the lowest Roche filling factor (0.20) of our survey targets.

We observed one transit observation of this planet on UT 2024 December 6. For this night, we found that the optimal covariate was the airmass. We confirmed that the binary companion should lie outside of our nominal 11 pixel (2.7\arcsec) aperture radius, even after accounting for the diffuser-controlled PSF radius of 1.5\arcsec.  We additionally found that the fixed limb darkening coefficients disagreed with the retrieved coefficients by more than $2\sigma$, and we therefore used free limb darkening coefficients in the final version of our fits. As discussed in \cite{Gascon2025}, the presence of extended tails in the outflow can modify the transit shape in a way that alters the preferred limb-darkening coefficients when fitting a standard transit shape to the light curve.  We discuss this possibility in more detail in \S\ref{sec:3D_outflows}.  Our optimized fits result in a radius ratio of $0.1306_{-0.0051}^{+0.0050}$. We compare our retrieved transit depth to the updated depth reported in \citep{KokoriTsiaras2023} and find a mid-transit excess absorption of $0.56_{-0.15}^{+0.15}\%$. We therefore conclude that there is statistically significant ($3.7\sigma$) evidence for a He$^*$ outflow.

\subsection{WASP-103~b}\label{sec:wasp103}

WASP-103~b is an ultra-short-period ($P=22.2$ hr, $T_{eq}=2508$~K) hot Jupiter (1.22 $M_J$, 1.44 $R_J$) orbiting an F8 star \citep[$6110\pm160$~K, $v\sin$$i_*$ \SI{10.6\pm0.9}{\kilo\meter\per\second};][]{GillonAnderson2014}. This star has a candidate companion with a projected separation of $0\arcsec.2$, $\Delta J=2.43\pm0.03$, and an effective temperature of \SI{4300}{\kelvin} \citep{Ngo2016}. We assess the effect of the unresolved candidate companion's flux on our retrieved mid-transit excess absorption at the end of this section. 
Observations of the tidally-deformed transit light curve led to a $3\sigma$ constraint on the planet's Love number \citep[$h_f = 1.59^{+0.45}_{-0.53}$;][]{BarrosAkinsanmi2022}, which is similar to that of Jupiter \citep[$1.565\pm0.006$;][]{DuranteParisi2020}. 
WASP-103~b's inflated radius, high Roche filling factor (0.51), and high equilibrium temperature \citep[2508~\unit{\kelvin};][]{GillonAnderson2014} make it an excellent target for atmospheric escape studies. 

We observed He$^*$ transits of WASP-103~b on UT May 16 2024, UT May 17 2024, and UT April 30 2025. We obtained optimal results when we included airmass as a detrending vector for the nights of UT May 16 2024 and UT May 17 2024, and no covariates for the night of UT April 30 2025. Our optimized individual fits result in radius ratios of $0.1184_{-0.0069}^{+0.0064}$, $0.1113_{-0.0015}^{+0.0030}$, $0.100_{-0.015}^{+0.013}$ for the first, second and third nights. We note that while the first and second night radius ratios are within approximately $1\sigma$, the third night radius ratio is discrepant and is likely a result of the significantly shorter baseline and higher noise on this night. 
For our final fits, we adopted fixed limb darkening coefficients as the calculated limb darkening coefficients were within $1\sigma$ of the retrieved limb darkening coefficients. Jointly fitting our observations resulted in a final radius ratio of $0.1092_{-0.0047}^{+0.0046}$.

\citet{CartierBeatty2017} measured resolved spectra of WASP-103 and its companion using the G141 grism on HST/WFC3, and reported a flux contamination ratio (flux of the contaminant over the flux of WASP-103) of 0.0923 in a 0.0325 $\mu$m wide bandpass centered at 1.0783 $\mu$m. We therefore adopted this value for the corresponding flux ratio in the He$^*$ feature and
rescaled our retrieved retrieved radius ratio of $0.1092_{-0.0047}^{+0.0046}$ to account for this contamination, 
resulting in a corrected radius ratio of $0.1141_{-0.0049}^{+0.0048}$. To find the corresponding mid-transit excess absorption, we compared our corrected radius ratio to the contaminant-corrected radius ratio reported in \citet{CartierBeatty2017} from a fit to the white-light curve ($1.1-1.7$ $\mu$m).
We find a mid-transit excess absorption of $-0.038_{-0.1097}^{+0.1121}\%$, consistent with zero.

\subsection{KELT-7~b}\label{sec:kelt7}

KELT-7~b is a 1.39 $M_J$, 1.60 $R_J$ hot Jupiter with a low Roche filling factor (0.26) on a close-in orbit ($P = 2.73$ days, $T_{eq}=2048$~K) around an early F star \citep[$T_\mathrm{eff}=6789\pm50$~K;][]{CannonPickering1918, BierylaCollins2015}.  Recent $3-5$~$\mu$m transmission spectroscopy with JWST NIRSpec indicates that this planet has an atmospheric metallicity of $1-16\times$ solar, although the limited wavelength coverage of these data means that this result is partially degenerate with the assumed cloud properties \citep{Ahrer2025}.   
We observed He$^*$ transits of KELT-7~b on UT December 9 2023, UT December 17 2023, and UT December 28 2023. We attempted to observe this target on one additional night (UT November 17 2023) but weather conditions were poor and we were unable to obtain useful data. Our UT December 17 2023 night was also affected by weather, which forced a closure near mid-transit with poor data quality that persisted throughout the night. We found that the transit was unrecoverable on this night, and therefore excluded it from our analysis. Our two remaining observations included a full transit on UT December 9 2023 and a partial transit on UT December 28 2023. The latter was affected by intermittent clouds during the early part of the observation, which we accounted for by excluding science frames where the flux of KELT-7 decreased below 0.6 of the median value. 

We found that the optimal detrending vectors for the night of UT December 9 2023 were the time-varying telluric water proxy and airmass, while we preferred no covariates for the night of UT December 28 2023. Our optimized fits resulted in radius ratios of $0.0854^{+0.0046}_{-0.0044}$ and $0.1113^{+0.0061}_{-0.0056}$  for the first and second nights, respectively. Comparing these radius ratios to the nominal ratio reported in the discovery paper \citep[$0.09097^{+0.00065}_{-0.00064}$;][]{BierylaCollins2015}, we found mid-transit excess absorption values of $-0.077^{+0.079}_{-0.080}\%$ and $0.53^{+0.14}_{-0.14}\%$ for the first and second night of data, respectively. We found that our calculated limb-darkening coefficient were discrepant ($>2\sigma$) with the joint distribution of retrieved quadratic limb darkening coefficients, and we therefore adopted the free limb darkening coefficients in the final version of our fit. This star has the highest $v$sin$i_*$ of all of the stars in our sample \citep[\SI{69.3\pm0.2}{\kilo\meter\per\second};][]{ZhouLatham2016}, and it is possible that this fast rotation causes its limb-darkening profile to deviate from the values predicted by 1D stellar models.  The joint fit resulted in a radius ratio of $0.0865^{+0.0033}_{-0.0032}$, which is approximately 1$\sigma$ consistent with  the nominal transit depth, and corresponds to a mid-transit excess absorption of $-0.086^{+0.063}_{-0.061}\%$.  We conclude that there is no evidence for a He$^*$ outflow from this planet, with a 95th-percentile
upper limit of $0.018\%$ on the excess absorption.

\subsection{WASP-12~b}\label{sec:wasp12}

WASP-12~b is a hot Jupiter (1.47 $M_J$, 1.97 $R_J$, $T_{eq}=2516$~K) orbiting a slowly rotating late F star ($6265\pm50$~K, $v$sin$i_*=$ $1.6^{+0.8}_{-0.4}~$\unit{\kilo\meter\per\second}) host every $1.09$ days \citep{HebbCollier-Cameron2009,AlbrechtWinn2012, LeonardiNascimbeni2024}. 
WASP-12~b is the only exoplanet that has been observationally confirmed to be undergoing tidally-induced orbital decay \citep{MaciejewskiDimitrov2016, PatraWinn2017, WeinbergSun2017, BaileyGoodman2019, YeeWinn2020, MaFuller2021}. Measurements of the tidal deformation of WASP-12~b have resulted in an estimated Love number ($h_2=1.55^{+0.45}_{-0.49}$) similar to those of Jupiter and WASP-103~b (see Section~\ref{sec:wasp103}), suggesting that these planets may have similar internal structures \citep{AkinsanmiBarros2024}.  This star is the primary in a hierarchical triple system, with a pair of early M dwarf companions located at a projected separation of $1\arcsec.06$ with $\Delta J$ of $3.81\pm0.05$ and $3.92\pm0.05$, respectively \citep{Bechter2014,Ngo2015}.  These companions are not resolved in our WIRC photometry, but contribute a negligible flux in the He$^*$ aperture. To confirm this, we 
used PHOENIX models to scale the reported $\Delta J$ value from Keck/NIRC2 to the Palomar/WIRC He$^*$ bandpass. We found a combined flux contamination ratio of 0.048 for the two nearby stars, corresponding to a change of 0.86$\sigma$ in our best-fit planet-star flux ratio.

 WASP-12~b's high Roche filling factor (0.55) and extreme stellar irradiation make it particularly susceptible to atmospheric loss. NUV observations reveal a deep transit signal consistent with Roche lobe overflow, with significant pre-ingress absorption indicative of extended escaping material that likely forms a torus around the star \citep{FossatiHaswell2010,LaiHelling2010,HaswellFossati2012,FossatiAyres2013,NicholsWynn2015}.  A separate detection of H$\alpha$ absorption by \cite{JensenCauley2018} is more difficult to interpret, as subsequent H$\alpha$ observations with other spectrographs were unable to recover the reported signal \citep{PaiAsnodkar2024,CzeslaLampon2024}. It has also been suggested that emission from escaping CO gas flowing from the planet onto the star might explain the shape of WASP-12~b's Spitzer \SI{4.5}{\micro\meter} phase curve, which has a modulation period that is half of the planet's orbital period \citep{BellZhang2019}.

Two previous He$^*$ transit observations of WASP-12~b found no traces of escaping helium \citep{KreidbergOklopcic2018, CzeslaLampon2024}. However, the \citet{KreidbergOklopcic2018} study used low resolution HST/WFC3 data, while the high resolution data obtained by \citet{CzeslaLampon2024} is at the sensitivity limit for CARMENES. We searched for photometric evidence of escaping helium using one transit observations of WASP-12~b obtained on UT February 20 2025. We found that our systematics model for this observation was optimized when we include the airmass as a detrending vector. We adopted fixed limb darkening coefficients in the final version of the fit, as the calculated limb darkening coefficients agreed well with the retrieved limb darkening coefficients. The radius ratio obtained in our optimized transit fit is $0.1434_{-0.0041}^{+0.0039}$. Comparing this value with the nominal radius ratio reported in \citet{KokoriTsiaras2023}, we find a mid-transit excess absorption of $0.73_{-0.13}^{+0.12}\%$ ($5.6\sigma$ detection significance).

\section{Mass Loss Modeling}\label{sec:massloss}

\subsection{1D Parker Wind Mass Loss Models}\label{sec:sunbather}
For our survey targets, we convert our measured He$^*$ excess absorptions into constraints on their atmospheric mass-loss rates using the \texttt{sunbather} package.  This package models the atmospheric outflows as one-dimensional isothermal Parker winds \citep{LinssenOklopcic2022, LinssenShih2024}. 
For all models, we assume the outflow is $90\%$ hydrogen and $10\%$ helium by number.  Although outflows can be fractionated, this is much less likely to occur for planets with relatively high temperatures and correspondingly fast mass loss rates \citep{SchulikOwen2025}, and this is therefore a reasonable assumption to make for the planets in our sample. 

In order to accurately model the outflows for our survey targets we must choose representative models for the stellar XUV and EUV spectra (X-ray: \SIrange[range-phrase={--}]{5}{100}{\angstrom} , EUV:\SIrange[range-phrase={--}]{100}{912}{\angstrom}), which drive the outflow by heating the uppermost layers of the atmosphere. Three of our survey targets have published measurements of their X-ray fluxes. For WASP-180~A, \citet{FosterPoppenhaeger2022} used an \textit{eROSITA} observation to reconstruct the star's EUV flux and calculated a corresponding XUV flux at the planet's surface of \SI{5.7e5}{\erg\per\second\per\centi\meter\squared}. We convert this into an XUV luminosity at the stellar surface of \SI{3.2e31}{\erg\per\second}.  This star has a companion with $T_\mathrm{eff}=5430\pm30$~K and a projected separation of 5\arcsec~\citep{TempleHellier2019} that is unresolved in the \textit{eROSITA} observation. Following the methodology in Appendix A of \cite{FosterPoppenhaeger2022}, we therefore multiplied the reported XUV flux by a factor of 0.5 to correct for the blended light from the second star.   
WASP-12 and KELT-7 have both been observed by \textit{XMM-Newton}, which measured an XUV luminosity of $(4.1\pm0.1)\times10^{28}~$\unit{\erg\per\second} for KELT-7 \citep{TaberneroZapateroOsorio2022} and an upper limit of \SI{6e28}{\erg\per\second} for the XUV luminosity of WASP-12 \citep{CzeslaLampon2024}.  The remaining three survey targets (HAT-P-8, WASP-93, WASP-103) have no reported XUV observations in the literature. 

\begin{figure}[h!]
\centering
\includegraphics[width=0.48\textwidth]{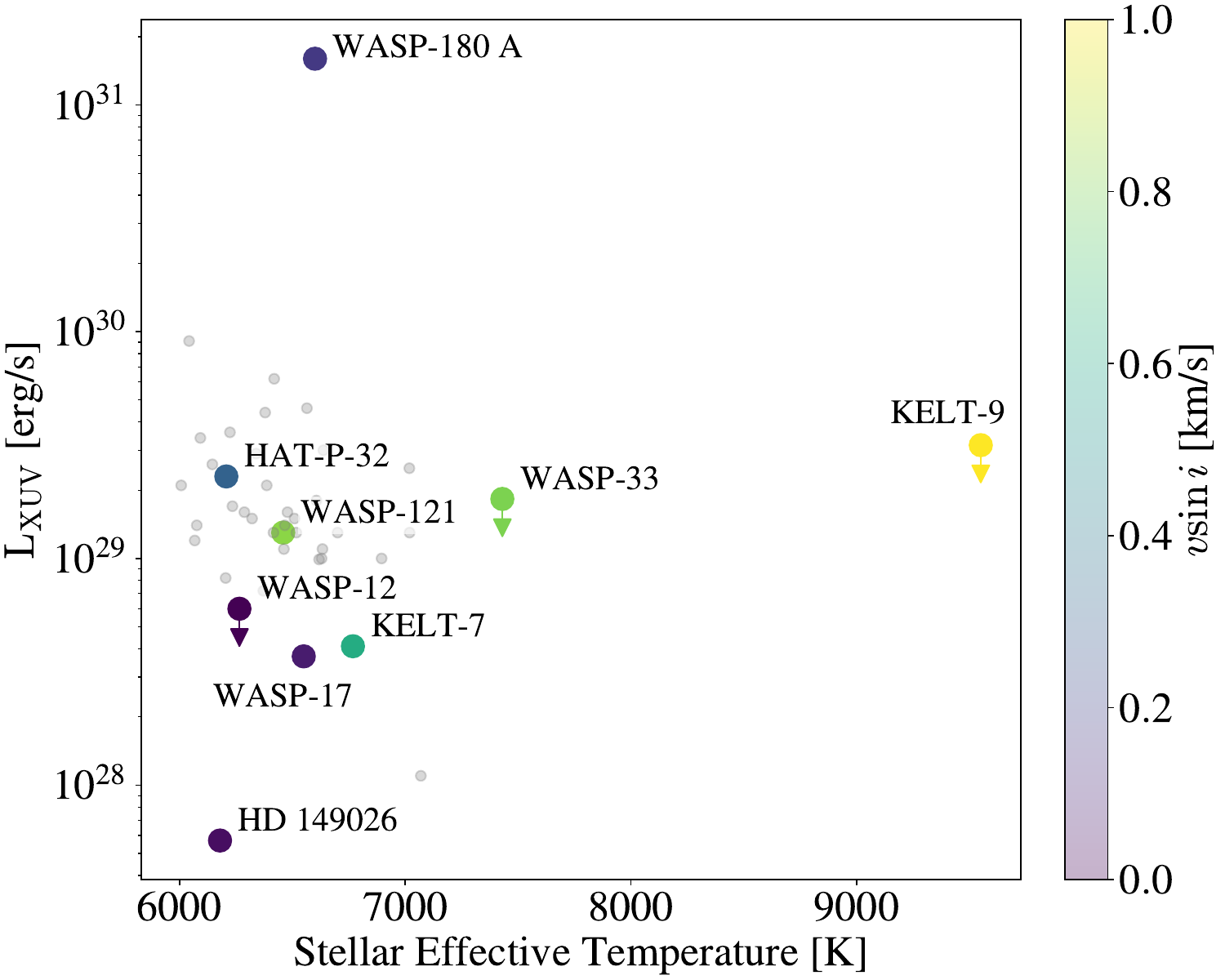}
\caption{Published XUV luminosity as a function of stellar effective temperature for early-type stars hosting planets with detected outflows and targets in our survey (see \S\ref{sec:intro} and \ref{sec:sunbather}). The XUV luminosities of the proxy MUSCLES stellar spectra used in our \texttt{sunbather} models are shown for reference. Shading corresponds to the projected stellar rotational velocities. The published XUV luminosities for WASP-12 and KELT-9 are upper bounds, which we plot here with downward pointing arrows. We note that WASP-33 has an unpublished \textit{XMM-Newton} observation, which we use to place an upper limit on the XUV luminosity of the star, as detailed in Appendix~\ref{wasp33appendix}. Gray points correspond to reported X-ray luminosities of F stars from \citet{ShimuraMitsuishi2025}.}
\label{fig:Lxuv}
\end{figure}

We model the high-energy stellar spectra for our survey targets by adopting proxy stellar spectra from the MUSCLES survey \citep{YoungbloodFrance2017} that are close matches in effective temperature and $v\sin i_*$ to the stellar hosts of our survey targets, although we note that the measured XUV luminosities of early-type stars with transiting hot Jupiters can vary by several orders of magnitude even for stars with similar effective temperatures and $v\sin i_*$ values (see Fig. \ref{fig:Lxuv}). 
For all survey targets other than WASP-103 and HAT-P-8, we adopt the MUSCLES spectrum of WASP-17 \citep[$T_\mathrm{eff}=6550\pm100$ K, $v{\sin}i_*=10.1^{+0.9}_{-0.8}$ \unit{\kilo\meter\per\second};][]{StassunCollins2017,BonomoDesidera2017} as the closest proxy. For WASP-103 and HAT-P-8, which have somewhat lower effective temperatures, we use the MUSCLES spectrum of HD 149026 \citep[$T_\mathrm{eff}=6179\pm15$ K, $v{\sin}i_*=6.0\pm0.5$ \unit{\kilo\meter\per\second};][]{StassunCollins2017,BonomoDesidera2017}. The temperature of WASP-12 is intermediate between those of our two proxy stars (HD 149026 and WASP-17), so we select WASP-17 as our proxy because it is a closer match in XUV luminosity to the upper limit reported for WASP-12 (see Fig.~\ref{fig:Lxuv}). For survey targets with published XUV luminosity constraints, we scale the XUV flux of the proxy spectra up or down to match the published XUV luminosity measurements. We note that the measured XUV luminosity of KELT-7 is already very close to the XUV luminosity of the proxy WASP-17, so no rescaling is needed.  For WASP-12, which has a published upper bound on its XUV luminosity, we confirm that our chosen proxy (WASP-17) has a luminosity that lies below this upper bound. For WASP-180~A, we scale the XUV luminosity of WASP-17 up to match the higher value measured for this star.  For the targets that do not have published XUV luminosity constraints (HAT-P-8, WASP-93, WASP-103), we carry out two separate fits.  In one fit we use the nominal XUV flux of the proxy star, and in the other fit we scale the XUV fluxes of the proxy stars up to match that of HAT-P-32 in order to investigate the role of a higher XUV flux on the retrieved mass loss rate. We adopt the flux of HAT-P-32 to simulate this upper bound as it is one of the few F stars with a precisely measured XUV flux that also has a published helium outflow, we note that WASP-121 is the only other such system, but it's XUV flux is slightly lower. We then use these two fits to quantify the effect of the uncertain stellar XUV luminosity on our retrieved mass loss constraints. 

Following the methodology outlined in \citet{SaidelVissapragada2025}, we define a grid of mass-loss rates ($\dot{M}$) and thermosphere temperatures ($T_0$) and compute a Parker wind profile at each $\dot{M}$-$T_0$ combination. We consider uniformly sampled values of $\dot{M}$ and $T_0$ between \SIrange[range-phrase={--}]{1e9}{1e13}{\gram\per\second} and \SIrange[range-phrase={--}]{5000}{11000}{\kelvin}, respectively. 
\texttt{sunbather} uses \texttt{Cloudy}  \citep{FerlandKorista1998, FerlandChatzikos2017} to iteratively eliminate grid models where the assumed outflow temperature and mass loss rate are inconsistent (i.e., the assumed mass loss rate requires a higher or lower outflow temperature to be physically self-consistent).
For the remaining self-consistent mass loss models, \texttt{sunbather} uses the \texttt{Cloudy}-computed 
He$^*$ density profile to calculate the corresponding helium absorption signal during transit. For each model, we then integrate this signal over the \SI{0.635}{\nano\meter} WIRC bandpass to calculate the predicted WIRC He$^*$ mid-transit excess absorption, which we compared to our measured values. We report the resulting best-fit mass-loss rates and thermosphere temperatures for each survey target in
Table~\ref{table:massloss}. 

We note that the Parker wind models created using \texttt{p-winds} within \texttt{sunbather} sometimes fail to converge for a subset of grid points, likely due to known instability issues within the code base for high gravity planets \citep{LinssenOklopcic2025}. This issue is particularly acute for WASP-103~b, where only four Parker wind models out of our entire $\dot{M}$-$T_0$ grid were successfully created and we therefore consider our retrieved mass-loss rates and temperatures to be less reliable for WASP-103. As a sanity check, we predict a mass loss rate for WASP-103~b using \texttt{ATES} \citep{Caldiroli2021}, which is a one-dimensional hydrodynamics code that simulates atmospheric escape at a steady-state mass loss rate based on the incoming stellar XUV flux. We generate a \texttt{p-winds} model corresponding to the \texttt{ATES}-predicted mass loss rate for this planet and find an excess absorption of $0.34\%$ in the Palomar/WIRC bandpass, which we can rule out with a $3.1\sigma$ confidence. Our \texttt{sunbather} runs for WASP-93~b also faced some convergence issues when constructing the Parker wind models, although the issue was less acute for this planet (40 grid models were successfully created, spanning a significant fraction of the parameter space).

\begin{table*}
\centering
\caption{Mass loss modeling summary and results for planets with He$^*$ detections and non-detections.}
\label{table:massloss}
\scriptsize
\begin{tabular}{ccccccc}
\toprule
Planet  & Spectrum & \multicolumn{2}{c}{\texttt{ATES} Predictions} &
\multicolumn{3}{c}{Retrieved}
\\
\cmidrule(lr){3-4}\cmidrule(lr){5-7}
&  & $\log\dot{M}$ {(\unit{\gram\per\second})} & $n_{eff}$ & $\log\dot{M}$ {(\unit{\gram\per\second})} & $T_0$ {(\unit{\kelvin})} & $n_{eff}$
\\
\midrule
HAT-P-8~b & HD 149026 & $9.96-10.05$ & $0.0070-0.0072$ & $9.2-10.9$ & $10800-11000$ & $0.0012-0.050$ \\
%
WASP-93~b & WASP-17 & $10.56-11.36$ &   $0.072-0.084$ & $9.10-11.2$ & $10300-11100$ & $0.0016-0.050$\\
%
WASP-180~A~b & WASP-17 & $12.00$ &  $0.033$ & $11.85^{+0.2}_{-0.2}$ & $10600^{+300}_{-900}$ & $2.2\times10^{-4}$  \\
WASP-103~b & HD 149026 & $12.23-12.60$ &  $0.17-0.27$ & $9.0-9.39$ & $10700-11100$ & $(1.0-1.6)\times10^{-4}$
\\
%
KELT-7~b & WASP-17 & $10.72$ & $0.14$ & $<11.58$ & $<10464$ & $0.034$\\
WASP-12~b & WASP-17 & $12.23$ & $0.34$ & $12.4^{+0.6}_{-0.5}$ & $9800^{+600}_{-900}$ & $0.51$ \\
\hline
Literature Measurements & & & & &  \\
\hline
HAT-P-32~b & HD 149026 & $11.54-12.74$ & $0.14-0.56$ & $12.03$, $12.56$  & $5750$, $14000$ & $0.027-0.092$
\\
HAT-P-67~b & HD 149026 & $11.81-12.49$ & $0.17-0.40$ & $13.3$ & $14000$ & $1.5$
\\
KELT-9~b & - & - & - & $12.8\pm0.3$ & $13200^{+800}_{-720}$ & -
\\
WASP-33~b & - & - & - & $11.8^{+0.6}_{-0.5}$ & $12200^{+1300}_{-1000}$ & - 
\\
WASP-94~A~b & HD 149026 & 10.89 & 0.71 & $10.86^{+0.14}_{-0.19}$ & $5000\pm1000$ & $0.65$ 
\\
WASP-121~b & WASP-17 & 12.52 & 0.20 & $12.7\pm0.1$ & \tablenotemark{a}  & -
\\
\bottomrule
\\
\multicolumn{7}{l}{\parbox{0.90\textwidth}{Note:
The ``spectrum'' column refers to the proxy MUSCLES stellar spectrum used. For planets in which multiple \texttt{sunbather} models were executed (HAT-P-8~b, WASP-93~b, WASP-103~b) we report the predicted and retrieved quantities as ranges reflecting the unscaled and scaled XUV flux model runs. For KELT-7~b, we report the 95th-percentile upper limit on the mass loss rate and temperature.
For planets with mass loss measurements from the literature, the ``Retrieved'' mass-loss rates and thermosphere temperatures are taken from the published studies cited in \S\ref{sec:intro}, some of which used different stellar models than the ones we use here to calculate their predicted mass loss rates.
\textsuperscript{a} \citet{Czesla2024} determined the mass loss rate for this planet by fitting a 3D outflow model with a spatially varying temperature structure.}}
\\
\end{tabular}
\end{table*}

\section{Discussion}\label{sec:disc}

\subsection{Are the Measured Mass Loss Rates Consistent with Energy-Limited Predictions?}

We compare our retrieved mass-loss rates 
to predicted energy-limited mass-loss rates calculated using the parameterization provided in Appendix A of \citet{CaldiroliHaardt2022}. Following the \texttt{ATES} parameterization, we also present heating efficiencies ($n_\mathrm{eff}$) for our predicted and retrieved mass-loss rates in Table~\ref{table:massloss}.  
For our survey targets with unknown stellar XUV fluxes (HAT-P-8~b, WASP-93~b, WASP-103~b), we report the range of allowed mass-loss rates and thermosphere temperatures in Table~\ref{table:massloss}. These ranges are calculated using our unscaled and scaled proxy stellar XUV \texttt{sunbather} model runs. In particular, the lower bounds of the allowed parameter ranges are obtained by taking the retrieved mass-loss rates and thermosphere temperatures from the unscaled models and subtracting their $1\sigma$ uncertainties. The upper bounds of the allowed temperature ranges are obtained by taking the retrieved mass-loss rates and thermosphere temperatures from the scaled XUV models and adding the $1\sigma$ uncertainties. For HAT-P-8~b and WASP-93~b, our retrieved mass-loss rate ranges agree well with the \texttt{ATES} model predictions, however our lower bounds on the retrieved mass loss rates are slightly lower than the lower bounds predicted by \texttt{ATES}. For WASP-103~b, our retrieved mass-loss rate ranges are much lower than the \texttt{ATES}-predicted mass-loss rates. We discuss possible sources of this discrepancy in Section~\ref{sec:RLOXUV}.

For the three survey targets with published XUV flux constraints (WASP-180~A~b, KELT-7~b, WASP-12~b), we find that our retrieved mass-loss rates are $1\sigma$ consistent with predictions for WASP-180~A~b and WASP-12~b, and within the 95th-percentile upper limit for KELT-17~b, highlighting the importance of using a measured XUV luminosity for the host star in these mass-loss models. We carry out a similar exercise using the published mass loss rates for HAT-P-32~b and HAT-P-67~b and find that they are in good agreement with the predicted mass-loss rates when the XUV flux of the proxy star is scaled up to match the published XUV flux measurement for HAT-P-32.  Similarly, for WASP-121~b, we find that the published mass-loss rate agrees well with model-predicted mass-loss rate when the proxy star is scaled to match the measured XUV flux of WASP-121. WASP-94~A~b's retrieved mass-loss rate is in good agreement with our energy-limited prediction, suggesting that the unknown stellar XUV flux of the host might be similar to that of our chosen proxy, HD 149026.  We therefore conclude that the measured mass-loss rates for all of these systems appear to be broadly in agreement with our energy-limited predictions, with the caveat that the uncertainties on the stellar XUV luminosities can be quite large for stars without direct measurements.

\subsection{Roche Lobe Overflow and Enhanced XUV Fluxes Drive Strong Outflows}\label{sec:RLOXUV}

\begin{figure*}
    \centering
    \includegraphics[width=\linewidth]{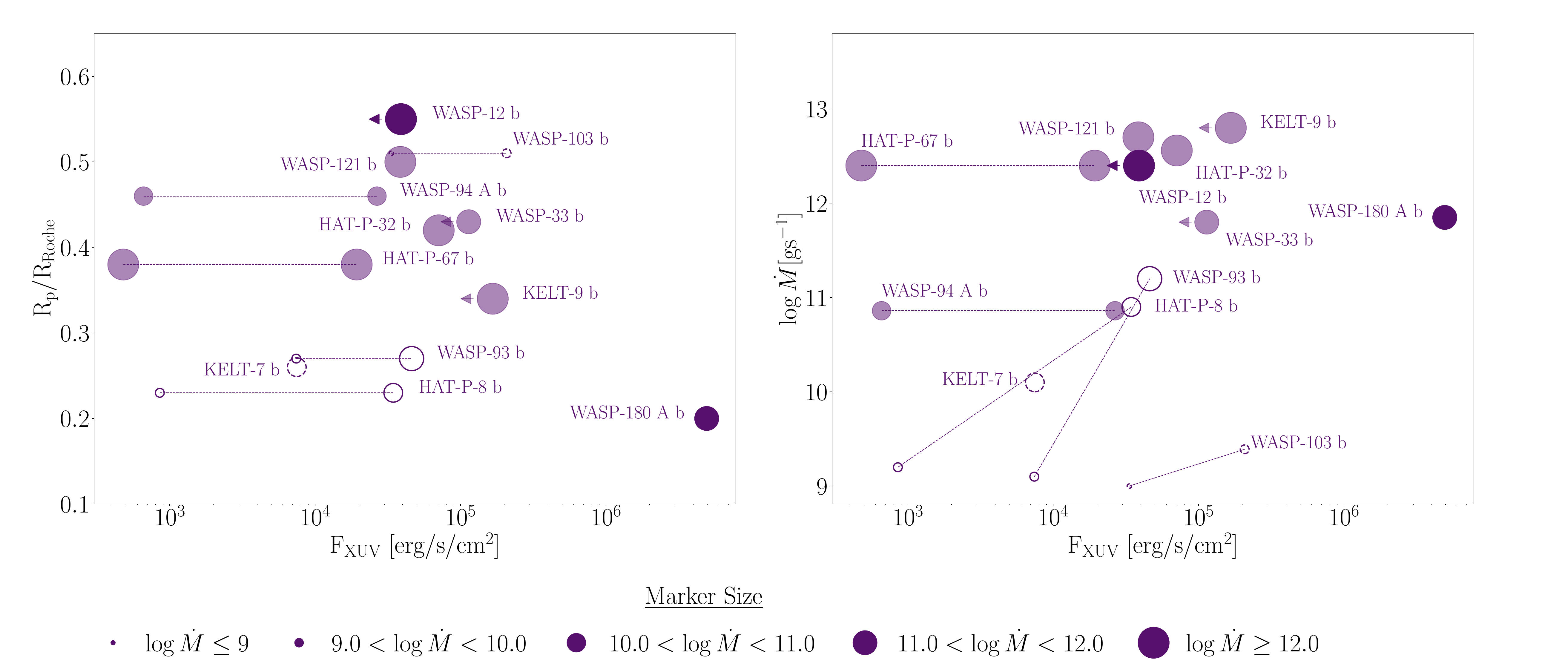}
    \caption{Left Panel: Roche filling factor as a function of XUV flux at the planet's orbit for survey targets and planets orbiting early-type stars with published mass loss measurements. Filled circles correspond to robust ($\geq3\sigma$) detections of atmospheric escape, open circles correspond to tentative ($\leq2\sigma$) detections of atmospheric escape, open circles with dashed edges indicate non-detections. Published detections are shown as a lighter shade of purple. Point size is proportional to the retrieved mass loss rate, indicated by the legend. The published XUV luminosities for WASP-12 and KELT-9 are upper bounds and therefore the XUV fluxes at the planets' orbits are also upper bounds, which we symbolize here with horizontal arrows. We similarly indicate that the XUV flux at the orbit of WASP-33 b is also an upper bound, which we derived from an unpublished \textit{XMM-Newton} observation, as detailed in Appendix~\ref{wasp33appendix}. We have indicated with arrows that the actual XUV flux may be higher or lower than this estimate. For F-type stars without published XUV luminosities (HAT-P-67, HAT-P-8, WASP-93, WASP-103, WASP-94 A), we plot two points connected by a dashed line.  The lower bound corresponds to a flux calculated using the proxy star XUV luminosity (see Table~\ref{table:massloss} for proxy star choice) and the upper bound corresponds to a flux calculated using the XUV luminosity of HAT-P-32. Palomar survey targets in this set (HAT-P-8, WASP-93, WASP-103) have point sizes corresponding to the retrieved mass-loss rates calculated at each luminosity value.  Right Panel: Retrieved mass loss rates as a function of XUV flux at the planet's orbit.  }
    \label{fig:RochevsXUV}
\end{figure*}

Of the six planets in our survey, WASP-12~b is the only target that has a measured atmospheric mass-loss rate greater than \SI{1e12}{\gram\per\second}, comparable to those of HAT-P-32~b, HAT-P-67~b, WASP-33~b, WASP-121~b, and KELT-9~b \citep{Wyttenbach2020,CzeslaLampon2022,ZhangMorley2023,BelloArufe2023AJ,Gully-SantiagoMorley2024,Czesla2024}. 
The remaining survey targets have much lower mass-loss rates, similar to the published constraints on the mass loss rates of WASP-48~b and WASP-94~b \citep{Bennett2023AJ,Mukherjee2025} and broadly in line with the average mass-loss rates reported for gas giant planets orbiting cooler stars.  We conclude that gas giants orbiting F stars do not experience systematically higher atmospheric mass-loss rates. Instead, our results suggest that mass loss rates greater than \SI{1e12}{\gram\per\second} are driven by properties that are specific to these systems. As shown in Fig.~\ref{fig:RochevsXUV} and Table \ref{table:stars}, the planets with the highest measured mass loss rates tend to have the highest Roche filling factors, indicating that RLO may play a significant role in driving these outflows.

WASP-103~b appears to be a notable exception to this trend. This planet has a Roche filling factor comparable to that of WASP-12~b, yet our retrieved upper limit on the mass loss rate for WASP-103~b is three orders of magnitude lower than the predicted value.  As noted earlier, there are no published measurements of the XUV flux of this star, although its effective temperature and $v{\sin}i_*$ are very close to those of our chosen proxy star, HD 149026. This suggests that either this star has a much lower XUV flux than HD 149026, or the fraction of metastable helium in the outflow is much lower than predicted. Similarly, WASP-94~A~b also has a high Roche filling factor, comparable to HAT-P-32~b, and yet it has a much lower mass loss rate, which might be a result of a much less XUV-luminous stellar host.

The importance of a high stellar XUV flux for driving outflows is supported by our detection of a He$^*$ outflow on WASP-180~A~b. Although this planet has a relatively moderate Roche filling factor of just 0.20, the host star has a higher XUV luminosity than any of the other stars in our sample with published measurements (see Fig. \ref{fig:Lxuv}).  This high XUV luminosity likely plays a significant role in generating the observed atmospheric outflow.  Although KELT-7 has a similar effective temperature and a much higher $v{\sin}i_*$ value than WASP-180~A, its XUV luminosity is approximately 3 orders of magnitude lower.  This underscores our conclusion in \S\ref{sec:sunbather} that effective temperature and $v{\sin}i_*$ alone are imperfect predictors of XUV luminosity and reinforces the importance of star-specific XUV measurements for interpreting atmospheric mass loss constraints. 

\begin{figure*}
    \centering
    \includegraphics[width=\linewidth]{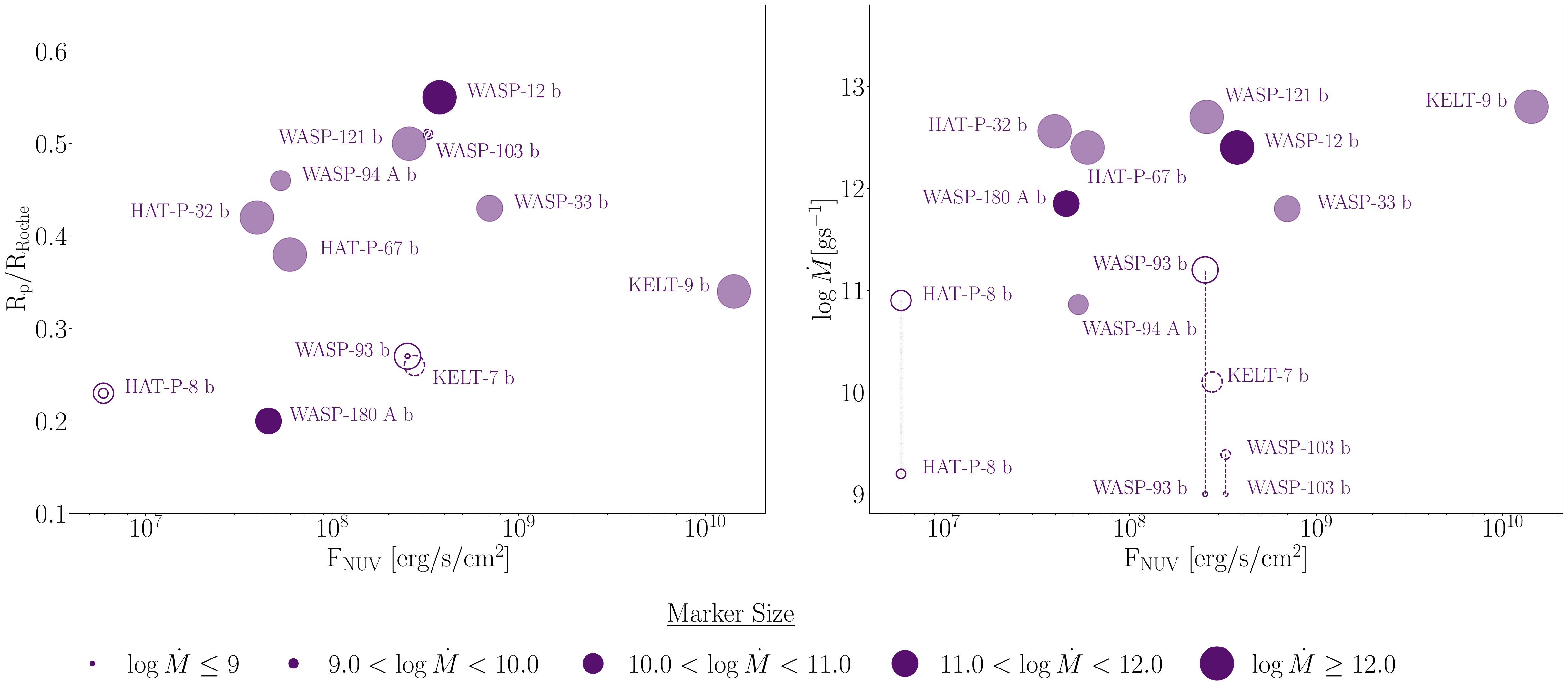}
    \caption{Left panel: Roche filling factor as a function of NUV flux at the planet's orbit for survey targets and planets orbiting early-type stars with published mass loss measurements. Right panel: Retrieved mass loss rates as a function of NUV flux at the planet's orbit. See the Fig. \ref{fig:RochevsXUV} caption for additional details.}
    \label{fig:RochevsNUV}
\end{figure*}

For all targets, we evaluate the role of NUV flux in driving the measured mass loss rates by calculating the NUV flux (\SIrange[range-phrase={--}]{1750}{3000}{\angstrom}) at the planets' orbits using PHOENIX model spectra \citep{Husser2013} and plotting the resulting values versus Roche filling factor in Fig.~\ref{fig:RochevsNUV}. For each star, we chose the PHOENIX grid model that was closest in effective temperature, surface gravity, and metallicity. We find that there is no clear correlation between NUV flux and the observed mass loss rates.  This is nicely illustrated by WASP-180 A b, which has a low Roche filling factor and a low NUV flux, yet has a high measured mass loss rate that can only be explained by its high XUV flux.  This suggests that Balmer-driven escape is unlikely to be the dominant escape mechanism for most of the planets in our sample.  

\subsection{Comparison to Published He$^*$ Observations of WASP-12~b}\label{sec:wasp12_discussion}

Of the six targets in our survey, only WASP-12~b has a published constraint on the He$^*$ excess absorption. As discussed in Section~\ref{sec:wasp12}, both HST/WFC3 and CARMENES did not detect He$^*$ absorption \citep{KreidbergOklopcic2018, CzeslaLampon2024} despite repeated NUV detections of an escaping atmosphere \citep{FossatiHaswell2010,HaswellFossati2012,NicholsWynn2015}. While the low resolution of HST/WFC3 could explain the non-detection reported by \citet{KreidbergOklopcic2018} the high resolution non-detection reported by \citet{CzeslaLampon2024} is puzzling given our strong excess He$^*$ measurement. Using the best-fit model from our \texttt{sunbather} fits, we find that the predicted peak excess absorption in the spectroscopically resolved He$^*$ line is $6\%$. We compare this peak absorption to the transmission spectrum presented in Figure~5 of \citet{CzeslaLampon2024} and find that this signal should have been readily detectable in their observations.  

\begin{figure}[h!]
\centering
\hspace*{-0.65cm}\includegraphics[width=0.48\textwidth]{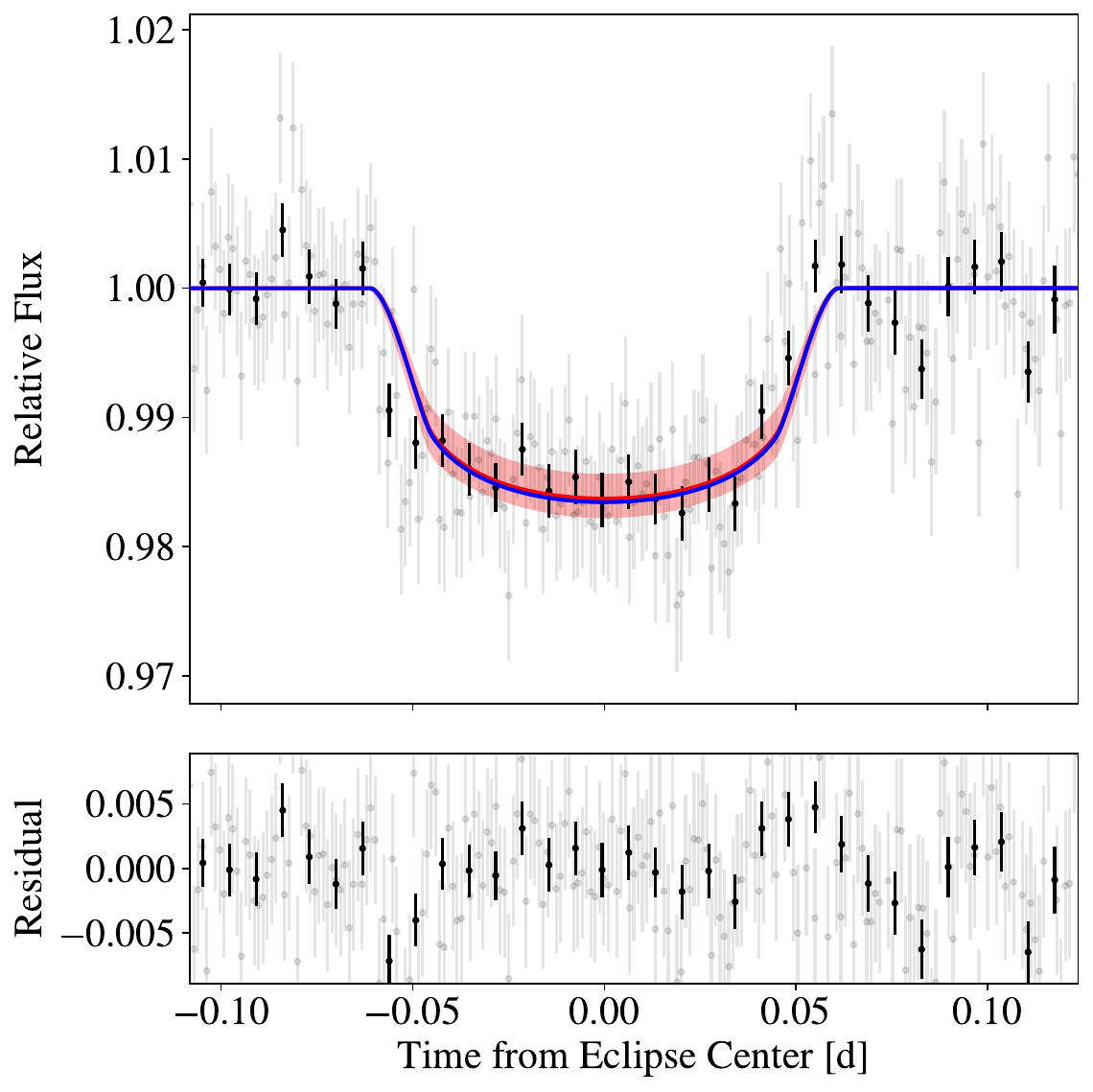}
\caption{Same as Figure~\ref{fig:lightcurves}, but for WASP-12~b with a a baseline clipped to match \cite{CzeslaLampon2024}. }
\label{fig:WASP12_clipped}
\end{figure}

Given our high measured mass loss rate for this planet, it is possible that it might have extended absorption prior to ingress or after egress, which could result in some self-subtraction of the measured helium signal if the chosen reference baseline significantly overlaps with this extended absorption.  The CARMENES observations presented in \cite{CzeslaLampon2024} have a time range of $-2.4$ to $+3.0$ hours around the transit center, whereas our observations extend from $-4.8$ to $+3.0$ hours around the transit center \citep[transit duration: $3.0010\pm0.0043$ hours; ][]{LeonardiNascimbeni2024}.  To investigate if this self-subtraction can bias the inferred signal, we clip our baseline to match the \cite{CzeslaLampon2024} observation and re-run our reduction and lightcurve fits for the WASP-12 b dataset. The clipped baseline observation (see Figure~\ref{fig:WASP12_clipped}) results in a non-detection of an excess absorption, with a mid-transit excess depth of $-0.00047_{-0.00211}^{+0.00203}$ corresponding to a 95th-percentile upper limit of 0.0029\%. This self-subtraction can therefore reconcile the non-detection in \cite{Czesla2024} with our Palomar observation.

Alternatively, the discrepancy between our strong excess He$^*$ measurement and the non-detection reported by \citet{CzeslaLampon2024} might be resolved if the WASP-12 system contains a time-varying circumstellar torus of gas fed by the strong outflow from WASP-12 b. WASP-12 b has a high inferred mass loss rate in our Palomar observation, suggesting that gas might accumulate around its host star faster than it can be dispersed by the stellar wind. 
The idea of a time-varying circumstellar torus has previously been proposed for the WASP-12 system \citep[e.g.,][]{HaswellFossati2012, DebrechtCarroll-Nellenback2018} in an attempt to explain the unexpected suppression of the stellar MgII H\&K and CaII H\&K lines \citep{HaswellFossati2012}, which is inconsistent with ISM absorption \citep{FossatiAyres2013} and cannot be explained by anomalously low stellar activity \citep{FossatiAyres2013, BonomoDesidera2017}. \citet{DebrechtCarroll-Nellenback2018} found that over the course of approximately 13 years, the density of the escaping gas in the circumstellar disk becomes high enough to reproduce the suppressed MgII H\&K lines from the \citet{HaswellFossati2012} observations. Similarly, if
 WASP-12 b's exosphere transits across a region of the stellar disk that is already obscured by the circumstellar torus, the exospheric He$^*$ absorption signal would be reduced, potentially rendering it undetectable \citep{Schreyer2025_submitted}. In order to reconcile the \citet{CzeslaLampon2024} non-detections from 2019 and 2020 with the detection presented here, the column density of the circumstellar torus must vary over years-long timescales, with a higher column density during the \citet{CzeslaLampon2024} observations. The time evolution of a circumstellar torus over such timescales is uncertain, and hydrodynamic models spanning these timescales are needed to fully explore this topic.  However, \citet{Schreyer2025_submitted} used simple arguments to calculate an approximate timescale for sufficient gas to accumulate and shield WASP-12~b from the XUV radiation that drives atmospheric escape, thereby halting the supply of gas to the torus and possibly causing it to become unstable. They found that this timescale is on the order of a year, suggesting that this mechanism can potentially explain the discrepancy between these two observations. Under \citet{DebrechtCarroll-Nellenback2018}'s models, this timescale is significantly longer, on the order of approximately 13 years. However, if we assume that the circumstellar torus reached its peak around the time of the \citet{HaswellFossati2012} observations in 2009 before becoming unstable and dissipating, the approximately 13 year timeline of disk re-accumulation aligns well with the \citet{CzeslaLampon2024} observations, potentially explaining their non-detection. A recent blow-off would result in relatively little accumulated disk material at the time of our observations, possibly explaining our subsequent detection.

\subsection{3D Outflow Geometry}\label{sec:3D_outflows}

While our photometric detections provide constraints on the mass-loss rates for our survey targets, the three-dimensional geometry of these outflows is not well constrained by our observations. As discussed in \citet{VissapragadaGreklek-McKeon2024}, extended tails of escaping atmospheric material can have low He$^*$ optical depths, making it difficult to detect them in the photometric Palomar/WIRC bandpass. Spectroscopically resolved observations of the He$^*$ line 
are much more sensitive to the presence of extended absorption pre- and post-transit \citep[e.g.,][]{SpakeOklopcic2021, CzeslaLampon2022}.

\begin{figure*}
    \centering
    \includegraphics[width=0.8\linewidth]{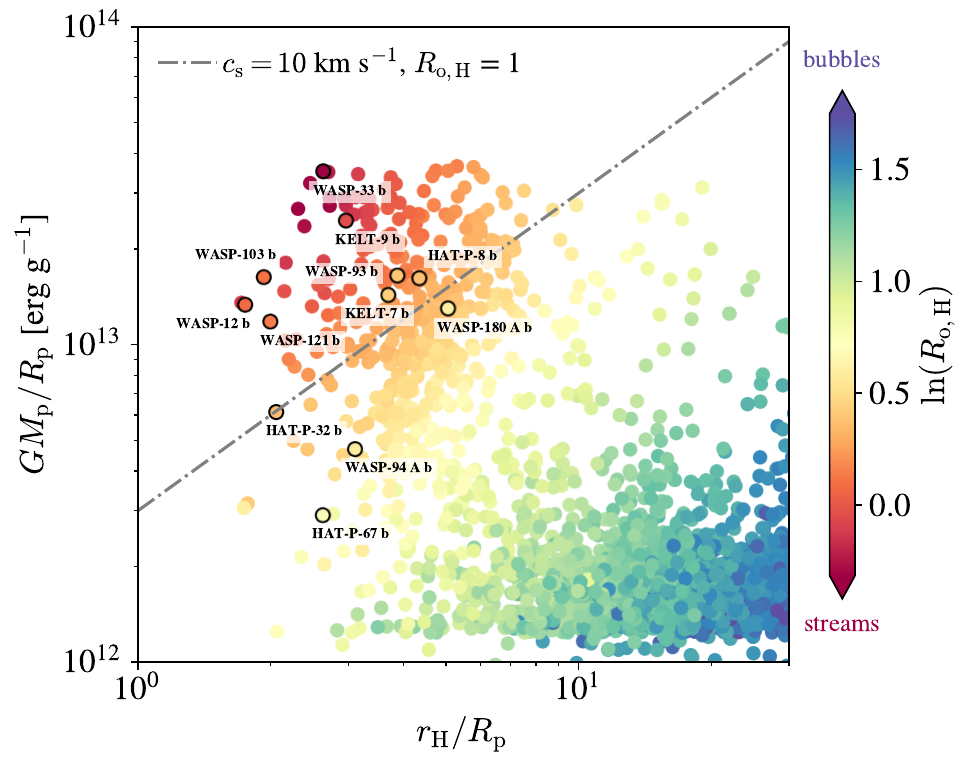}
    \caption{Figure adapted from \citet{MacLeodOklopcic2025} and compiled using \texttt{PWMorphology}. Colored circles represent confirmed exoplanets with radii greater than $1.6R_{\oplus}$ from the \texttt{sunset} catalog shown as a function of gravitational potential and the ratio of the Hill to planet radius. Survey targets as well as HAT-P-32~b, HAT-P-67~b, KELT-9~b, WASP-33~b, WASP-94~A~b, and WASP-121~b are indicated with circles outlined in black and labeled appropriately. The different colors represent the predicted hill sphere Rossby number $\ln(R_{o,H})$, where $\ln(R_{o,H})>0$ indicates a bubble-like outflow geometry and $\ln(R_{o,H})<0$ indicates a stream or tail-like outflow geometry. We note that values of $\ln(R_{o,H})$ for HAT-P-32~b, HAT-P-67~b, KELT-9~b, WASP-33~b, WASP-94~A~b, and WASP-121~b are calculated using the \texttt{sunset} photoevaporative outflow models for these planets.   
     The dashed grey line indicates photoevaporative outflows with sound speeds of 10~km~s$^{-1}$ and $R_{o,H}=1$. }
    \label{fig:stvsbub}
\end{figure*}

In the absence of direct observational constraints, we can still make predictions for the outflow morphologies of the planets in our sample. \citet{MacLeodOklopcic2025} showed that a planet's predicted Rossby number (defined as a velocity $v$ over the product of a length scale $L$ with an orbital angular speed $\Omega$: $v/\Omega L$) can be used to predict the morphologies of planetary outflows. This study adopts the Hill sphere ($r_H$) as the relevant length scale and the sound speed ($c_s$) as the relevant velocity:

\begin{equation}
    R_{o,H} = \frac{c_s}{\Omega r_H}.
\end{equation}

When $R_{o,H} \gg 1$, outflowing gas quickly expands  radially away from the planet in nearly every direction, creating a large scale-height outflow that resembles a bubble \citep[see top panel of Figure~1 in][]{MacLeodOklopcic2025}. For $R_{o,H}\lesssim 1$, 
the Coriolis acceleration deviates 
the outflow before it can significantly expand around the planet, the outflow is then sheared into tidal tails by the differential rotation of the Keplerian orbits, resulting in a stream or tail-like outflow morphology \citep[see bottom panel of Figure~1 in][]{MacLeodOklopcic2025}.

We use estimated outflow sound speeds from \citet{LinssenOklopcic2025} to compute $R_{o,H}$ for our survey targets (Figure~\ref{fig:stvsbub}) and find that all of them should have stream-like morphologies. If these planets have outflows sheared into tidal tails, there could be additional He$^*$ absorption outside of the nominal transit window that is undetectable in our Palomar/WIRC observations. If there is significant absorption outside of the transit window, this can potentially bias the excess absorption results towards smaller values or in the most severe cases, non-detections. Our 3D outflow predictions (Fig.~\ref{fig:stvsbub}) indicate that of our survey targets, WASP-103~b and WASP-12~b are most likely to have large extended tails of escaping planetary material that can lead and trail the planet. In fact, the $R_{o,H}$ values of these two survey targets are larger than WASP-121~b, where the escaping gas tail extends over approximately 60\% of the planet's orbit \citep{AllartCoulombe2025}.  It is therefore possible that extended absorption during the baseline of our observations could be artificially lowering our He$^*$ transit depths for both planets. However, we note that for both WASP-121~b and HAT-P-32~b, the excess absorption signal was still easily detectable in single-night ground-based transit observations despite this self-subtraction \citep{CzeslaLampon2022,Czesla2024}. 
In the case of WASP-103~b, the baseline of our 
Palomar observations covered approximately 40\% of the planet's orbit. This suggests that in order to fully mask the predicted He$^*$ signal, this planet would need to host an optically thick gas torus similar to the one invoked to explain published non-detections of mass loss from WASP-12~b as discussed in \S\ref{sec:wasp12_discussion}.
In the future, spectroscopically resolved observations with significant pre- and post-transit baseline could provide tighter constraints on the presence and extent of tail-like morphologies in these systems. Spectroscopic observations can also provide useful information on the kinematics of the outflow (from the line shifts) and the temperature of the outflow (from the line width), providing a more complete picture of the outflow dynamics.

\section{Conclusions}\label{sec:conc}

In this work, we present the first survey of atmospheric escape on planets orbiting F stars. Using Palomar/WIRC, we obtained He$^*$ transits of six close-in gas giants over the course of two years. We obtain two strong ($>3\sigma$) detections of atmospheric escape for WASP-12~b and WASP-180~A~b, two tentative ($2-3\sigma$) detections of atmospheric escape for HAT-P-8~b, and WASP-93~b, and place upper bounds on the atmospheric escape rates of WASP-103~b and KELT-7~b. We fit our measured He$^*$ excess absorption values with an energetically self-consistent 1D Parker wind model using the \texttt{sunbather} package, and find that WASP-12~b and WASP-180~A~b have corresponding mass loss rates of $\log\dot{M}=12.4^{+0.6}_{-0.5}$ (\unit{\gram\per\second}) and $\log\dot{M}=11.85^{+0.2}_{-0.2}$ (\unit{\gram\per\second}) respectively,
comparable to the fast outflows reported in the literature for most other planets orbiting early-type stars. 
Our mass loss constraints for our other four survey targets (HAT-P-8~b, WASP-93~b, WASP-103~b, and KELT-7~b) are similar to published measurements for WASP-48~b and WASP-94~A~b, which also orbit early-type stars, and are broadly in line with measured mass loss rates for gas giants orbiting cooler stars. 
This suggests that the strong outflows reported in the literature for planets orbiting early-type stars are not representative of all early-type systems. We do not see any evidence for a correlation between the measured mass loss rate and the NUV flux received by the planet, which suggests that Balmer-driven escape is not the dominant driver of atmospheric escape for most planets in our sample. 
However, the high NUV stellar fluxes that power Balmer-driven escape also depopulate He$^*$.  As a result, He$^*$ observations might not be the optimal tracer to use for this particular escape process. Instead, transmission spectroscopy of Balmer absorption lines is likely more suitable for quantifying the efficacy of NUV-powered escape on a population level. Notably, the mass loss measurement for KELT-9~b, which has the highest NUV flux of all the planets in our sample, was obtained using these lines \citep{Wyttenbach2020}.

We combine our sample of retrieved mass loss rates with mass loss measurements from the literature to explore the factors that determine outflow strength in these systems.  We find that the planets with the highest Roche filling factors tend to have the highest measured mass loss rates, although our non-detection of an outflow for WASP-103~b is a puzzling exception to this trend.  WASP-180~A~b has a relatively low Roche filling factor and an anomalously high measured XUV luminosity, and our detection of an outflow on this planet demonstrates the importance of XUV luminosity as a second variable controlling outflow strength.  We find that effective temperature and $v{\sin}i_*$ are not strongly correlated with XUV luminosity for the stars in our sample, underscoring the importance of direct measurements of XUV luminosities for the host stars of planets with detected outflows.  Notably, WASP-103 does not have a measured XUV luminosity and our non-detection of an outflow in this system could be explained by a lower-than-expected XUV flux. 

Our photometric He$^*$ observations are relatively insensitive to the presence of extended outflow structures similar to those observed for HAT-P-32~b and HAT-P-67~b \citep{ZhangMorley2023,Gully-SantiagoMorley2024}. Instead, we use a simple scaling law to predict the 3D outflow geometries for all of our survey targets. We find that all of them should have tail-like outflow morphologies, some of which may be detectable in spectroscopically resolved observations with extended baselines.

Our detection of a strong outflow from WASP-12~b is surprising, as this planet's spectroscopically resolved He$^*$ signal was observed in 2019 and 2020 with CARMENES, resulting in a non-detection \citep{CzeslaLampon2024}. The mass loss model that provides the best fit to our Palomar/WIRC He$^*$ excess absorption has a peak excess absorption of 6\%, which should have been easily detectable with CARMENES.  If the outflow from this planet forms an extended torus around the star, the discrepancy between our strong detection and the CARMENES non-detection 
might potentially be explained by temporal variability in the disk properties. If the planet was observed during a period where the gas density in the torus was relatively high, it could obscure the He$^*$ outflow from the planet. This could be tested by long-term monitoring of the stellar He$^*$ line equivalent width over a period of multiple years. This planet's He$^*$ absorption will be re-measured as part of upcoming JWST NIRISS/SOSS transit observations, providing another data point on this timeseries.

\begin{acknowledgments}

This material is based upon work supported by the National Science Foundation Graduate Research Fellowship Program under Grant No.~DGE‐1745301. Any opinions, findings, and conclusions or recommendations expressed in this material are those of the author(s) and do not necessarily reflect the views of the National Science Foundation. This research made use of \texttt{nep-des} (available in \url{https://github.com/castro-gzlz/nep-des}). Co-author WGL gratefully acknowledges support from the Department of Defense's National Defense Science \& Engineering Graduate (NDSEG) Fellowship. WGL also thanks the LSST-DA Data Science Fellowship Program, which is funded by LSST-DA, the Brinson Foundation, the WoodNext Foundation, and the Research Corporation for Science Advancement Foundation; his participation in the program has benefited this work.

We thank James Owen and Ruth Murray-Clay for their informative discussion. We also thank the Palomar Observatory telescope and support operators for their support of this work, with special thanks to Tom Barlow, Carolyn Heffner, Isaac
Wilson, Diana Roderick, Paul Neid, Kathleen Koviak, Rigel Rafto, and John Stone.
\end{acknowledgments}

\facilities{ADS, Exoplanet Archive, Hale}

\software{\texttt{ATES} \citep{Caldiroli2021}, \texttt{ldtk} \citep{ParviainenAigrain2015}, \texttt{nep-des} \citep{Castro-GonzalezBourrier2024}, \texttt{photutils} \citep{BradleySipocz2023}, \texttt{p-winds} \citep{DosSantosVidotto2022}, \texttt{PWMorphology} \citep{MacLeodOklopcic2025},
\texttt{pymc3} \citep{SalvatierWiecki2016}, \texttt{sunbather} \citep{LinssenShih2024}, \texttt{Xspec} \citep{Arnaud1996}         }

\appendix

\section{Systematics Model} 

\begin{figure}
    \centering
    \includegraphics[width=\linewidth]{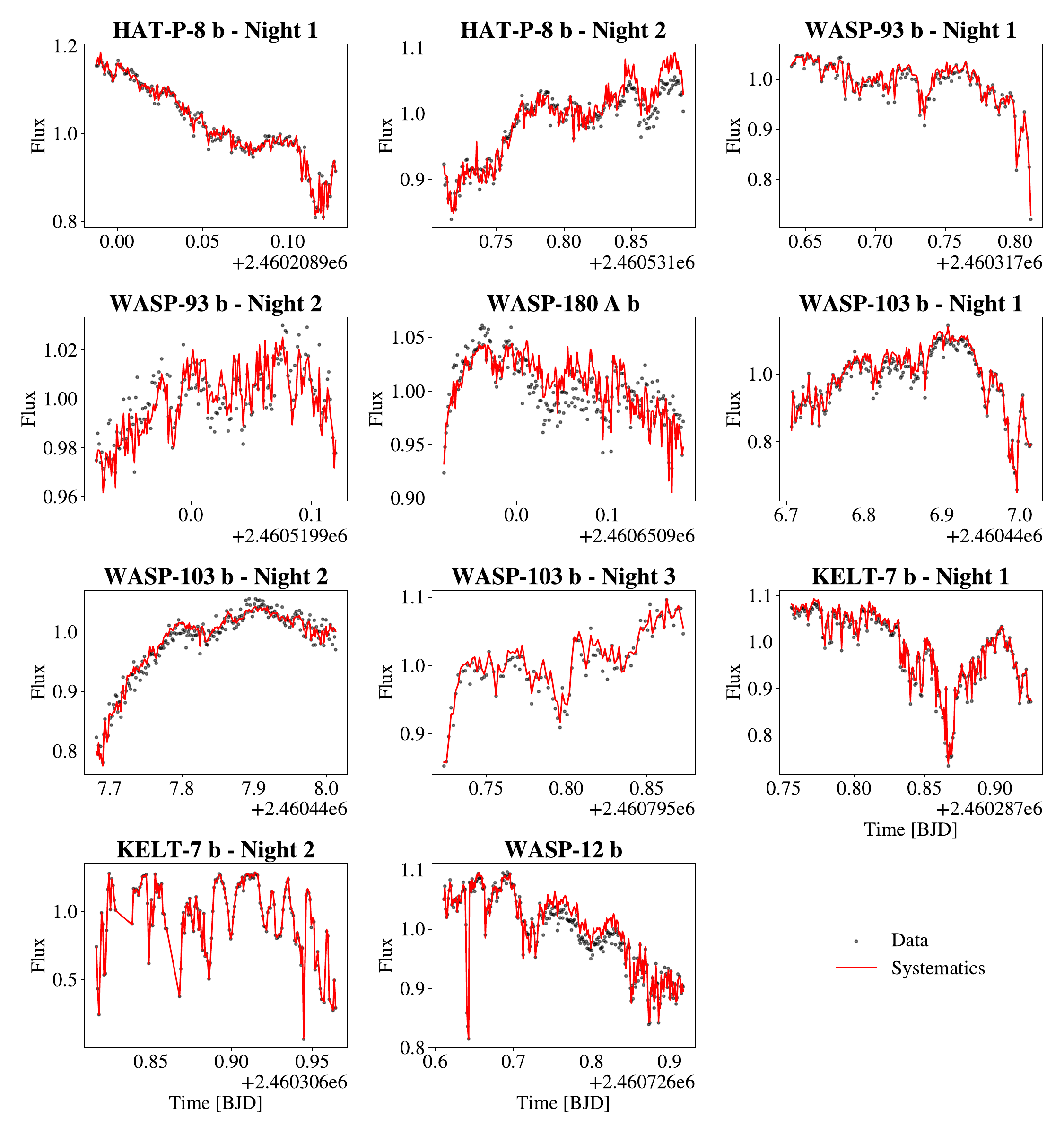}
    \caption{White light curves and systematics models for all survey targets, with each night of observation displayed separately. Black points correspond to the sigma clipped flux in the optimal aperture. Red lines represent the systematics model, which is comprised of a linear trend in time, a linear combination of comparison star light curves, and a combination of additional covariates (airmass, the distance from the median centroid, the psf width, the time varying telluric water absorption proxy, no covariates) that minimize the BIC (see Section~\ref{sec:lightcurve} for more information).}
    \label{fig:systematics}
\end{figure}

\begin{figure}
    \centering
    \includegraphics[width=\linewidth]{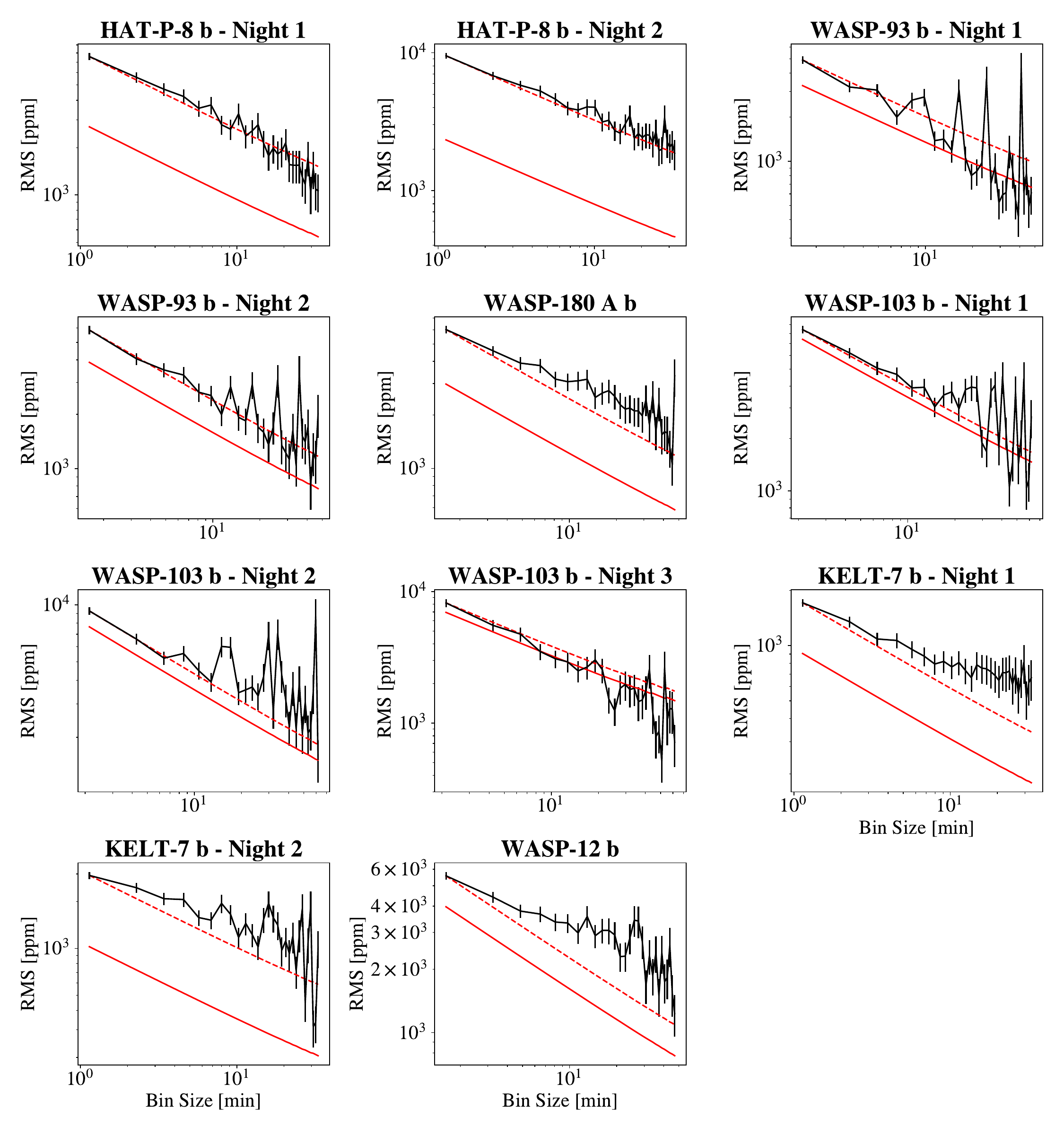}
    \caption{RMS versus bins plots for all survey targets, with each night of observation displayed separately. Each plot contains the rms of the binned residuals (black points), the noise expected from photon noise statistics (red solid line), and the expected noise rescaled to match the the rms of the unbinned residuals (dashed red line).}
    \label{fig:allandevi}
\end{figure}

In Figure~\ref{fig:systematics}, we display white light curves and the corresponding systematics models for each survey target. For targets in which multiple observations occurred, we display each night separately. We also present RMS versus bins plots for all observations in Figure~\ref{fig:allandevi}. 

\section{Limb Darkening Coefficients}\label{limbdarkappendix}

\begin{figure}
    \centering
    \includegraphics[width=0.9\linewidth]{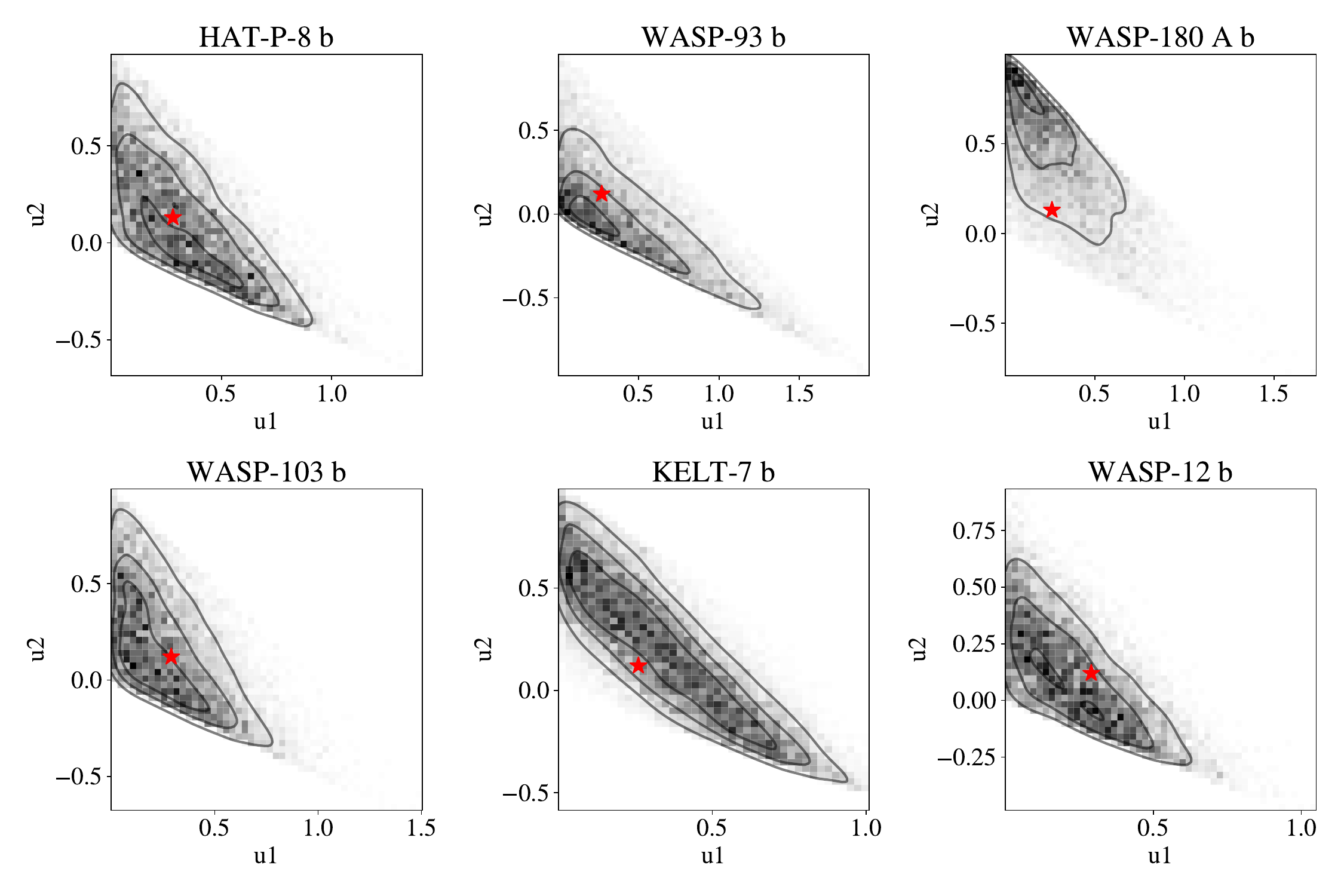}
    \caption{Posteriors for the quadratic limb darkening coefficients for all targets in this survey.  1, 2, and $3\sigma$ confidence levels are plotted as contours. The red star indicates the calculated limb darkening coefficients from \texttt{ldtk} \citep{Husser2013, ParviainenAigrain2015}.}
    \label{fig:LDCdistrib}
\end{figure}

The posterior distributions on the limb darkening coefficients for all stars in this survey are shown in Figure~\ref{fig:LDCdistrib} along with the calculated limb darkening coefficients from \texttt{ldtk} \citep{Husser2013, ParviainenAigrain2015}.

\section{WASP-33 X-ray Luminosity Reduction \& Analysis}\label{wasp33appendix}

The WASP-33 system was serendipitously observed by \textit{XMM-Newton} on 2017 Jan 31 (ObsID: 0785120201; PI: Agueros) for 8\,ks. A5V star WASP-33A was in, but near the edge of, the field of view of two of the European Photon Imaging Cameras (EPIC): pn and MOS1. We used the Scientific Analysis System (SAS) 22.1.0 to reduce the data. As determined through standard analysis processes\footnote{See the SAS Threads: \url{https://www.cosmos.esa.int/web/xmm-newton/sas-threads}}, high Solar particle background affected substantial portions of the observation, leaving us with only 2.6\,ks and 5.3\,ks of good time intervals for the pn and MOS1, respectively.

In the EPIC-pn image, there is potentially a very faint increase in the count rate at the location of WASP-33A, though no such increase is visible in the image taken with (the less sensitive) MOS1. Using a 15\,arcsec radius extraction region, we determined a 0.2--2.4\,keV count rate in EPIC-pn of $\left(1.7\pm1.0\right)\times10^{-3}$\,s$^{-1}$, and therefore not a statistically significant detection. For reference, a detection with this count rate, assuming a single APEC model \citep{Smith2001} with a temperature of 0.3\,keV and a hydrogen column density of $4.8\times10^{18}$\,cm$^{-2}$ \citep{Youngblood2025}, would imply a flux of $\sim2\times10^{-15}$\,erg\,s$^{-1}$\,cm$^{-2}$.

Given the inconclusive detection, we instead placed an upper limit on the flux. In order to do that, we first placed 95\% confidence upper limits on the count rate in each of the pn and MOS1 cameras using the framework of \citet{Kraft1991}. To combine the information from the two cameras together and convert from count rate to flux, we used equation 10 of \citet{Ruiz2022}. For the encircled energy fraction (EEF) parameter, we adopted values of 0.68 and 0.69 for pn and MOS1, respectively, taken from table 1 of the same study.

The energy to count conversion factor (ECF) for each detector depends on a number of factors, most notably the underlying stellar spectrum and its intersection with the instruments' responses as a function of energy. However, other relevant factors include the date of the observation (as instrument response may change over time), the position of the object on the detector, and the choice of optical blocking filter. As such, rather than using tabulated ECF values\footnote{e.g. those available at \url{https://www.cosmos.esa.int/web/xmm-newton/epic-upper-limits}}, we used a manual two step process for obtaining ECFs to use with equation 10 of \citet{Ruiz2022}.

First, we analyzed pn and MOS1 data of Altair (ObsID: 0502360101; PI: Robrade), another A-type star that should therefore provide a realistic expectation for the X-ray spectral shape of WASP-33A. Using \texttt{Xspec} \citep{Arnaud1996}, we simultaneously fitted a two temperature APEC model to the pn and MOS1 spectra, together with a TBABS model for interstellar absorption \citep{Wilms2000}, using a hydrogen column density of $1.26\times10^{18}$\,cm$^{-2}$ \citep{Youngblood2025}. Abundances were held at Solar \citep{Asplund2009}. 

Second, we then took the best fit model from Altair, combined it with response files we generated from the WASP-33 observation, and used \texttt{Xspec's} ``fakeit" command to create synthetic spectra for each of pn and MOS1. This simulates a situation where our proxy Altair had been observed at the same time and at the same position on the detectors as WASP-33A was in our observations. We divided the 0.2--2.4\,keV count rates of this faked data by the flux corresponding to the underlying model in the same band to obtain realistic ECFs for the WASP-33A data.

Bringing this all together and applying equation 10 of \citet{Ruiz2022}, we obtain a final 95\% confidence upper limit on the flux of $2.9\times10^{-14}$\,erg\,s$^{-1}$\,cm$^{-2}$. Based on the \textit{Gaia} DR3 parallax \citep{GaiaDR3}, this would imply an upper limit on the X-ray Luminosity, $L_{\rm X}$ of $5.3\times10^{28}$\,erg\,s$^{-1}$. We use Eqn.~1 in \citet{GeorgeKing2018} to place a constraint on the extreme ultra-violet and XUV luminosity, which we use to estimate an XUV flux, $F_{XUV}$ of $1.1\times10^{5}$\,erg\,s$^{-1}$\,cm$^{-2}$ at the planet's orbit. Finally, using the bolometric luminosity, $L_{\rm bol}$, of WASP-33A from the TESS Input Catalog v8 \citep{Stassun2019}, we find $L_{\rm X}/L_{\rm bol} < 2.1\times10^{-6}$. This upper limit agrees with the few unambiguous detections of X-rays from A-type stars made to date \citep{Guenther2022}, which are generally in the range $L_{\rm X}/L_{\rm bol} \sim 10^{-6}-10^{-8}$. WASP-33A has an effective temperature (7300\,K) below the $\sim8300$\,K limit where coronal and chromospheric emission is thought to switch off due to the lack of a convection zone \citep{Simon2002}, and thus would perhaps be expected to reside in this range, around or somewhat below our upper limit.

This may not be the full story though, as WASP-33A has two potential stellar companions. A co-moving G star, labeled WASP-33C, has been identified at 49\,arcsec from the primary \citep{Mugrauer2019}. It shows a faint detection by eye in the EPIC-pn image, but is sufficiently separated from star A that contamination is negligible. However, a closer likely companion, WASP-33B (so named because of its earlier identification), has been observed at just 2\,arcsec from the primary \citep{Moya2011,Adams2013,Wollert2015,Ngo2016}, though some of these studies did not rule out the possibility of the source being a background extragalactic object. Regardless, this separation is considerably smaller than the psf of the \textit{XMM-Newton} EPIC detectors, particularly for objects close to the edge of the field of view. Thus, our upper limit is in reality that of the combined X-ray brightness of these two objects. 

Would WASP-33B be expected to brighter or dimmer in X-rays than star A? If it is a stellar companion, its color-derived $T_{\rm eff}\sim3200$\,K \citep{Ngo2016} implies a spectral type of around M4 and $L_{\rm bol} \sim 0.007$\,L$_{\odot}$ \citep{Pecaut2013}. Furthermore, \citet{Moya2011} estimated a system age of less than 500\,Myr. If true, this is an age for which a mid-M star should still be in the saturated regime of X-ray emission: i.e. $L_{\rm X}/L_{\rm bol} \sim 10^{-3}$ \citep[e.g.][]{Engle2024}. Under this assumption, its X-ray flux at Earth would be around $10^{-14}$\,erg\,s$^{-1}$, slightly below our upper flux limit. This is a similar brightness to the upper end of expected fluxes for WASP-33A based on other A-type stars, but if the primary's $L_{\rm X}/L_{\rm bol}$ were towards the $10^{-8}$ lower end, then WASP-33B could be considerably brighter in X-rays despite its vastly dimmer $L_{\rm bol}$. A much deeper observation with \textit{Chandra}, whose superior spatial resolution would be able to separate stars A and B, could help detect either or both of the stars, distinguish between these possibilities, and provide a much tighter constraint on the XUV irradiation of WASP-33Ab.

\bibliography{citations}{}

\begin{thebibliography}{}
\expandafter\ifx\csname natexlab\endcsname\relax\def\natexlab#1{#1}\fi
\providecommand{\url}[1]{\href{#1}{#1}}
\providecommand{\dodoi}[1]{doi:~\href{http://doi.org/#1}{\nolinkurl{#1}}}
\providecommand{\doeprint}[1]{\href{http://ascl.net/#1}{\nolinkurl{http://ascl.net/#1}}}
\providecommand{\doarXiv}[1]{\href{https://arxiv.org/abs/#1}{\nolinkurl{https://arxiv.org/abs/#1}}}

\bibitem[{E.~R. {Adams} {et~al.}(2013){Adams}, {Dupree}, {Kulesa}, \& {McCarthy}}]{Adams2013}
{Adams}, E.~R., {Dupree}, A.~K., {Kulesa}, C., \& {McCarthy}, D. 2013, \bibinfo{title}{{Adaptive Optics Images. II. 12 Kepler Objects of Interest and 15 Confirmed Transiting Planets},} \aj, 146, 9, \dodoi{10.1088/0004-6256/146/1/9}

\bibitem[{B.~C. {Addison} {et~al.}(2016){Addison}, {Tinney}, {Wright}, \& {Bayliss}}]{AddisonTinney2016}
{Addison}, B.~C., {Tinney}, C.~G., {Wright}, D.~J., \& {Bayliss}, D. 2016, \bibinfo{title}{{Spin-orbit Alignment for Three Transiting Hot Jupiters: WASP-103b, WASP-87b, and WASP-66b},} The Astrophysical Journal, 823, 29, \dodoi{10.3847/0004-637X/823/1/29}

\bibitem[{E.-M. {Ahrer} {et~al.}(2025){Ahrer}, {Fairman}, {Kirk}, {Wakeford}, {Barstow}, {Penzlin}, {Alderson}, {Booth}, {Christie}, {Claringbold}, {Esparza-Borges}, {Gasc{\'o}n}, {L{\'o}pez-Morales}, {Mayne}, {McCormack}, {Meech}, {Molli{\`e}re}, {Owen}, {Panwar}, {Sergeev}, {Valentine}, {Wheatley}, \& {Zamyatina}}]{Ahrer2025}
{Ahrer}, E.-M., {Fairman}, C., {Kirk}, J., {et~al.} 2025, \bibinfo{title}{{BOWIE-ALIGN: Weak spectral features in KELT-7b's JWST NIRSpec/G395H transmission spectrum imply a high cloud deck or a low-metallicity atmosphere},} arXiv e-prints, arXiv:2509.12479.
\newblock \doarXiv{2509.12479}

\bibitem[{R.~L. {Akeson} {et~al.}(2013){Akeson}, {Chen}, {Ciardi}, \& et~al.}]{AkesonChen2013}
{Akeson}, R.~L., {Chen}, X., {Ciardi}, D., \& et~al. 2013, \bibinfo{title}{{The NASA Exoplanet Archive: Data and Tools for Exoplanet Research},} Publications of the Astronomical Society of the Pacific, 125, 989, \dodoi{10.1086/672273}

\bibitem[{B. {Akinsanmi} {et~al.}(2024){Akinsanmi}, {Barros}, {Lendl}, \& et~al.}]{AkinsanmiBarros2024}
{Akinsanmi}, B., {Barros}, S.~C.~C., {Lendl}, M., \& et~al. 2024, \bibinfo{title}{{The tidal deformation and atmosphere of WASP-12 b from its phase curve★},} Astronomy and Astrophysics, 685, A63, \dodoi{10.1051/0004-6361/202348502}

\bibitem[{S. {Albrecht} {et~al.}(2012){Albrecht}, {Winn}, {Johnson}, \& et~al.}]{AlbrechtWinn2012}
{Albrecht}, S., {Winn}, J.~N., {Johnson}, J.~A., \& et~al. 2012, \bibinfo{title}{{Obliquities of Hot Jupiter Host Stars: Evidence for Tidal Interactions and Primordial Misalignments},} The Astrophysical Journal, 757, 18, \dodoi{10.1088/0004-637X/757/1/18}

\bibitem[{R. {Allart} {et~al.}(2025){Allart}, {Coulombe}, {Carteret}, \& et~al.}]{AllartCoulombe2025}
{Allart}, R., {Coulombe}, L.-P., {Carteret}, Y., \& et~al. 2025, \bibinfo{title}{{A complex structure of escaping helium spanning more than half the orbit of the ultra-hot Jupiter WASP-121 b},} Nature Communications, 16, 10822, \dodoi{10.1038/s41467-025-66628-5}

\bibitem[{K.~R. {Anderson} {et~al.}(2016){Anderson}, {Storch}, \& {Lai}}]{AndersonStorch2016}
{Anderson}, K.~R., {Storch}, N.~I., \& {Lai}, D. 2016, \bibinfo{title}{{Formation and stellar spin-orbit misalignment of hot Jupiters from Lidov-Kozai oscillations in stellar binaries},} Monthly Notices of the Royal Astronomical Society, 456, 3671, \dodoi{10.1093/mnras/stv2906}

\bibitem[{K.~A. {Arnaud}(1996){Arnaud}}]{Arnaud1996}
{Arnaud}, K.~A. 1996, in Astronomical Society of the Pacific Conference Series, Vol. 101, Astronomical Data Analysis Software and Systems V, ed. G.~H. {Jacoby} \& J.~{Barnes}, 17

\bibitem[{M. {Asplund} {et~al.}(2009){Asplund}, {Grevesse}, {Sauval}, \& {Scott}}]{Asplund2009}
{Asplund}, M., {Grevesse}, N., {Sauval}, A.~J., \& {Scott}, P. 2009, \bibinfo{title}{{The Chemical Composition of the Sun},} \araa, 47, 481, \dodoi{10.1146/annurev.astro.46.060407.145222}

\bibitem[{A. {Bailey} \& J. {Goodman}(2019){Bailey} \& {Goodman}}]{BaileyGoodman2019}
{Bailey}, A., \& {Goodman}, J. 2019, \bibinfo{title}{{Understanding WASP-12b},} Monthly Notices of the Royal Astronomical Society, 482, 1872, \dodoi{10.1093/mnras/sty2805}

\bibitem[{S.~C.~C. {Barros} {et~al.}(2022){Barros}, {Akinsanmi}, {Bou{\'e}}, \& et~al.}]{BarrosAkinsanmi2022}
{Barros}, S.~C.~C., {Akinsanmi}, B., {Bou{\'e}}, G., \& et~al. 2022, \bibinfo{title}{{Detection of the tidal deformation of WASP-103b at 3 {\ensuremath{\sigma}} with CHEOPS},} Astronomy and Astrophysics, 657, A52, \dodoi{10.1051/0004-6361/202142196}

\bibitem[{E.~B. {Bechter} {et~al.}(2014){Bechter}, {Crepp}, {Ngo}, {Knutson}, {Batygin}, {Hinkley}, {Muirhead}, {Johnson}, {Howard}, {Montet}, {Matthews}, \& {Morton}}]{Bechter2014}
{Bechter}, E.~B., {Crepp}, J.~R., {Ngo}, H., {et~al.} 2014, \bibinfo{title}{{WASP-12b and HAT-P-8b are Members of Triple Star Systems},} \apj, 788, 2, \dodoi{10.1088/0004-637X/788/1/2}

\bibitem[{T.~J. {Bell} {et~al.}(2019){Bell}, {Zhang}, {Cubillos}, \& et~al.}]{BellZhang2019}
{Bell}, T.~J., {Zhang}, M., {Cubillos}, P.~E., \& et~al. 2019, \bibinfo{title}{{Mass loss from the exoplanet WASP-12b inferred from Spitzer phase curves},} Monthly Notices of the Royal Astronomical Society, 489, 1995, \dodoi{10.1093/mnras/stz2018}

\bibitem[{A. {Bello-Arufe} {et~al.}(2023){Bello-Arufe}, {Knutson}, {Mendon{\c{c}}a}, {Zhang}, {Cabot}, {Rathcke}, {Ulla}, {Vissapragada}, \& {Buchhave}}]{BelloArufe2023AJ}
{Bello-Arufe}, A., {Knutson}, H.~A., {Mendon{\c{c}}a}, J.~M., {et~al.} 2023, \bibinfo{title}{{Transmission Spectroscopy of the Lowest-density Gas Giant: Metals and a Potential Extended Outflow in HAT-P-67b},} \aj, 166, 69, \dodoi{10.3847/1538-3881/acd935}

\bibitem[{K.~A. {Bennett} {et~al.}(2023){Bennett}, {Redfield}, {Oklop{\v{c}}i{\'c}}, {Carleo}, {Ninan}, \& {Endl}}]{Bennett2023AJ}
{Bennett}, K.~A., {Redfield}, S., {Oklop{\v{c}}i{\'c}}, A., {et~al.} 2023, \bibinfo{title}{{Nondetection of Helium in the Hot Jupiter WASP-48b},} \aj, 165, 264, \dodoi{10.3847/1538-3881/acd34b}

\bibitem[{A. {Bieryla} {et~al.}(2015){Bieryla}, {Collins}, {Beatty}, \& et~al.}]{BierylaCollins2015}
{Bieryla}, A., {Collins}, K., {Beatty}, T.~G., \& et~al. 2015, \bibinfo{title}{{KELT-7b: A Hot Jupiter Transiting a Bright V = 8.54 Rapidly Rotating F-star},} The Astronomical Journal, 150, 12, \dodoi{10.1088/0004-6256/150/1/12}

\bibitem[{A.~S. {Bonomo} {et~al.}(2017){Bonomo}, {Desidera}, {Benatti}, \& et~al.}]{BonomoDesidera2017}
{Bonomo}, A.~S., {Desidera}, S., {Benatti}, S., \& et~al. 2017, \bibinfo{title}{{The GAPS Programme with HARPS-N at TNG . XIV. Investigating giant planet migration history via improved eccentricity and mass determination for 231 transiting planets},} Astronomy and Astrophysics, 602, A107, \dodoi{10.1051/0004-6361/201629882}

\bibitem[{V. {Bourrier} {et~al.}(2020){Bourrier}, {Ehrenreich}, {Lendl}, {Cretignier}, {Allart}, {Dumusque}, {Cegla}, {Su{\'a}rez-Mascare{\~n}o}, {Wyttenbach}, {Hoeijmakers}, {Melo}, {Kuntzer}, {Astudillo-Defru}, {Giles}, {Heng}, {Kitzmann}, {Lavie}, {Lovis}, {Murgas}, {Nascimbeni}, {Pepe}, {Pino}, {Segransan}, \& {Udry}}]{Bourrier2020}
{Bourrier}, V., {Ehrenreich}, D., {Lendl}, M., {et~al.} 2020, \bibinfo{title}{{Hot Exoplanet Atmospheres Resolved with Transit Spectroscopy (HEARTS). III. Atmospheric structure of the misaligned ultra-hot Jupiter WASP-121b},} \aap, 635, A205, \dodoi{10.1051/0004-6361/201936640}

\bibitem[{L. {Bradley} {et~al.}(2023){Bradley}, {Sip{\H{o}}cz}, {Robitaille}, \& et~al.}]{BradleySipocz2023}
{Bradley}, L., {Sip{\H{o}}cz}, B., {Robitaille}, T., \& et~al. 2023, \bibinfo{title}{{astropy/photutils: 1.7.0},}, 1.7.0 Zenodo, \dodoi{10.5281/zenodo.7804137}

\bibitem[{A. {Caldiroli} {et~al.}(2022){Caldiroli}, {Haardt}, {Gallo}, \& et~al.}]{CaldiroliHaardt2022}
{Caldiroli}, A., {Haardt}, F., {Gallo}, E., \& et~al. 2022, \bibinfo{title}{{Irradiation-driven escape of primordial planetary atmospheres. II. Evaporation efficiency of sub-Neptunes through hot Jupiters},} Astronomy and Astrophysics, 663, A122, \dodoi{10.1051/0004-6361/202142763}

\bibitem[{A. {Caldiroli} {et~al.}(2021){Caldiroli}, {Haardt}, {Gallo}, {Spinelli}, {Malsky}, \& {Rauscher}}]{Caldiroli2021}
{Caldiroli}, A., {Haardt}, F., {Gallo}, E., {et~al.} 2021, \bibinfo{title}{{Irradiation-driven escape of primordial planetary atmospheres. I. The ATES photoionization hydrodynamics code},} \aap, 655, A30, \dodoi{10.1051/0004-6361/202141497}

\bibitem[{A.~J. {Cannon} \& E.~C. {Pickering}(1918){Cannon} \& {Pickering}}]{CannonPickering1918}
{Cannon}, A.~J., \& {Pickering}, E.~C. 1918, \bibinfo{title}{{The Henry Draper catalogue 0h, 1h, 2h, and 3h},} Annals of Harvard College Observatory, 91, 1

\bibitem[{K.~M.~S. {Cartier} {et~al.}(2017){Cartier}, {Beatty}, {Zhao}, \& et~al.}]{CartierBeatty2017}
{Cartier}, K. M.~S., {Beatty}, T.~G., {Zhao}, M., \& et~al. 2017, \bibinfo{title}{{Near-infrared Emission Spectrum of WASP-103b Using Hubble Space Telescope/Wide Field Camera 3},} The Astronomical Journal, 153, 34, \dodoi{10.3847/1538-3881/153/1/34}

\bibitem[{N. {Casasayas-Barris} {et~al.}(2019){Casasayas-Barris}, {Pall{\'e}}, {Yan}, {Chen}, {Kohl}, {Stangret}, {Parviainen}, {Helling}, {Watanabe}, {Czesla}, {Fukui}, {Monta{\~n}{\'e}s-Rodr{\'\i}guez}, {Nagel}, {Narita}, {Nortmann}, {Nowak}, {Schmitt}, \& {Zapatero Osorio}}]{CasasayasBarris2019}
{Casasayas-Barris}, N., {Pall{\'e}}, E., {Yan}, F., {et~al.} 2019, \bibinfo{title}{{Atmospheric characterization of the ultra-hot Jupiter MASCARA-2b/KELT-20b. Detection of CaII, FeII, NaI, and the Balmer series of H (H{\ensuremath{\alpha}}, H{\ensuremath{\beta}}, and H{\ensuremath{\gamma}}) with high-dispersion transit spectroscopy},} \aap, 628, A9, \dodoi{10.1051/0004-6361/201935623}

\bibitem[{A. {Castro-Gonz{\'a}lez} {et~al.}(2024){Castro-Gonz{\'a}lez}, {Bourrier}, \& et~al.}]{Castro-GonzalezBourrier2024}
{Castro-Gonz{\'a}lez}, A., {Bourrier}, V., \& et~al. 2024, \bibinfo{title}{{Mapping the exo-Neptunian landscape: A ridge between the desert and savanna},} Astronomy and Astrophysics, 689, A250, \dodoi{10.1051/0004-6361/202450957}

\bibitem[{A. {Chakrabarty} \& S. {Sengupta}(2019){Chakrabarty} \& {Sengupta}}]{ChakrabartySengupta2019}
{Chakrabarty}, A., \& {Sengupta}, S. 2019, \bibinfo{title}{{Precise Photometric Transit Follow-up Observations of Five Close-in Exoplanets: Update on Their Physical Properties},} The Astronomical Journal, 158, 39, \dodoi{10.3847/1538-3881/ab24dd}

\bibitem[{J.~L. {Christiansen} {et~al.}(2025){Christiansen}, {McElroy}, {Harbut}, \& et~al.}]{ChristiansenMcElroy2025}
{Christiansen}, J.~L., {McElroy}, D.~L., {Harbut}, M., \& et~al. 2025, \bibinfo{title}{{The NASA Exoplanet Archive and Exoplanet Follow-up Observing Program: Data, Tools, and Usage},} arXiv e-prints, arXiv:2506.03299, \dodoi{10.48550/arXiv.2506.03299}

\bibitem[{A. {Collier Cameron} {et~al.}(2010){Collier Cameron}, {Guenther}, {Smalley}, {McDonald}, {Hebb}, {Andersen}, {Augusteijn}, {Barros}, {Brown}, {Cochran}, {Endl}, {Fossey}, {Hartmann}, {Maxted}, {Pollacco}, {Skillen}, {Telting}, {Waldmann}, \& {West}}]{CollierCameron2010}
{Collier Cameron}, A., {Guenther}, E., {Smalley}, B., {et~al.} 2010, \bibinfo{title}{{Line-profile tomography of exoplanet transits - II. A gas-giant planet transiting a rapidly rotating A5 star},} \mnras, 407, 507, \dodoi{10.1111/j.1365-2966.2010.16922.x}

\bibitem[{S. {Czesla} {et~al.}(2024{\natexlab{a}}){Czesla}, {Lamp{\'o}n}, {Cont}, \& et~al.}]{CzeslaLampon2024}
{Czesla}, S., {Lamp{\'o}n}, M., {Cont}, D., \& et~al. 2024{\natexlab{a}}, \bibinfo{title}{{The elusive atmosphere of WASP-12 b. High-resolution transmission spectroscopy with CARMENES},} Astronomy and Astrophysics, 683, A67, \dodoi{10.1051/0004-6361/202348107}

\bibitem[{S. {Czesla} {et~al.}(2022){Czesla}, {Lamp{\'o}n}, {Sanz-Forcada}, \& et~al.}]{CzeslaLampon2022}
{Czesla}, S., {Lamp{\'o}n}, M., {Sanz-Forcada}, J., \& et~al. 2022, \bibinfo{title}{{H{\ensuremath{\alpha}} and He I absorption in HAT-P-32 b observed with CARMENES. Detection of Roche lobe overflow and mass loss},} Astronomy and Astrophysics, 657, A6, \dodoi{10.1051/0004-6361/202039919}

\bibitem[{S. {Czesla} {et~al.}(2024{\natexlab{b}}){Czesla}, {Nail}, {Lavail}, {Cont}, {Nortmann}, {Lesjak}, {Rengel}, {Boldt-Christmas}, {Shulyak}, {Seemann}, {Schneider}, {Hatzes}, {Kochukhov}, {Piskunov}, {Reiners}, {Wilson}, \& {Yan}}]{Czesla2024}
{Czesla}, S., {Nail}, F., {Lavail}, A., {et~al.} 2024{\natexlab{b}}, \bibinfo{title}{{The overflowing atmosphere of WASP-121 b: High-resolution He I {\ensuremath{\lambda}}10833 transmission spectroscopy with VLT/CRIRES$^{+}$},} \aap, 692, A230, \dodoi{10.1051/0004-6361/202451003}

\bibitem[{A. {Debrecht} {et~al.}(2018){Debrecht}, {Carroll-Nellenback}, {Frank}, \& et~al.}]{DebrechtCarroll-Nellenback2018}
{Debrecht}, A., {Carroll-Nellenback}, J., {Frank}, A., \& et~al. 2018, \bibinfo{title}{{Generation of a circumstellar gas disc by hot Jupiter WASP-12b},} Monthly Notices of the Royal Astronomical Society, 478, 2592, \dodoi{10.1093/mnras/sty1164}

\bibitem[{L. {Delrez} {et~al.}(2016){Delrez}, {Santerne}, {Almenara}, {Anderson}, {Collier-Cameron}, {D{\'\i}az}, {Gillon}, {Hellier}, {Jehin}, {Lendl}, {Maxted}, {Neveu-VanMalle}, {Pepe}, {Pollacco}, {Queloz}, {S{\'e}gransan}, {Smalley}, {Smith}, {Triaud}, {Udry}, {Van Grootel}, \& {West}}]{Delrez2016}
{Delrez}, L., {Santerne}, A., {Almenara}, J.~M., {et~al.} 2016, \bibinfo{title}{{WASP-121 b: a hot Jupiter close to tidal disruption transiting an active F star},} \mnras, 458, 4025, \dodoi{10.1093/mnras/stw522}

\bibitem[{L.~A. {Dos Santos}(2023){Dos Santos}}]{DosSantos2023}
{Dos Santos}, L.~A. 2023, in IAU Symposium, Vol. 370, Winds of Stars and Exoplanets, ed. A.~A. {Vidotto}, L.~{Fossati}, \& J.~S. {Vink}, 56--71, \dodoi{10.1017/S1743921322004239}

\bibitem[{L.~A. {Dos Santos} {et~al.}(2022){Dos Santos}, {Vidotto}, {Vissapragada}, \& et~al.}]{DosSantosVidotto2022}
{Dos Santos}, L.~A., {Vidotto}, A.~A., {Vissapragada}, S., \& et~al. 2022, \bibinfo{title}{{p-winds: An open-source Python code to model planetary outflows and upper atmospheres},} Astronomy and Astrophysics, 659, A62, \dodoi{10.1051/0004-6361/202142038}

\bibitem[{D. {Durante} {et~al.}(2020){Durante}, {Parisi}, {Serra}, \& et~al.}]{DuranteParisi2020}
{Durante}, D., {Parisi}, M., {Serra}, D., \& et~al. 2020, \bibinfo{title}{{Jupiter's Gravity Field Halfway Through the Juno Mission},} Geophysical Research Letters, 47, e86572, \dodoi{10.1029/2019GL086572}

\bibitem[{S.~G. {Engle}(2024){Engle}}]{Engle2024}
{Engle}, S.~G. 2024, \bibinfo{title}{{Living with a Red Dwarf: X-Ray, UV, and Ca II Activity-Age Relationships of M Dwarfs},} \apj, 960, 62, \dodoi{10.3847/1538-4357/ad0840}

\bibitem[{G.~J. {Ferland} {et~al.}(2017){Ferland}, {Chatzikos}, \& et~al.}]{FerlandChatzikos2017}
{Ferland}, G.~J., {Chatzikos}, M., \& et~al. 2017, \bibinfo{title}{{The 2017 Release Cloudy},} Revista Mexicana de Astronomia y Astrofisica, 53, 385, \dodoi{10.48550/arXiv.1705.10877}

\bibitem[{G.~J. {Ferland} {et~al.}(1998){Ferland}, {Korista}, \& et~al.}]{FerlandKorista1998}
{Ferland}, G.~J., {Korista}, K.~T., \& et~al. 1998, \bibinfo{title}{{CLOUDY 90: Numerical Simulation of Plasmas and Their Spectra},} Publications of the Astronomical Society of the Pacific, 110, 761, \dodoi{10.1086/316190}

\bibitem[{D. {Foreman-Mackey} {et~al.}(2021){Foreman-Mackey}, {Luger}, {Agol}, \& et~al.}]{Foreman-MackeyLuger2021}
{Foreman-Mackey}, D., {Luger}, R., {Agol}, E., \& et~al. 2021, \bibinfo{title}{{exoplanet: Gradient-based probabilistic inference for exoplanet data \& other astronomical time series},} The Journal of Open Source Software, 6, 3285, \dodoi{10.21105/joss.03285}

\bibitem[{D. {Foreman-Mackey} {et~al.}(2024){Foreman-Mackey}, {Luger}, {Agol}, \& et~al.}]{Foreman-MackeyLuger2024}
{Foreman-Mackey}, D., {Luger}, R., {Agol}, E., \& et~al. 2024, \bibinfo{title}{{exoplanet: Gradient-based probabilistic inference for exoplanet data \& other astronomical time series},}, v0.6.0 Zenodo, \dodoi{10.5281/zenodo.1998447}

\bibitem[{L. {Fossati} {et~al.}(2013){Fossati}, {Ayres}, {Haswell}, \& et~al.}]{FossatiAyres2013}
{Fossati}, L., {Ayres}, T.~R., {Haswell}, C.~A., \& et~al. 2013, \bibinfo{title}{{Absorbing Gas around the WASP-12 Planetary System},} The Astrophysical Journal, 766, L20, \dodoi{10.1088/2041-8205/766/2/L20}

\bibitem[{L. {Fossati} {et~al.}(2010){Fossati}, {Haswell}, {Froning}, \& et~al.}]{FossatiHaswell2010}
{Fossati}, L., {Haswell}, C.~A., {Froning}, C.~S., \& et~al. 2010, \bibinfo{title}{{Metals in the Exosphere of the Highly Irradiated Planet WASP-12b},} The Astrophysical Journal, 714, L222, \dodoi{10.1088/2041-8205/714/2/L222}

\bibitem[{G. {Foster} \& K. {Poppenhaeger}(2022){Foster} \& {Poppenhaeger}}]{FosterPoppenhaeger2022}
{Foster}, G., \& {Poppenhaeger}, K. 2022, \bibinfo{title}{{Identifying interesting planetary systems for future X-ray observations},} Astronomische Nachrichten, 343, e20007, \dodoi{10.1002/asna.20220007}

\bibitem[{B.~J. {Fulton} {et~al.}(2017){Fulton}, {Petigura}, {Howard}, \& et~al.}]{FultonPetigura2017}
{Fulton}, B.~J., {Petigura}, E.~A., {Howard}, A.~W., \& et~al. 2017, \bibinfo{title}{{The California-Kepler Survey. III. A Gap in the Radius Distribution of Small Planets},} The Astronomical Journal, 154, 109, \dodoi{10.3847/1538-3881/aa80eb}

\bibitem[{ {Gaia Collaboration} {et~al.}(2023){Gaia Collaboration}, {Vallenari}, {Brown}, {Prusti}, {de Bruijne}, {Arenou}, {Babusiaux}, {Biermann}, {Creevey}, {Ducourant}, {Evans}, {Eyer}, {Guerra}, {Hutton}, {Jordi}, {Klioner}, {Lammers}, {Lindegren}, {Luri}, {Mignard}, {Panem}, {Pourbaix}, {Randich}, {Sartoretti}, {Soubiran}, {Tanga}, {Walton}, {Bailer-Jones}, {Bastian}, {Drimmel}, {Jansen}, {Katz}, {Lattanzi}, {van Leeuwen}, {Bakker}, {Cacciari}, {Casta{\~n}eda}, {De Angeli}, {Fabricius}, {Fouesneau}, {Fr{\'e}mat}, {Galluccio}, {Guerrier}, {Heiter}, {Masana}, {Messineo}, {Mowlavi}, {Nicolas}, {Nienartowicz}, {Pailler}, {Panuzzo}, {Riclet}, {Roux}, {Seabroke}, {Sordo}, {Th{\'e}venin}, {Gracia-Abril}, {Portell}, {Teyssier}, {Altmann}, {Andrae}, {Audard}, {Bellas-Velidis}, {Benson}, {Berthier}, {Blomme}, {Burgess}, {Busonero}, {Busso}, {C{\'a}novas}, {Carry}, {Cellino}, {Cheek}, {Clementini}, {Damerdji}, {Davidson}, {de Teodoro}, {Nu{\~n}ez Campos}, {Delchambre}, {Dell'Oro}, {Esquej},
  {Fern{\'a}ndez-Hern{\'a}ndez}, {Fraile}, {Garabato}, {Garc{\'\i}a-Lario}, {Gosset}, {Haigron}, {Halbwachs}, {Hambly}, {Harrison}, {Hern{\'a}ndez}, {Hestroffer}, {Hodgkin}, {Holl}, {Jan{\ss}en}, {Jevardat de Fombelle}, {Jordan}, {Krone-Martins}, {Lanzafame}, {L{\"o}ffler}, {Marchal}, {Marrese}, {Moitinho}, {Muinonen}, {Osborne}, {Pancino}, {Pauwels}, {Recio-Blanco}, {Reyl{\'e}}, {Riello}, {Rimoldini}, {Roegiers}, {Rybizki}, {Sarro}, {Siopis}, {Smith}, {Sozzetti}, {Utrilla}, {van Leeuwen}, {Abbas}, {{\'A}brah{\'a}m}, {Abreu Aramburu}, {Aerts}, {Aguado}, {Ajaj}, {Aldea-Montero}, {Altavilla}, {{\'A}lvarez}, {Alves}, {Anders}, {Anderson}, {Anglada Varela}, {Antoja}, {Baines}, {Baker}, {Balaguer-N{\'u}{\~n}ez}, {Balbinot}, {Balog}, {Barache}, {Barbato}, {Barros}, {Barstow}, {Bartolom{\'e}}, {Bassilana}, {Bauchet}, {Becciani}, {Bellazzini}, {Berihuete}, {Bernet}, {Bertone}, {Bianchi}, {Binnenfeld}, {Blanco-Cuaresma}, {Blazere}, {Boch}, {Bombrun}, {Bossini}, {Bouquillon}, {Bragaglia}, {Bramante}, {Breedt},
  {Bressan}, {Brouillet}, {Brugaletta}, {Bucciarelli}, {Burlacu}, {Butkevich}, {Buzzi}, {Caffau}, {Cancelliere}, {Cantat-Gaudin}, {Carballo}, {Carlucci}, {Carnerero}, {Carrasco}, {Casamiquela}, {Castellani}, {Castro-Ginard}, {Chaoul}, {Charlot}, {Chemin}, {Chiaramida}, {Chiavassa}, {Chornay}, {Comoretto}, {Contursi}, {Cooper}, {Cornez}, {Cowell}, {Crifo}, {Cropper}, {Crosta}, {Crowley}, {Dafonte}, {Dapergolas}, {David}, {David}, {de Laverny}, {De Luise}, \& {De March}}]{GaiaDR3}
{Gaia Collaboration}, {Vallenari}, A., {Brown}, A.~G.~A., {et~al.} 2023, \bibinfo{title}{{Gaia Data Release 3. Summary of the content and survey properties},} \aap, 674, A1, \dodoi{10.1051/0004-6361/202243940}

\bibitem[{A. {Garc{\'\i}a Mu{\~n}oz} \& P.~C. {Schneider}(2019){Garc{\'\i}a Mu{\~n}oz} \& {Schneider}}]{GarciaMunozSchneider2019}
{Garc{\'\i}a Mu{\~n}oz}, A., \& {Schneider}, P.~C. 2019, \bibinfo{title}{{Rapid Escape of Ultra-hot Exoplanet Atmospheres Driven by Hydrogen Balmer Absorption},} The Astrophysical Journal, 884, L43, \dodoi{10.3847/2041-8213/ab498d}

\bibitem[{C. {Gasc{\'o}n} {et~al.}(2025){Gasc{\'o}n}, {L{\'o}pez-Morales}, {Vissapragada}, {MacLeod}, {Wakeford}, {Grant}, {Ribas}, \& {Anglada-Escud{\'e}}}]{Gascon2025}
{Gasc{\'o}n}, C., {L{\'o}pez-Morales}, M., {Vissapragada}, S., {et~al.} 2025, \bibinfo{title}{{Modeling tails of escaping gas in exoplanet atmospheres with Harmonica},} arXiv e-prints, arXiv:2508.14846, \dodoi{10.48550/arXiv.2508.14846}

\bibitem[{B.~S. {Gaudi} {et~al.}(2017){Gaudi}, {Stassun}, {Collins}, {Beatty}, {Zhou}, {Latham}, {Bieryla}, {Eastman}, {Siverd}, {Crepp}, {Gonzales}, {Stevens}, {Buchhave}, {Pepper}, {Johnson}, {Colon}, {Jensen}, {Rodriguez}, {Bozza}, {Novati}, {D'Ago}, {Dumont}, {Ellis}, {Gaillard}, {Jang-Condell}, {Kasper}, {Fukui}, {Gregorio}, {Ito}, {Kielkopf}, {Manner}, {Matt}, {Narita}, {Oberst}, {Reed}, {Scarpetta}, {Stephens}, {Yeigh}, {Zambelli}, {Fulton}, {Howard}, {James}, {Penny}, {Bayliss}, {Curtis}, {Depoy}, {Esquerdo}, {Gould}, {Joner}, {Kuhn}, {Labadie-Bartz}, {Lund}, {Marshall}, {McLeod}, {Pogge}, {Relles}, {Stockdale}, {Tan}, {Trueblood}, \& {Trueblood}}]{Gaudi2017}
{Gaudi}, B.~S., {Stassun}, K.~G., {Collins}, K.~A., {et~al.} 2017, \bibinfo{title}{{A giant planet undergoing extreme-ultraviolet irradiation by its hot massive-star host},} \nat, 546, 514, \dodoi{10.1038/nature22392}

\bibitem[{A. {Gelman} \& D.~B. {Rubin}(1992){Gelman} \& {Rubin}}]{GelmanRubin1992}
{Gelman}, A., \& {Rubin}, D.~B. 1992, \bibinfo{title}{{Inference from Iterative Simulation Using Multiple Sequences},} Statistical Science, 7, 457, \dodoi{10.1214/ss/1177011136}

\bibitem[{M. {Gillon} {et~al.}(2014){Gillon}, {Anderson}, {Collier-Cameron}, \& et~al.}]{GillonAnderson2014}
{Gillon}, M., {Anderson}, D.~R., {Collier-Cameron}, A., \& et~al. 2014, \bibinfo{title}{{WASP-103 b: a new planet at the edge of tidal disruption},} Astronomy and Astrophysics, 562, L3, \dodoi{10.1051/0004-6361/201323014}

\bibitem[{M. {Gully-Santiago} {et~al.}(2024){Gully-Santiago}, {Morley}, {Luna}, \& et~al.}]{Gully-SantiagoMorley2024}
{Gully-Santiago}, M., {Morley}, C.~V., {Luna}, J., \& et~al. 2024, \bibinfo{title}{{A Large and Variable Leading Tail of Helium in a Hot Saturn Undergoing Runaway Inflation},} The Astronomical Journal, 167, 142, \dodoi{10.3847/1538-3881/ad1ee8}

\bibitem[{H.~M. {G{\"u}nther} {et~al.}(2022){G{\"u}nther}, {Melis}, {Robrade}, {Schneider}, {Wolk}, \& {Yadav}}]{Guenther2022}
{G{\"u}nther}, H.~M., {Melis}, C., {Robrade}, J., {et~al.} 2022, \bibinfo{title}{{Coronal and Chromospheric Emission in A-type Stars},} \aj, 164, 8, \dodoi{10.3847/1538-3881/ac6ef6}

\bibitem[{C.~A. {Haswell} {et~al.}(2012){Haswell}, {Fossati}, {Ayres}, \& et~al.}]{HaswellFossati2012}
{Haswell}, C.~A., {Fossati}, L., {Ayres}, T., \& et~al. 2012, \bibinfo{title}{{Near-ultraviolet Absorption, Chromospheric Activity, and Star-Planet Interactions in the WASP-12 system},} The Astrophysical Journal, 760, 79, \dodoi{10.1088/0004-637X/760/1/79}

\bibitem[{K.~L. {Hay} {et~al.}(2016){Hay}, {Collier-Cameron}, {Doyle}, \& et~al.}]{HayCollier-Cameron2016}
{Hay}, K.~L., {Collier-Cameron}, A., {Doyle}, A.~P., \& et~al. 2016, \bibinfo{title}{{WASP-92b, WASP-93b and WASP-118b: three new transiting close-in giant planets},} Monthly Notices of the Royal Astronomical Society, 463, 3276, \dodoi{10.1093/mnras/stw2090}

\bibitem[{L. {Hebb} {et~al.}(2009){Hebb}, {Collier-Cameron}, {Loeillet}, \& et~al.}]{HebbCollier-Cameron2009}
{Hebb}, L., {Collier-Cameron}, A., {Loeillet}, B., \& et~al. 2009, \bibinfo{title}{{WASP-12b: The Hottest Transiting Extrasolar Planet Yet Discovered},} The Astrophysical Journal, 693, 1920, \dodoi{10.1088/0004-637X/693/2/1920}

\bibitem[{H.~J. {Hoeijmakers} {et~al.}(2019){Hoeijmakers}, {Ehrenreich}, {Kitzmann}, {Allart}, {Grimm}, {Seidel}, {Wyttenbach}, {Pino}, {Nielsen}, {Fisher}, {Rimmer}, {Bourrier}, {Cegla}, {Lavie}, {Lovis}, {Patzer}, {Stock}, {Pepe}, \& {Heng}}]{Hoeijmakers2019}
{Hoeijmakers}, H.~J., {Ehrenreich}, D., {Kitzmann}, D., {et~al.} 2019, \bibinfo{title}{{A spectral survey of an ultra-hot Jupiter. Detection of metals in the transmission spectrum of KELT-9 b},} \aap, 627, A165, \dodoi{10.1051/0004-6361/201935089}

\bibitem[{M.~D. {Hoffman} \& A. {Gelman}(2011){Hoffman} \& {Gelman}}]{HoffmanGelman2011}
{Hoffman}, M.~D., \& {Gelman}, A. 2011, \bibinfo{title}{{The No-U-Turn Sampler: Adaptively Setting Path Lengths in Hamiltonian Monte Carlo},} arXiv e-prints, arXiv:1111.4246, \dodoi{10.48550/arXiv.1111.4246}

\bibitem[{T.~O. {Husser} {et~al.}(2013){Husser}, {Wende-von Berg}, {Dreizler}, {Homeier}, {Reiners}, {Barman}, \& {Hauschildt}}]{Husser2013}
{Husser}, T.~O., {Wende-von Berg}, S., {Dreizler}, S., {et~al.} 2013, \bibinfo{title}{{A new extensive library of PHOENIX stellar atmospheres and synthetic spectra},} \aap, 553, A6, \dodoi{10.1051/0004-6361/201219058}

\bibitem[{B. {Jackson} {et~al.}(2017){Jackson}, {Arras}, {Penev}, {Peacock}, \& {Marchant}}]{Jackson2017}
{Jackson}, B., {Arras}, P., {Penev}, K., {Peacock}, S., \& {Marchant}, P. 2017, \bibinfo{title}{{A New Model of Roche Lobe Overflow for Short-period Gaseous Planets and Binary Stars},} \apj, 835, 145, \dodoi{10.3847/1538-4357/835/2/145}

\bibitem[{A.~G. {Jensen} {et~al.}(2018){Jensen}, {Cauley}, {Redfield}, {Cochran}, \& {Endl}}]{JensenCauley2018}
{Jensen}, A.~G., {Cauley}, P.~W., {Redfield}, S., {Cochran}, W.~D., \& {Endl}, M. 2018, \bibinfo{title}{{Hydrogen and Sodium Absorption in the Optical Transmission Spectrum of WASP-12b},} The Astronomical Journal, 156, 154, \dodoi{10.3847/1538-3881/aadca7}

\bibitem[{A.~G. {Jensen} {et~al.}(2012){Jensen}, {Redfield}, {Endl}, {Cochran}, {Koesterke}, \& {Barman}}]{Jensen2012}
{Jensen}, A.~G., {Redfield}, S., {Endl}, M., {et~al.} 2012, \bibinfo{title}{{A Detection of H{\ensuremath{\alpha}} in an Exoplanetary Exosphere},} \apj, 751, 86, \dodoi{10.1088/0004-637X/751/2/86}

\bibitem[{E.~M.~R. {Kempton} \& H.~A. {Knutson}(2024){Kempton} \& {Knutson}}]{Kempton2024}
{Kempton}, E. M.~R., \& {Knutson}, H.~A. 2024, \bibinfo{title}{{Transiting Exoplanet Atmospheres in the Era of JWST},} Reviews in Mineralogy and Geochemistry, 90, 411, \dodoi{10.2138/rmg.2024.90.12}

\bibitem[{G.~W. {King} {et~al.}(2018){King}, {Wheatley}, {Salz}, {Bourrier}, {Czesla}, {Ehrenreich}, {Kirk}, {Lecavelier des Etangs}, {Louden}, {Schmitt}, \& {Schneider}}]{GeorgeKing2018}
{King}, G.~W., {Wheatley}, P.~J., {Salz}, M., {et~al.} 2018, \bibinfo{title}{{The XUV environments of exoplanets from Jupiter-size to super-Earth},} \mnras, 478, 1193, \dodoi{10.1093/mnras/sty1110}

\bibitem[{D.~M. {Kipping}(2013){Kipping}}]{Kipping2013}
{Kipping}, D.~M. 2013, \bibinfo{title}{{Efficient, uninformative sampling of limb darkening coefficients for two-parameter laws},} Monthly Notices of the Royal Astronomical Society, 435, 2152, \dodoi{10.1093/mnras/stt1435}

\bibitem[{A. {Kokori} {et~al.}(2023){Kokori}, {Tsiaras}, {Edwards}, \& et~al.}]{KokoriTsiaras2023}
{Kokori}, A., {Tsiaras}, A., {Edwards}, B., \& et~al. 2023, \bibinfo{title}{{ExoClock Project. III. 450 New Exoplanet Ephemerides from Ground and Space Observations},} The Astrophysical Journal Supplement Series, 265, 4, \dodoi{10.3847/1538-4365/ac9da4}

\bibitem[{T.~T. {Koskinen} {et~al.}(2022){Koskinen}, {Lavvas}, {Huang}, {Bergsten}, {Fernandes}, \& {Young}}]{Koskinen2022}
{Koskinen}, T.~T., {Lavvas}, P., {Huang}, C., {et~al.} 2022, \bibinfo{title}{{Mass Loss by Atmospheric Escape from Extremely Close-in Planets},} \apj, 929, 52, \dodoi{10.3847/1538-4357/ac4f45}

\bibitem[{R.~P. {Kraft}(1967){Kraft}}]{Kraft1967}
{Kraft}, R.~P. 1967, \bibinfo{title}{{Studies of Stellar Rotation. V. The Dependence of Rotation on Age among Solar-Type Stars},} The Astrophysical Journal, 150, 551, \dodoi{10.1086/149359}

\bibitem[{R.~P. {Kraft} {et~al.}(1991){Kraft}, {Burrows}, \& {Nousek}}]{Kraft1991}
{Kraft}, R.~P., {Burrows}, D.~N., \& {Nousek}, J.~A. 1991, \bibinfo{title}{{Determination of Confidence Limits for Experiments with Low Numbers of Counts},} \apj, 374, 344, \dodoi{10.1086/170124}

\bibitem[{L. {Kreidberg} \& A. {Oklop{\v{c}}i{\'c}}(2018){Kreidberg} \& {Oklop{\v{c}}i{\'c}}}]{KreidbergOklopcic2018}
{Kreidberg}, L., \& {Oklop{\v{c}}i{\'c}}, A. 2018, \bibinfo{title}{{Non-detection of a Helium Exosphere for the Hot Jupiter WASP-12b},} Research Notes of the American Astronomical Society, 2, 44, \dodoi{10.3847/2515-5172/aac887}

\bibitem[{D. {Lai} {et~al.}(2010){Lai}, {Helling}, \& {van den Heuvel}}]{LaiHelling2010}
{Lai}, D., {Helling}, C., \& {van den Heuvel}, E.~P.~J. 2010, \bibinfo{title}{{Mass Transfer, Transiting Stream, and Magnetopause in Close-in Exoplanetary Systems with Applications to WASP-12},} The Astrophysical Journal, 721, 923, \dodoi{10.1088/0004-637X/721/2/923}

\bibitem[{D.~W. {Latham} {et~al.}(2009){Latham}, {Bakos}, {Torres}, \& et~al.}]{LathamBakos2009}
{Latham}, D.~W., {Bakos}, G.~{\'A}., {Torres}, G., \& et~al. 2009, \bibinfo{title}{{Discovery of a Transiting Planet and Eight Eclipsing Binaries in HATNet Field G205},} The Astrophysical Journal, 704, 1107, \dodoi{10.1088/0004-637X/704/2/1107}

\bibitem[{P. {Leonardi} {et~al.}(2024){Leonardi}, {Nascimbeni}, {Granata}, \& et~al.}]{LeonardiNascimbeni2024}
{Leonardi}, P., {Nascimbeni}, V., {Granata}, V., \& et~al. 2024, \bibinfo{title}{{TASTE. V. A new ground-based investigation of orbital decay in the ultra-hot Jupiter WASP-12b},} Astronomy and Astrophysics, 686, A84, \dodoi{10.1051/0004-6361/202348363}

\bibitem[{W.~G. {Levine} {et~al.}(2024){Levine}, {Vissapragada}, {Feinstein}, \& et~al.}]{LevineVissapragada2024}
{Levine}, W.~G., {Vissapragada}, S., {Feinstein}, A.~D., \& et~al. 2024, \bibinfo{title}{{Exoplanet Aeronomy: A Case Study of WASP-69 b's Variable Thermosphere},} The Astronomical Journal, 168, 65, \dodoi{10.3847/1538-3881/ad5354}

\bibitem[{D. {Linssen} {et~al.}(2025){Linssen}, {Oklop{\v{c}}i{\'c}}, \& {MacLeod}}]{LinssenOklopcic2025}
{Linssen}, D., {Oklop{\v{c}}i{\'c}}, A., \& {MacLeod}, M. 2025, \bibinfo{title}{{sunset: A database of synthetic atmospheric-escape transmission spectra for nearly every transiting planet},} Astronomy and Astrophysics, 698, A112, \dodoi{10.1051/0004-6361/202452431}

\bibitem[{D. {Linssen} {et~al.}(2024){Linssen}, {Shih}, {MacLeod}, \& {Oklop{\v{c}}i{\'c}}}]{LinssenShih2024}
{Linssen}, D., {Shih}, J., {MacLeod}, M., \& {Oklop{\v{c}}i{\'c}}, A. 2024, \bibinfo{title}{{The open-source sunbather code: Modeling escaping planetary atmospheres and their transit spectra},} Astronomy and Astrophysics, 688, A43, \dodoi{10.1051/0004-6361/202450240}

\bibitem[{D.~C. {Linssen} {et~al.}(2022){Linssen}, {Oklop{\v{c}}i{\'c}}, \& {MacLeod}}]{LinssenOklopcic2022}
{Linssen}, D.~C., {Oklop{\v{c}}i{\'c}}, A., \& {MacLeod}, M. 2022, \bibinfo{title}{{Constraining planetary mass-loss rates by simulating Parker wind profiles with Cloudy},} Astronomy and Astrophysics, 667, A54, \dodoi{10.1051/0004-6361/202243830}

\bibitem[{R. {Luger} {et~al.}(2019){Luger}, {Agol}, {Foreman-Mackey}, \& et~al.}]{LugerAgol2019}
{Luger}, R., {Agol}, E., {Foreman-Mackey}, D., \& et~al. 2019, \bibinfo{title}{{starry: Analytic Occultation Light Curves},} The Astronomical Journal, 157, 64, \dodoi{10.3847/1538-3881/aae8e5}

\bibitem[{L. {Ma} \& J. {Fuller}(2021){Ma} \& {Fuller}}]{MaFuller2021}
{Ma}, L., \& {Fuller}, J. 2021, \bibinfo{title}{{Orbital Decay of Short-period Exoplanets via Tidal Resonance Locking},} The Astrophysical Journal, 918, 16, \dodoi{10.3847/1538-4357/ac088e}

\bibitem[{G. {Maciejewski} {et~al.}(2016){Maciejewski}, {Dimitrov}, {Fern{\'a}ndez}, \& et~al.}]{MaciejewskiDimitrov2016}
{Maciejewski}, G., {Dimitrov}, D., {Fern{\'a}ndez}, M., \& et~al. 2016, \bibinfo{title}{{Departure from the constant-period ephemeris for the transiting exoplanet WASP-12},} Astronomy and Astrophysics, 588, L6, \dodoi{10.1051/0004-6361/201628312}

\bibitem[{M. {MacLeod} {et~al.}(2025){MacLeod}, {Oklop{\v{c}}i{\'c}}, {Nail}, \& {Linssen}}]{MacLeodOklopcic2025}
{MacLeod}, M., {Oklop{\v{c}}i{\'c}}, A., {Nail}, F., \& {Linssen}, D. 2025, \bibinfo{title}{{Streams and Bubbles: Tidal Shaping of Hydrodynamic Planetary Outflows},} The Astrophysical Journal, 988, 63, \dodoi{10.3847/1538-4357/ade0b7}

\bibitem[{C. {Moutou} {et~al.}(2011){Moutou}, {D{\'\i}az}, {Udry}, \& et~al.}]{MoutouDiaz2011}
{Moutou}, C., {D{\'\i}az}, R.~F., {Udry}, S., \& et~al. 2011, \bibinfo{title}{{Spin-orbit inclinations of the exoplanetary systems HAT-P-8b, HAT-P-9b, HAT-P-16b, and HAT-P-23b},} Astronomy and Astrophysics, 533, A113, \dodoi{10.1051/0004-6361/201116760}

\bibitem[{A. {Moya} {et~al.}(2011){Moya}, {Bouy}, {Marchis}, {Vicente}, \& {Barrado}}]{Moya2011}
{Moya}, A., {Bouy}, H., {Marchis}, F., {Vicente}, B., \& {Barrado}, D. 2011, \bibinfo{title}{{High spatial resolution imaging of the star with a transiting planet WASP-33},} \aap, 535, A110, \dodoi{10.1051/0004-6361/201116889}

\bibitem[{M. {Mugrauer}(2019){Mugrauer}}]{Mugrauer2019}
{Mugrauer}, M. 2019, \bibinfo{title}{{Search for stellar companions of exoplanet host stars by exploring the second ESA-Gaia data release},} \mnras, 490, 5088, \dodoi{10.1093/mnras/stz2673}

\bibitem[{S. {Mukherjee} {et~al.}(2025){Mukherjee}, {Sing}, {Fu}, {Stevenson}, {Schmidt}, {Baskett}, {McCreery}, {Allen}, {Bennett}, {Christie}, {Gasc{\'o}n}, {Goyal}, {H{\'e}brard}, {Lothringer}, {L{\'o}pez-Morales}, {Lustig-Yaeger}, {May}, {Mayorga}, {Mayne}, {Ramos Rosado}, {Reggiani}, {Rustamkulov}, {Schlaufman}, {Sotzen}, {Thorngren}, {Wang}, \& {Zamyatina}}]{Mukherjee2025}
{Mukherjee}, S., {Sing}, D.~K., {Fu}, G., {et~al.} 2025, \bibinfo{title}{{Cloudy mornings and clear evenings on a giant extrasolar world},} arXiv e-prints, arXiv:2505.10910, \dodoi{10.48550/arXiv.2505.10910}

\bibitem[{ {NASA Exoplanet Archive}(2024){NASA Exoplanet Archive}}]{ps}
{NASA Exoplanet Archive}. 2024, \bibinfo{title}{Planetary Systems,}, Version: 2024-01-23 12:42 NExScI-Caltech/IPAC, \dodoi{10.26133/NEA12}

\bibitem[{M. {Neveu-VanMalle} {et~al.}(2014){Neveu-VanMalle}, {Queloz}, {Anderson}, {Charbonnel}, {Collier Cameron}, {Delrez}, {Gillon}, {Hellier}, {Jehin}, {Lendl}, {Maxted}, {Pepe}, {Pollacco}, {S{\'e}gransan}, {Smalley}, {Smith}, {Southworth}, {Triaud}, {Udry}, \& {West}}]{NeveuVanMalle2014}
{Neveu-VanMalle}, M., {Queloz}, D., {Anderson}, D.~R., {et~al.} 2014, \bibinfo{title}{{WASP-94 A and B planets: hot-Jupiter cousins in a twin-star system},} \aap, 572, A49, \dodoi{10.1051/0004-6361/201424744}

\bibitem[{H. {Ngo} {et~al.}(2015){Ngo}, {Knutson}, {Hinkley}, {Crepp}, {Bechter}, {Batygin}, {Howard}, {Johnson}, {Morton}, \& {Muirhead}}]{Ngo2015}
{Ngo}, H., {Knutson}, H.~A., {Hinkley}, S., {et~al.} 2015, \bibinfo{title}{{Friends of Hot Jupiters. II. No Correspondence between Hot-jupiter Spin-Orbit Misalignment and the Incidence of Directly Imaged Stellar Companions},} \apj, 800, 138, \dodoi{10.1088/0004-637X/800/2/138}

\bibitem[{H. {Ngo} {et~al.}(2016){Ngo}, {Knutson}, {Hinkley}, {Bryan}, {Crepp}, {Batygin}, {Crossfield}, {Hansen}, {Howard}, {Johnson}, {Mawet}, {Morton}, {Muirhead}, \& {Wang}}]{Ngo2016}
{Ngo}, H., {Knutson}, H.~A., {Hinkley}, S., {et~al.} 2016, \bibinfo{title}{{Friends of Hot Jupiters. IV. Stellar Companions Beyond 50 au Might Facilitate Giant Planet Formation, but Most are Unlikely to Cause Kozai-Lidov Migration},} \apj, 827, 8, \dodoi{10.3847/0004-637X/827/1/8}

\bibitem[{J.~D. {Nichols} {et~al.}(2015){Nichols}, {Wynn}, {Goad}, \& et~al.}]{NicholsWynn2015}
{Nichols}, J.~D., {Wynn}, G.~A., {Goad}, M., \& et~al. 2015, \bibinfo{title}{{Hubble Space Telescope Observations of the NUV Transit of WASP-12b},} The Astrophysical Journal, 803, 9, \dodoi{10.1088/0004-637X/803/1/9}

\bibitem[{A. {Oklop{\v{c}}i{\'c}}(2019){Oklop{\v{c}}i{\'c}}}]{Oklopcic2019}
{Oklop{\v{c}}i{\'c}}, A. 2019, \bibinfo{title}{{Helium Absorption at 1083 nm from Extended Exoplanet Atmospheres: Dependence on Stellar Radiation},} The Astrophysical Journal, 881, 133, \dodoi{10.3847/1538-4357/ab2f7f}

\bibitem[{A. {Oklop{\v{c}}i{\'c}} \& C.~M. {Hirata}(2018){Oklop{\v{c}}i{\'c}} \& {Hirata}}]{OklopcicHirata2018}
{Oklop{\v{c}}i{\'c}}, A., \& {Hirata}, C.~M. 2018, \bibinfo{title}{{A New Window into Escaping Exoplanet Atmospheres: 10830 {\r{A}} Line of Helium},} The Astrophysical Journal, 855, L11, \dodoi{10.3847/2041-8213/aaada9}

\bibitem[{J. {Orell-Miquel} {et~al.}(2025){Orell-Miquel}, {Sampson}, {Morley}, {Cochran}, {Duvvuri}, {Krolikowski}, {Mahadevan}, \& {Tran}}]{OrellMiquel2025}
{Orell-Miquel}, J., {Sampson}, K., {Morley}, C.~V., {et~al.} 2025, \bibinfo{title}{{HET/HPF observations of Helium in warm, hot, and ultra-hot Jupiters},} arXiv e-prints, arXiv:2509.06847.
\newblock \doarXiv{2509.06847}

\bibitem[{J.~E. {Owen}(2019){Owen}}]{Owen2019}
{Owen}, J.~E. 2019, \bibinfo{title}{{Atmospheric Escape and the Evolution of Close-In Exoplanets},} Annual Review of Earth and Planetary Sciences, 47, 67, \dodoi{10.1146/annurev-earth-053018-060246}

\bibitem[{J.~E. {Owen} \& D. {Lai}(2018){Owen} \& {Lai}}]{OwenLai2018}
{Owen}, J.~E., \& {Lai}, D. 2018, \bibinfo{title}{{Photoevaporation and high-eccentricity migration created the sub-Jovian desert},} Monthly Notices of the Royal Astronomical Society, 479, 5012, \dodoi{10.1093/mnras/sty1760}

\bibitem[{A. {Pai Asnodkar} {et~al.}(2024){Pai Asnodkar}, {Wang}, {Broome}, {Huang}, {Johnson}, {Ilyin}, {Strassmeier}, \& {Jensen}}]{PaiAsnodkar2024}
{Pai Asnodkar}, A., {Wang}, J., {Broome}, M., {et~al.} 2024, \bibinfo{title}{{PEPSI's non-detection of escaping hydrogen and metal lines adds to the enigma of WASP-12 b},} \mnras, 535, 1829, \dodoi{10.1093/mnras/stae2441}

\bibitem[{K. {Paragas} {et~al.}(2021){Paragas}, {Vissapragada}, {Knutson}, \& et~al.}]{ParagasVissapragada2021}
{Paragas}, K., {Vissapragada}, S., {Knutson}, H.~A., \& et~al. 2021, \bibinfo{title}{{Metastable Helium Reveals an Extended Atmosphere for the Gas Giant HAT-P-18b},} The Astrophysical Journal, 909, L10, \dodoi{10.3847/2041-8213/abe706}

\bibitem[{H. {Parviainen} \& S. {Aigrain}(2015){Parviainen} \& {Aigrain}}]{ParviainenAigrain2015}
{Parviainen}, H., \& {Aigrain}, S. 2015, \bibinfo{title}{{LDTK: Limb Darkening Toolkit},} Monthly Notices of the Royal Astronomical Society, 453, 3821, \dodoi{10.1093/mnras/stv1857}

\bibitem[{J.~A. {Patel} \& N. {Espinoza}(2022){Patel} \& {Espinoza}}]{PatelEspinoza2022}
{Patel}, J.~A., \& {Espinoza}, N. 2022, \bibinfo{title}{{Empirical Limb-darkening Coefficients and Transit Parameters of Known Exoplanets from TESS},} The Astronomical Journal, 163, 228, \dodoi{10.3847/1538-3881/ac5f55}

\bibitem[{K.~C. {Patra} {et~al.}(2017){Patra}, {Winn}, {Holman}, \& et~al.}]{PatraWinn2017}
{Patra}, K.~C., {Winn}, J.~N., {Holman}, M.~J., \& et~al. 2017, \bibinfo{title}{{The Apparently Decaying Orbit of WASP-12b},} The Astronomical Journal, 154, 4, \dodoi{10.3847/1538-3881/aa6d75}

\bibitem[{M.~J. {Pecaut} \& E.~E. {Mamajek}(2013){Pecaut} \& {Mamajek}}]{Pecaut2013}
{Pecaut}, M.~J., \& {Mamajek}, E.~E. 2013, \bibinfo{title}{{Intrinsic Colors, Temperatures, and Bolometric Corrections of Pre-main-sequence Stars},} \apjs, 208, 9, \dodoi{10.1088/0067-0049/208/1/9}

\bibitem[{J. {P{\'e}rez-Gonz{\'a}lez} {et~al.}(2024){P{\'e}rez-Gonz{\'a}lez}, {Greklek-McKeon}, {Vissapragada}, \& et~al.}]{Perez-GonzalezGreklek-McKeon2024}
{P{\'e}rez-Gonz{\'a}lez}, J., {Greklek-McKeon}, M., {Vissapragada}, S., \& et~al. 2024, \bibinfo{title}{{Detection of an Atmospheric Outflow from the Young Hot Saturn TOI-1268b},} The Astronomical Journal, 167, 214, \dodoi{10.3847/1538-3881/ad34b6}

\bibitem[{C. {Petrovich}(2015){Petrovich}}]{Petrovich2015}
{Petrovich}, C. 2015, \bibinfo{title}{{Steady-state Planet Migration by the Kozai-Lidov Mechanism in Stellar Binaries},} The Astrophysical Journal, 799, 27, \dodoi{10.1088/0004-637X/799/1/27}

\bibitem[{J.~G. {Rogers} {et~al.}(2024){Rogers}, {Owen}, \& {Schlichting}}]{Rogers2024}
{Rogers}, J.~G., {Owen}, J.~E., \& {Schlichting}, H.~E. 2024, \bibinfo{title}{{Under the light of a new star: evolution of planetary atmospheres through protoplanetary disc dispersal and boil-off},} \mnras, 529, 2716, \dodoi{10.1093/mnras/stae563}

\bibitem[{A. {Ruiz} {et~al.}(2022){Ruiz}, {Georgakakis}, {Gerakakis}, {Saxton}, {Kretschmar}, {Akylas}, \& {Georgantopoulos}}]{Ruiz2022}
{Ruiz}, A., {Georgakakis}, A., {Gerakakis}, S., {et~al.} 2022, \bibinfo{title}{{The RapidXMM upper limit server: X-ray aperture photometry of the XMM-Newton archival observations},} \mnras, 511, 4265, \dodoi{10.1093/mnras/stac272}

\bibitem[{M. {Saidel} {et~al.}(2025){Saidel}, {Vissapragada}, {Spake}, \& et~al.}]{SaidelVissapragada2025}
{Saidel}, M., {Vissapragada}, S., {Spake}, J., \& et~al. 2025, \bibinfo{title}{{Atmospheric Mass Loss from TOI-1259 A b, a Gas Giant Planet with a White Dwarf Companion},} The Astronomical Journal, 169, 104, \dodoi{10.3847/1538-3881/ada49f}

\bibitem[{J. {Salvatier} {et~al.}(2016){Salvatier}, {Wiecki}, \& {Fonnesbeck}}]{SalvatierWiecki2016}
{Salvatier}, J., {Wiecki}, T.~V., \& {Fonnesbeck}, C. 2016, \bibinfo{title}{{PyMC3: Python probabilistic programming framework},}, Astrophysics Source Code Library, record ascl:1610.016

\bibitem[{M. {Salz} {et~al.}(2016){Salz}, {Czesla}, {Schneider}, \& {Schmitt}}]{SalzCzesla2016}
{Salz}, M., {Czesla}, S., {Schneider}, P.~C., \& {Schmitt}, J.~H.~M.~M. 2016, \bibinfo{title}{{Simulating the escaping atmospheres of hot gas planets in the solar neighborhood},} Astronomy and Astrophysics, 586, A75, \dodoi{10.1051/0004-6361/201526109}

\bibitem[{J. {Sanz-Forcada} {et~al.}(2025){Sanz-Forcada}, {L{\'o}pez-Puertas}, {Lamp{\'o}n}, {Czesla}, {Nortmann}, {Caballero}, {Zapatero Osorio}, {Amado}, {Murgas}, {Orell-Miquel}, {Pall{\'e}}, {Quirrenbach}, {Reiners}, {Ribas}, {S{\'a}nchez-L{\'o}pez}, \& {Solano}}]{SanzForcada2025}
{Sanz-Forcada}, J., {L{\'o}pez-Puertas}, M., {Lamp{\'o}n}, M., {et~al.} 2025, \bibinfo{title}{{Connection between planetary He I {\ensuremath{\lambda}}10 830 {\r{A}} absorption and extreme-ultraviolet emission of planet-host stars},} \aap, 693, A285, \dodoi{10.1051/0004-6361/202451680}

\bibitem[{W. {Schmidt} {et~al.}(1994){Schmidt}, {Knoelker}, \& {Westendorp Plaza}}]{SchmidtKnoelker1994}
{Schmidt}, W., {Knoelker}, M., \& {Westendorp Plaza}, C. 1994, \bibinfo{title}{{Limb observations of the HeI 1083.0 NM line.},} Astronomy and Astrophysics, 287, 229

\bibitem[{E. {Schreyer} \& R. {Murray-Clay}(2025,){Schreyer} \& {Murray-Clay}}]{Schreyer2025_submitted}
{Schreyer}, E., \& {Murray-Clay}, R. 2025, submitted to the Monthly Notices of the Royal Astronomical Society

\bibitem[{M. {Schulik} \& J.~E. {Owen}(2025){Schulik} \& {Owen}}]{SchulikOwen2025}
{Schulik}, M., \& {Owen}, J.~E. 2025, \bibinfo{title}{{Using the helium triplet as a tracer of the physics of giant planet outflows},} Monthly Notices of the Royal Astronomical Society, 542, 927, \dodoi{10.1093/mnras/staf775}

\bibitem[{G. {Schwarz}(1978){Schwarz}}]{Schwarz1978}
{Schwarz}, G. 1978, \bibinfo{title}{{Estimating the Dimension of a Model},} Annals of Statistics, 6, 461

\bibitem[{T. {Shimura} {et~al.}(2025){Shimura}, {Mitsuishi}, {Kunitomo}, \& et~al.}]{ShimuraMitsuishi2025}
{Shimura}, T., {Mitsuishi}, I., {Kunitomo}, M., \& et~al. 2025, \bibinfo{title}{{X-ray emission of F-type stars and its analogy with G-type stars},} arXiv e-prints, arXiv:2509.18677, \dodoi{10.48550/arXiv.2509.18677}

\bibitem[{T. {Simon} {et~al.}(2002){Simon}, {Ayres}, {Redfield}, \& {Linsky}}]{Simon2002}
{Simon}, T., {Ayres}, T.~R., {Redfield}, S., \& {Linsky}, J.~L. 2002, \bibinfo{title}{{Limits on Chromospheres and Convection among the Main-Sequence A Stars},} \apj, 579, 800, \dodoi{10.1086/342941}

\bibitem[{R.~K. {Smith} {et~al.}(2001){Smith}, {Brickhouse}, {Liedahl}, \& {Raymond}}]{Smith2001}
{Smith}, R.~K., {Brickhouse}, N.~S., {Liedahl}, D.~A., \& {Raymond}, J.~C. 2001, \bibinfo{title}{{Collisional Plasma Models with APEC/APED: Emission-Line Diagnostics of Hydrogen-like and Helium-like Ions},} \apjl, 556, L91, \dodoi{10.1086/322992}

\bibitem[{J.~J. {Spake} {et~al.}(2021){Spake}, {Oklop{\v{c}}i{\'c}}, \& {Hillenbrand}}]{SpakeOklopcic2021}
{Spake}, J.~J., {Oklop{\v{c}}i{\'c}}, A., \& {Hillenbrand}, L.~A. 2021, \bibinfo{title}{{The Posttransit Tail of WASP-107b Observed at 10830 {\r{A}}},} The Astronomical Journal, 162, 284, \dodoi{10.3847/1538-3881/ac178a}

\bibitem[{K.~G. {Stassun} {et~al.}(2017){Stassun}, {Collins}, \& {Gaudi}}]{StassunCollins2017}
{Stassun}, K.~G., {Collins}, K.~A., \& {Gaudi}, B.~S. 2017, \bibinfo{title}{{Accurate Empirical Radii and Masses of Planets and Their Host Stars with Gaia Parallaxes},} The Astronomical Journal, 153, 136, \dodoi{10.3847/1538-3881/aa5df3}

\bibitem[{K.~G. {Stassun} {et~al.}(2019){Stassun}, {Oelkers}, {Paegert}, {Torres}, {Pepper}, {De Lee}, {Collins}, {Latham}, {Muirhead}, {Chittidi}, {Rojas-Ayala}, {Fleming}, {Rose}, {Tenenbaum}, {Ting}, {Kane}, {Barclay}, {Bean}, {Brassuer}, {Charbonneau}, {Ge}, {Lissauer}, {Mann}, {McLean}, {Mullally}, {Narita}, {Plavchan}, {Ricker}, {Sasselov}, {Seager}, {Sharma}, {Shiao}, {Sozzetti}, {Stello}, {Vanderspek}, {Wallace}, \& {Winn}}]{Stassun2019}
{Stassun}, K.~G., {Oelkers}, R.~J., {Paegert}, M., {et~al.} 2019, \bibinfo{title}{{The Revised TESS Input Catalog and Candidate Target List},} \aj, 158, 138, \dodoi{10.3847/1538-3881/ab3467}

\bibitem[{G. {Stefansson} {et~al.}(2017){Stefansson}, {Mahadevan}, {Hebb}, \& et~al.}]{StefanssonMahadevan2017}
{Stefansson}, G., {Mahadevan}, S., {Hebb}, L., \& et~al. 2017, \bibinfo{title}{{Toward Space-like Photometric Precision from the Ground with Beam-shaping Diffusers},} The Astrophysical Journal, 848, 9, \dodoi{10.3847/1538-4357/aa88aa}

\bibitem[{N.~I. {Storch} {et~al.}(2017){Storch}, {Lai}, \& {Anderson}}]{StorchLai2017}
{Storch}, N.~I., {Lai}, D., \& {Anderson}, K.~R. 2017, \bibinfo{title}{{Dynamics of stellar spin driven by planets undergoing Lidov-Kozai migration: paths to spin-orbit misalignment},} Monthly Notices of the Royal Astronomical Society, 465, 3927, \dodoi{10.1093/mnras/stw3018}

\bibitem[{H.~M. {Tabernero} {et~al.}(2022){Tabernero}, {Zapatero Osorio}, {Allende Prieto}, \& et~al.}]{TaberneroZapateroOsorio2022}
{Tabernero}, H.~M., {Zapatero Osorio}, M.~R., {Allende Prieto}, C., \& et~al. 2022, \bibinfo{title}{{HORuS transmission spectroscopy and revised planetary parameters of KELT-7 b},} Monthly Notices of the Royal Astronomical Society, 515, 1247, \dodoi{10.1093/mnras/stac1759}

\bibitem[{L.~Y. {Temple} {et~al.}(2019){Temple}, {Hellier}, {Anderson}, \& et~al.}]{TempleHellier2019}
{Temple}, L.~Y., {Hellier}, C., {Anderson}, D.~R., \& et~al. 2019, \bibinfo{title}{{WASP-180Ab: Doppler tomography of a hot Jupiter orbiting the primary star in a visual binary},} Monthly Notices of the Royal Astronomical Society, 490, 2467, \dodoi{10.1093/mnras/stz2632}

\bibitem[{J.~D. {Turner} {et~al.}(2016){Turner}, {Pearson}, {Biddle}, {Smart}, {Zellem}, {Teske}, {Hardegree-Ullman}, {Griffith}, {Leiter}, {Cates}, {Nieberding}, {Smith}, {Thompson}, {Hofmann}, {Berube}, {Nguyen}, {Small}, {Guvenen}, {Richardson}, {McGraw}, {Raphael}, {Crawford}, {Robertson}, {Tombleson}, {Carleton}, {Towner}, {Walker-LaFollette}, {Hume}, {Watson}, {Jones}, {Lichtenberger}, {Hoglund}, {Cook}, {Crossen}, {Jorgensen}, {Romine}, {Thompson}, {Villegas}, {Wilson}, {Sanford}, {Taylor}, \& {Henz}}]{Turner2016}
{Turner}, J.~D., {Pearson}, K.~A., {Biddle}, L.~I., {et~al.} 2016, \bibinfo{title}{{Ground-based near-UV observations of 15 transiting exoplanets: constraints on their atmospheres and no evidence for asymmetrical transits},} \mnras, 459, 789, \dodoi{10.1093/mnras/stw574}

\bibitem[{S. {Vissapragada} {et~al.}(2024){Vissapragada}, {Greklek-McKeon}, {Linssen}, \& et~al.}]{VissapragadaGreklek-McKeon2024}
{Vissapragada}, S., {Greklek-McKeon}, M., {Linssen}, D., \& et~al. 2024, \bibinfo{title}{{Helium in the Extended Atmosphere of the Warm Superpuff TOI-1420b},} The Astronomical Journal, 167, 199, \dodoi{10.3847/1538-3881/ad3241}

\bibitem[{S. {Vissapragada} {et~al.}(2020{\natexlab{a}}){Vissapragada}, {Jontof-Hutter}, {Shporer}, \& et~al.}]{VissapragadaJontof-Hutter2020}
{Vissapragada}, S., {Jontof-Hutter}, D., {Shporer}, A., \& et~al. 2020{\natexlab{a}}, \bibinfo{title}{{Diffuser-assisted Infrared Transit Photometry for Four Dynamically Interacting Kepler Systems},} The Astronomical Journal, 159, 108, \dodoi{10.3847/1538-3881/ab65c8}

\bibitem[{S. {Vissapragada} {et~al.}(2022){Vissapragada}, {Knutson}, {Greklek-McKeon}, \& et~al.}]{VissapragadaKnutson2022}
{Vissapragada}, S., {Knutson}, H.~A., {Greklek-McKeon}, M., \& et~al. 2022, \bibinfo{title}{{The Upper Edge of the Neptune Desert Is Stable Against Photoevaporation},} The Astronomical Journal, 164, 234, \dodoi{10.3847/1538-3881/ac92f2}

\bibitem[{S. {Vissapragada} {et~al.}(2020{\natexlab{b}}){Vissapragada}, {Knutson}, {Jovanovic}, \& et~al.}]{VissapragadaKnutson2020}
{Vissapragada}, S., {Knutson}, H.~A., {Jovanovic}, N., \& et~al. 2020{\natexlab{b}}, \bibinfo{title}{{Constraints on Metastable Helium in the Atmospheres of WASP-69b and WASP-52b with Ultranarrowband Photometry},} The Astronomical Journal, 159, 278, \dodoi{10.3847/1538-3881/ab8e34}

\bibitem[{Y.~H. {Wang} {et~al.}(2019){Wang}, {Wang}, {Hinse}, \& et~al.}]{WangWang2019}
{Wang}, Y.~H., {Wang}, S., {Hinse}, T.~C., \& et~al. 2019, \bibinfo{title}{{Transiting Exoplanet Monitoring Project (TEMP). V. Transit Follow Up for HAT-P-9b, HAT-P-32b, and HAT-P-36b},} The Astronomical Journal, 157, 82, \dodoi{10.3847/1538-3881/aaf6b6}

\bibitem[{N.~N. {Weinberg} {et~al.}(2017){Weinberg}, {Sun}, {Arras}, \& {Essick}}]{WeinbergSun2017}
{Weinberg}, N.~N., {Sun}, M., {Arras}, P., \& {Essick}, R. 2017, \bibinfo{title}{{Tidal Dissipation in WASP-12},} The Astrophysical Journal, 849, L11, \dodoi{10.3847/2041-8213/aa9113}

\bibitem[{J. {Wilms} {et~al.}(2000){Wilms}, {Allen}, \& {McCray}}]{Wilms2000}
{Wilms}, J., {Allen}, A., \& {McCray}, R. 2000, \bibinfo{title}{{On the Absorption of X-Rays in the Interstellar Medium},} \apj, 542, 914, \dodoi{10.1086/317016}

\bibitem[{M. {W{\"o}llert} \& W. {Brandner}(2015){W{\"o}llert} \& {Brandner}}]{Wollert2015}
{W{\"o}llert}, M., \& {Brandner}, W. 2015, \bibinfo{title}{{A Lucky Imaging search for stellar sources near 74 transit hosts},} \aap, 579, A129, \dodoi{10.1051/0004-6361/201526525}

\bibitem[{A. {Wyttenbach} {et~al.}(2020){Wyttenbach}, {Molli{\`e}re}, {Ehrenreich}, {Cegla}, {Bourrier}, {Lovis}, {Pino}, {Allart}, {Seidel}, {Hoeijmakers}, {Nielsen}, {Lavie}, {Pepe}, {Bonfils}, \& {Snellen}}]{Wyttenbach2020}
{Wyttenbach}, A., {Molli{\`e}re}, P., {Ehrenreich}, D., {et~al.} 2020, \bibinfo{title}{{Mass-loss rate and local thermodynamic state of the KELT-9 b thermosphere from the hydrogen Balmer series},} \aap, 638, A87, \dodoi{10.1051/0004-6361/201937316}

\bibitem[{S. {Yal{\c{c}}{\i}nkaya} {et~al.}(2024){Yal{\c{c}}{\i}nkaya}, {Esmer}, {Ba{\c{s}}t{\"u}rk}, \& et~al.}]{YalcinkayaEsmer2024}
{Yal{\c{c}}{\i}nkaya}, S., {Esmer}, E.~M., {Ba{\c{s}}t{\"u}rk}, {\"O}., \& et~al. 2024, \bibinfo{title}{{Looking for timing variations in the transits of 16 exoplanets},} Monthly Notices of the Royal Astronomical Society, 530, 2475, \dodoi{10.1093/mnras/stae854}

\bibitem[{D. {Yan} {et~al.}(2021){Yan}, {Guo}, {Huang}, \& {Xing}}]{Yan2021}
{Yan}, D., {Guo}, J., {Huang}, C., \& {Xing}, L. 2021, \bibinfo{title}{{Atmosphere Escape Inferred from Modeling the H{\ensuremath{\alpha}} Transmission Spectrum of WASP-121b},} \apjl, 907, L47, \dodoi{10.3847/2041-8213/abda41}

\bibitem[{D. {Yan} {et~al.}(2024){Yan}, {Guo}, {Seon}, {L{\'o}pez-Puertas}, {Czesla}, \& {Lamp{\'o}n}}]{Yan2024}
{Yan}, D., {Guo}, J., {Seon}, K.-i., {et~al.} 2024, \bibinfo{title}{{A possibly solar metallicity atmosphere escaping from HAT-P-32b revealed by H{\ensuremath{\alpha}} and He absorption},} \aap, 686, A208, \dodoi{10.1051/0004-6361/202348210}

\bibitem[{D. {Yan} {et~al.}(2022){Yan}, {Seon}, {Guo}, {Chen}, \& {Li}}]{Yan2022}
{Yan}, D., {Seon}, K.-i., {Guo}, J., {Chen}, G., \& {Li}, L. 2022, \bibinfo{title}{{Modeling the H{\ensuremath{\alpha}} and He 10830 Transmission Spectrum of WASP-52b},} \apj, 936, 177, \dodoi{10.3847/1538-4357/ac8793}

\bibitem[{F. {Yan} \& T. {Henning}(2018){Yan} \& {Henning}}]{Yan2018}
{Yan}, F., \& {Henning}, T. 2018, \bibinfo{title}{{An extended hydrogen envelope of the extremely hot giant exoplanet KELT-9b},} Nature Astronomy, 2, 714, \dodoi{10.1038/s41550-018-0503-3}

\bibitem[{S.~W. {Yee} {et~al.}(2020){Yee}, {Winn}, {Knutson}, \& et~al.}]{YeeWinn2020}
{Yee}, S.~W., {Winn}, J.~N., {Knutson}, H.~A., \& et~al. 2020, \bibinfo{title}{{The Orbit of WASP-12b Is Decaying},} The Astrophysical Journal, 888, L5, \dodoi{10.3847/2041-8213/ab5c16}

\bibitem[{A. {Youngblood} {et~al.}(2017){Youngblood}, {France}, \& et~al.}]{YoungbloodFrance2017}
{Youngblood}, A., {France}, K., \& et~al. 2017, \bibinfo{title}{{The MUSCLES Treasury Survey. IV. Scaling Relations for Ultraviolet, Ca II K, and Energetic Particle Fluxes from M Dwarfs},} The Astrophysical Journal, 843, 31, \dodoi{10.3847/1538-4357/aa76dd}

\bibitem[{A. {Youngblood} {et~al.}(2025){Youngblood}, {France}, {Koskinen}, {Mason}, {Redfield}, {Wood}, {Bourrier}, {dos Santos}, {Johns-Krull}, {King}, {Linsky}, \& {Peacock}}]{Youngblood2025}
{Youngblood}, A., {France}, K., {Koskinen}, T., {et~al.} 2025, \bibinfo{title}{{Toward a 2D H I Map of the Local Interstellar Medium},} arXiv e-prints, arXiv:2509.08125, \dodoi{10.48550/arXiv.2509.08125}

\bibitem[{Z. {Zhang} {et~al.}(2023){Zhang}, {Morley}, {Gully-Santiago}, \& et~al.}]{ZhangMorley2023}
{Zhang}, Z., {Morley}, C.~V., {Gully-Santiago}, M., \& et~al. 2023, \bibinfo{title}{{Giant tidal tails of helium escaping the hot Jupiter HAT-P-32 b},} Science Advances, 9, eadf8736, \dodoi{10.1126/sciadv.adf8736}

\bibitem[{G. {Zhou} {et~al.}(2017){Zhou}, {Bakos}, {Hartman}, \& et~al.}]{ZhouBakos2017}
{Zhou}, G., {Bakos}, G.~{\'A}., {Hartman}, J.~D., \& et~al. 2017, \bibinfo{title}{{HAT-P-67b: An Extremely Low Density Saturn Transiting an F-subgiant Confirmed via Doppler Tomography},} The Astronomical Journal, 153, 211, \dodoi{10.3847/1538-3881/aa674a}

\bibitem[{G. {Zhou} {et~al.}(2016){Zhou}, {Latham}, {Bieryla}, \& et~al.}]{ZhouLatham2016}
{Zhou}, G., {Latham}, D.~W., {Bieryla}, A., \& et~al. 2016, \bibinfo{title}{{Spin-orbit alignment for KELT-7b and HAT-P-56b via Doppler tomography with TRES},} Monthly Notices of the Royal Astronomical Society, 460, 3376, \dodoi{10.1093/mnras/stw1107}

\end{thebibliography}
\bibliographystyle{aasjournalv7}

\end{document}